\documentclass[11pt, fleqn]{article}
\pdfoutput=1

\usepackage{geometry}
\usepackage{comment}

\usepackage{jheppub}
\usepackage{amsmath,amssymb,amsthm}
\usepackage{bm}
\usepackage{slashed}
\usepackage{graphicx}
\usepackage[T1]{fontenc}
\usepackage{fix-cm}

\makeatletter
\newcommand*\bigcdot{\mathpalette\bigcdot@{.5}}
\newcommand*\bigcdot@[2]{\mathbin{\vcenter{\hbox{\scalebox{#2}{$\m@th#1\bullet$}}}}}
\makeatother

\newlength{\dummysp}
\usepackage[font=scriptsize]{caption}
\usepackage{mwe}

\def\R{{\mathbb R}}
\def\S{{\mathbb S}}
\def\Z{{\mathbb Z}}

\title{String tensions in deformed Yang-Mills theory}
\author{Erich Poppitz and M. Erfan Shalchian T.}
\affiliation{Department of Physics, University of Toronto, 60 St. George St., Toronto, ON, M5S 1A7, Canada}
\emailAdd{poppitz@physics.utoronto.ca}
\emailAdd{eshalchian@physics.utoronto.ca, erfanshalchian@gmail.com}

\abstract{

{\flushleft{W}}e study  k-strings in deformed Yang-Mills (dYM)  with SU(N) gauge group in the semiclassically calculable regime on $\R^3 \times \S^1$. Their tensions T$_{\text{k}}$ are computed in two ways: numerically,    for $2$ $\le$ N $\le$ $10$, and via an analytic approach using a re-summed perturbative expansion. The latter serves both as a consistency check on the numerical results and as a tool to analytically study the large-N limit. We find that dYM k-string ratios T$_{\text{k}}$/T$_{\text{1}}$ do not obey the well-known sine- or Casimir-scaling laws. Instead, we show that the ratios T$_{\text{k}}$/T$_{\text{1}}$ are bound above by a square root of Casimir scaling, previously found to hold for stringlike solutions of the MIT Bag Model. The reason behind this similarity is that dYM dynamically realizes, in    a theoretically controlled setting, the main model assumptions of the Bag Model. 
We also compare confining strings in dYM and in other four-dimensional theories with abelian confinement, notably Seiberg-Witten theory, and show that the unbroken $\Z_N$ center symmetry in dYM leads to different properties of k-strings in the two theories; for example, a ``baryon vertex" exists in dYM but not in softly-broken Seiberg-Witten theory. Our results also indicate that, at large values of N, k-strings in dYM do not become free.

}

\begin{document}

\maketitle
\section{Introduction }

Systematic ways to study the long-distance behaviour of nonabelian gauge theories, where nonperturbative phenomena set in---confinement, the generation of mass gap, and the breaking of chiral symmetries---are hard to come by. Up to date, there are only a few examples in continuum quantum field theory where theoretically-controlled analytic methods allow one to make progress. Many of those examples, such as Seiberg-Witten theory, require various amounts of supersymmetry and utilize its power. 

In the past 10 years, a new direction of research into nonperturbative dynamics, applicable to a wider class of gauge theories, not necessarily supersymmetric,   has emerged \cite{Unsal:2007jx,01}: the study of gauge theories compactified\footnote{Hereafter, as most of our studies are Euclidean, we shall denote the spacetime manifold simply by $\R^3 \times \S^1$, but we use $\R^{1,2}\times \S^1$ here in order to stress that $\S^1$ is a spatial circle and the object of our study is not  finite-temperature theory.} on $\R^{1,2} \times \S^1$. 
 The control parameter is the size of the $\S^1$-circle L. When L is taken such that  NL$\Lambda \ll 1$, where N is the number of colours of  an  SU(N)  gauge  theory and $\Lambda$  its dynamical scale, it allows---as we shall review here for the theory we study---for  semiclassical  weak-coupling  calculability. It has led to new insight into  a variety of nonperturbative phenomena and has spawned new  areas of research. A comprehensive list of references is, at this point, too long to include here and we recommend the recent review article  \cite{Dunne:2016nmc}  instead. 

This paper  studies  confining strings in deformed Yang-Mills theory (dYM). dYM is a deformation of pure Yang-Mills theory, whose nonperturbative dynamics is calculable at small L. It is also believed that the dynamics is continuously connected to the large-L limit of $\R^4$,  in particular that the theory exhibits confinement and has a nonzero\footnote{Apart for the large-N limit, see below.} mass gap for every size of $\S^1$. The confining mechanism in dYM is a generalization of the  three dimensional  Polyakov mechanism of confinement \cite{12}, but owing to the locally four-dimensional nature of the theory many of its properties are quite distinct. As we further discuss, many features of dYM on $\R^3 \times \S^1$ can be traced back to the unbroken global center symmetry.

\begin{figure}
\centering
\begin{minipage}{0.9\textwidth}
	\includegraphics[width=\textwidth]{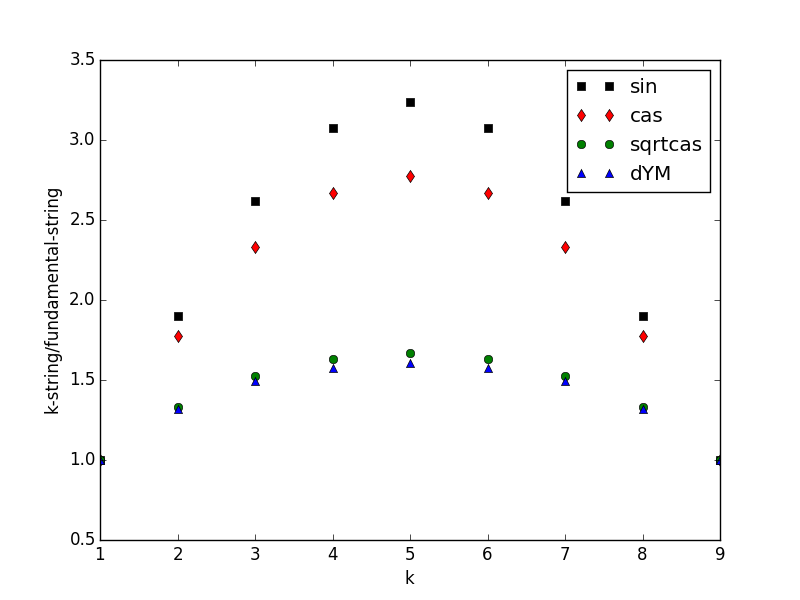}
	\caption{Comparison of different SU(10) k-string ratios scaling laws with dYM k-string ratios, labeled by ``dYM'' (blue triangles).  
The Sine law is labeled by ``sin,''    the Casimir scaling by ``cas,''  and  scaling with the Square root of the Casimir by  ``sqrtcas.'' From the known theoretical models predicting different scalings of k-string tensions, the ones we find in dYM are closest to the MIT Bag Model ``Square root of Casimir'' k-string tension law. We argue that it gives an upper bound on dYM k-string ratios. }{\label{fig:0}}
\end{minipage}
\end{figure}

The properties we set out to study here are the $N$-ality dependence of the string tensions and their behaviour in the  large-N limit. Renewed motivation to study the large-N limit of dYM arose from a recent  intriguing observation \cite{Cherman:2016jtu}: in the double-scaling limit L $\rightarrow 0$, N $\rightarrow \infty$, with fixed LN$\Lambda$, the four-dimensional theory on $\R^3 \times \S^1_{L \rightarrow 0}$ dynamically generates a latticized dimension whose size grows  with N.  This phenomenon has superficial similarities to T-duality in string theory and is not usually expected in quantum field theory. 
Originally, the emergence of a discretized dimension and its properties were studied in a  $\R^3 \times \S^1$ compactification and double scaling limit of ${\cal{N}}=1$ super-Yang-Mills (sYM) theory.   We show here that, as already observed in sYM, in dYM string tensions also stay finite in the large-N limit while the mass gap vanishes.

Most of this rather long paper is devoted to  a review of dYM and to a detailed explanation of the various methods we have developed; a guide to the paper is at the end of this  Section. 

The expert reader interested in the physics and not in the technical details should proceed to our ``Summary of results''  Section \ref{summary}, and to the more extended discussion in  Section \ref{sec:5}.

\subsection{Summary of results}
\label{summary}

Here we summarize  our main results, concerning both the confining string properties and the technical tools  developed for their study:
\begin{enumerate}
\item
{\it k-string tension ratios:}  In the regime of parameters studied in this work, in particular  NL$\Lambda \ll 1$, the asymptotic string tensions in dYM depend only on the $N$-ality of the representation. We argue in  Section \ref{sec:5} that the lowest tension stable strings between sources of $N$-ality k are  sourced by quarks with charges in the highest weight of the k-index antisymmetric representation, see (\ref{eq:54}).\footnote{There is a plethora of metastable strings that can also be studied using the tools developed here. An evaluation of their tensions and decay rates is left for future work. See Appendix E for a calculation of some metastable string tensions at leading order.}  Their tensions are hence referred to as the ``k-string tensions.'' 

Denoting by T$_{\text k}$ the k-string tension, on Fig.~\ref{fig:0} we show the ratio T$_{\text k}$/T$_1$ for SU(10), the largest group we studied numerically. The string tension ratio in dYM is compared to other known and much studied scaling laws, such as the Sine law and the Casimir law. It is clear from the figure that k-string tension ratios in dYM are different and do, instead,  come closest to a less-known scaling, found long ago in the MIT Bag Model of the Yang-Mills vacuum: the ``Square root of Casimir'' scaling \cite{13}.  In  Section \ref{bagmodelsection}, we argue that the relation between the two is
\begin{equation}
\label{dymscaling}
\left(T_k \over T_1\right)_{\text{dYM}} \le \;\sqrt{ k (N-k) \over N}~,
\end{equation}
where the r.h.s. is the square root of the ratio of quadratic Casimirs of the k-index antisymmetric representation and the fundamental representation. 
The  reason behind the  similarity is  that the model assumptions of the MIT Bag, that inside the bag the QCD chromoelectric fields can be treated classically and that the vacuum abhors chromoelectric flux, are  realized almost verbatim---albeit for the Cartan components only---by  the calculable confinement  in dYM.

\item {\it Large-N limit and $1\over N$ corrections to string tensions:} As already mentioned, string tensions stay finite at large N and fixed LN$\Lambda \ll 1$, as we show using various tools in  Section   \ref{sec:largeN}. Further, as can be inferred qualitatively from Figure \ref{fig:0}, and quantitatively from the analysis of  Section \ref{sec:largeN},  k-strings in dYM are not free at large N. We show that  
\begin{eqnarray}
\label{dymscaling2}
{\text{T}_2 \over \text{T}_1} &=& 1.347 \pm 0.001 + (-2.7 \pm 0.2) ({1 \over N})^2 + ...,\nonumber \\
{\text{T}_3 \over \text{T}_1} &=& 1.570 \pm 0.001 + (-7.5 \pm 0.2) ({1 \over N})^2 + ... ,
\end{eqnarray} instead of approaching the free-string values T$_\text{k}$ = kT$_1$. 

The large-N limit leading to the above behaviour   is taken {\it after} the large-RT limit (RT is the Wilson loop area). As the discussion there shows, assuming large-N factorization does not always imply that k-strings are free and the way the large-RT and large-N limits are taken has to be treated with care, as we 
discuss in detail in  Section \ref{sec:largeN}.\footnote{An important additional subtlety is that the values of N for which  the relations (\ref{dymscaling2}) have been derived, while numerically large, are bounded above by an exponentially large number $N \ll 2 \pi e^{c\over \lambda}$, where  $\lambda \sim |\log \Lambda N L|^{-1}$ is the arbitrarily small 't Hooft coupling  and $c$ is an ${\cal{O}}(1)$ coefficient. Preliminary estimates suggest that the  effect of the W-boson induced  mixing on the string tensions (whose neglect is the source of the upper bound on N, see  Section \ref{sec:largeN}) will not  qualitatively change the large-$N$ limit. However,  we prefer to defer further discussion until the relevant calculations for dYM have been performed.  }

\item {\it Comparing abelian confinements:} We compare the properties of confining strings in dYM and in Seiberg-Witten theory \cite{Seiberg:1994rs}, another four dimensional theory with   calculable abelian confinement. We argue that the unbroken $\Z_N$ center symmetry in dYM has dramatic implications for the meson and baryon spectra. In particular  there is a 		``baryon vertex'' in dYM, leading to ``Y''-type baryons, while only linear baryons exist in Seiberg-Witten theory \cite{15}. Thus, owing to the unbroken center symmetry, in many ways confinement in dYM is closer to the one in ``real world'' YM theory.\footnote{Some of these points were, without elaboration, made earlier in \cite{Anber:2015kea}. We also note that the glueball spectra in  dYM, as well as the mesonic  and baryonic spectra with quarks  added as in \cite{Cherman:2016hcd},  exhibit many intriguing properties and are   the subject of the more quantitative recent study  \cite{Aitken:2017ayq}.} For a   discussion of these issues, see    Section \ref{sec:compare} and  Figs.~\ref{fig:dymstring} and \ref{fig:SWstrings}.

\item {\it ``Perturbative evaluation'' of string tensions:} A technical tool to calculate string tensions analytically is developed in  Section \ref{sec:4}. We call it ``perturbative,'' as it utilizes a resummed all-order expansion and, at every step, requires the use of only Gaussian integrals. This method serves as a check on the computationally very intensive numerical methods that were employed in the numerical study. It also allows  the large-N limit to be taken analytically, subject to the limitations discussed above, and permits us to discuss the subtleties regarding the order of limits that lead to (\ref{dymscaling2}).  This method can be generalized to perform a path integral expansion about a saddle point boundary value problem  (e.g.~a transition amplitude in quantum mechanics) using perturbation theory (Gaussian integrals) only. Further applications of these tools is the subject of work in progress \cite{63}.
\end{enumerate}

 \subsection{Open issues for future studies}
 
As already stressed, one of our motivations is to study the peculiar large-N limit of dYM confinement, similar to the large-N limit of sYM from ref.~\cite{Cherman:2016jtu}, which shows many intriguing features that (at least superficially) resemble stringy properties. We have not yet fully addressed this limit in dYM, as there is the upper bound on N discussed above. We believe that this restriction on N is technical and more work is required to remove it. 

Our study here also only briefly touched on the spatial structure of confining k-strings, noting that, upon increasing N, they become more ``fuzzy'' due to the  decreasing mass of many of the dual photons, but retain a finite string tension due to the (also large) number of dual photons of finite mass. This spatial structure may have to do with their interacting nature and would be interesting to investigate further. 

Further, in this paper, we ignored the $\theta$-angle dependence of the $k$-strings. The  topological angle dependence in Yang-Mills theory has received  renewed recent attention, see 
e.g.~\cite{Thomas:2011ee,Unsal:2012zj,Anber:2013sga,Bhoonah:2014gpa,Gaiotto:2017yup,Tanizaki:2017bam,Kikuchi:2017pcp,Gaiotto:2017tne,Anber:2017rch}. As seen in  some of the aforementioned work, the corresponding physics in dYM is also very rich and worth of future studies. 

There are also the many intriguing observations of  \cite{Aitken:2017ayq} on the nature of the dual photon, glueball, etc., bound state spectra in dYM (at arbitrary N) that  await better understanding. 
Finally, there is the question about the (still conjectural) continuity of dYM from the calculable small $\Lambda$NL regime  to the regime of   large $\Lambda$NL. To this end, it would be desirable to study this theory on the lattice; for some lattice studies of related theories, see \cite{Cossu:2009sq,Vairinhos:2011gv,Bergner:2014dua,Bergner:2015cqa}.

\subsection{Organization of this paper}

 Section \ref{sec:2} is devoted to a review of dYM theory.\footnote{The reader already familiar with dYM and interested in our numerical and analytic metnods can proceed to  Sections \ref{numericsection} and \ref{sec:4} and the discussion in  Section \ref{sec:5}.} In  Section \ref{sec:2.1} we review how dYM theory on $\R^3\times \S^1$ avoids a deconfinement transition at small L. The perturbative spectrum of dYM is discussed in  Section \ref{sec:2.2.1} and the nonperturbative minimal action monopole-instanton solutions---in  Section \ref{sec:2.2.2}. The action of a dilute gas of monopoles is discussed in at length in  Section \ref{2.3}, with emphasis on details that often not emphasized in the literature. The derivation of the string tension action, used to calculate the semiclassical string tensions is given in  Section \ref{sec:2.4.1}.

{\flushleft{S}}ection \ref{numericsection} is devoted to a numerical study of the k-string tensions in dYM. The action and its discretization are studied in  Sections \ref{sec:3.1} and \ref{sec:3.3}. The minimization procedure, the numerical methods, and the error analysis are described in  Section \ref{sec:minimize}. The numerical results for the k-string tensions for gauge groups up to SU(10) are summarized in Table \ref{table:1}, see  Section \ref{sec:3.4}.

{\flushleft{S}}ection \ref{sec:4} presents an analytic perturbative procedure to calculate the string tensions. We begin by explaining the main ideas with fewer technical details. In  Sections \ref{su2analyticsection} and \ref{suNanalyticsection} we give the detailed calculations for SU(2) and general SU(N) gauge groups, respectively. The results are tabulated in Appendix \ref{sec:C}, demonstrating the precision of this procedure, which also serves as a check on the numerical results of  Section \ref{numericsection}. 

{\flushleft{S}}ection \ref{sec:5} contains the discussion of our results from various points of view, including relations to other models of confinement and the behaviour of k-strings in the large-N limit:
\smallskip

In  Section \ref{sec:5.1.1} we argue that the lowest, among all weights of any given representation, semiclassical asymptotic string tensions in dYM depends only on $N$-ality $k$ of the representation and is the one obtained for quark sources with charges in the highest weight (and its $\Z_N$ orbit) of the $k$-index antisymmetric representation. 

In  Section \ref{sec:compare} we compare confinement in dYM with confinement in Seiberg-Witten theory and point out that the unbroken $\Z_N$ center symmetry in dYM is responsible for the major differences, which make abelian confinement in dYM closer---in many aspects---to confinement in the nonabelian regime.  

In  Section \ref{bagmodelsection} we point out the similarity, already discussed around  eq.~(\ref{dymscaling}), of the k-string tension ratios in dYM to the ones in the MIT Bag Model and discuss the physical reasons. 

In  Section \ref{sec:5.1.4} we compare the k-string tension scaling laws to other scaling laws considered in various theoretical models.

In  Section \ref{sec:largeN}, we discuss the abelian large-N limit. The leading large-N terms in the k-string tension ratios, eq.~(\ref{dymscaling2}) above, are derived in \ref{sec:5.2.1}. The fact that large-N factorization does not always imply that k-strings become free at large-N is discussed in  Section \ref{sec:5.2.2}. The analytic methods of  Section \ref{sec:4} prove indispensable in being able to track the importance of the way the large-N and large area limits are taken.

\section{Review of dYM theory}

\label{sec:2}

In this  Section we will have a brief review of dYM theory. The emphasis is on topics usually not covered in detail the literature and on topics that will be needed for the rest of the paper.

\subsection{Confinement of charges in deformed Yang-Mills theory for all $\S^1$-circle sizes }

\label{sec:2.1}

Consider four-dimensional Yang-Mills theory in the Euclidean formulation with one of its dimensions compactified on the circle:
\begin{equation}
\label{eq:2.1}
S = \int_{{\rm \R}^3 \times \S^1}d^4 x{1 \over 2g^2} \text{tr} F^2_{\mu\nu}(x)~.
\end{equation} We set the $\theta$-angle to zero in this paper, leaving the study of the $\theta$-dependence of strings' properties for the future. 
Here,  $T^a$ ($a=1, ..., N^2 -1$) refer to the Hermitean generators of the group $SU(N)$, $F_{\mu\nu} = F^a_{\mu\nu}T^a$, $\text{tr} (T^aT^b) = {1\over 2} \delta^{ab}$.
The compactification circle $\S^1$ in pure Yang-Mills theory can either be considered as a spatial dimension of size $L$, or as a temporal one with $L = {1 / T}$ being   the inverse temperature $T$. It is known, see e.g. \cite{04}, that above a critical temperature $T_c = {1\over L_c}$ Yang-Mills theory loses confinement (i.e.~the static potential between two heavy probe quarks no longer shows a linearly rising behaviour as a function of distance between the quarks). The transition from a confining to a non-confining phase, in theories with gauge groups that have a nontrivial center, is accompanied by the breaking of the  center-symmetry.\footnote{\label{center}Center symmetry transformations are global symmetries that can be loosely thought as ``gauge'' transformations periodic up to the centre of the gauge group. For example, for an $SU(2)$ gauge group, the center-symmetry transformation periodic up to the nontrivial $\Z_2$ center element $z=-1$ can be represented by  $U_{-1}(\text{x},x_4) = \text{exp}(i {\pi \over L} x_4 \sigma_3)$, with $U_{-1}(\text{x},0) = - U_{-1}(\text{x},L)$ with $\sigma_3$ the third Pauli matrix and $x_4$---the $\S^1$ coordinate. See \cite{04} for a proper definition of center symmetry as a global symmetry on the lattice   and \cite{Gaiotto:2014kfa} for a  continuum point of view. }  The critical size $L_c$ is approximately of order $\Lambda^{-1}$, with $\Lambda$ the $\overline{MS}$ strong scale of the theory. Different studies give an estimate of $200$ MeV $<T_c <300$ MeV for $SU(2)$ Yang-Mills theory in four dimensions. In what follows we shall  deform Yang-Mills theory in a way that preserves confinement of charges for any circle size $L$. Due to asymptotic freedom the coupling constant is small at the compactification scale ${1 \over L}$ for small circle sizes $(L \ll \Lambda^{-1}$; as we argue below, the precise condition for $SU(N)$ gauge theories turns out to be $\Lambda L N \ll 1$). This deformation would enable us to have a model of confinement that we can study analytically in the limit of a small circle size $L$.

The expectation value of the trace of the Polyakov loop, $P(\text{x}) = \text{tr} \;{\cal{P}}\text{exp}(-i \oint_{S^1} dx_4 \text{A}_4 (\text{x},x_4))$ (where ${\cal{P}}$ denotes path ordering) serves as an order parameter for confinement \cite{04}:
\begin{flalign}
\label{eq:2.2}
&\langle P(\text{x})\rangle = 0 \ \ \   \text{confined phase with an infinite energy for an isolated free quark}, \nonumber \\
&\langle P(\text{x})\rangle \ne 0 \ \ \  \text{deconfined phase with a finite energy for an isolated free quark}. &&
\end{flalign}
On the other hand, the Polyakov loop is not invariant under a center-symmetry transformation and picks up a center element $z$, i.e.~$U_zP(\text{x})U^{\dagger}_{z} = z P(\text{x})$, where we used the notation of Footnote \ref{center}. Therefore for a center-symmetric vacuum $| 0 \rangle$ we have:
\begin{flalign}
\label{eq:2.3}
\langle 0 | P(\text{x})|0\rangle = \langle0|U^{\dagger}_zU_z P(\text{x})U^{\dagger}_zU_z|0 \rangle = z\langle 0 | P(\text{x})|0 \rangle \Longrightarrow \langle P(\text{x}) \rangle = 0 , \ \ z \ne 1 ,&&
\end{flalign}
indicating that a center-symmetric phase is a confined phase.

In order to show that Yang-Mills theory deconfines at high temperatures we need to show that the expectation value of the Polyakov loop at high temperatures is nonzero. The Polyakov loop is gauge invariant and the eigenvalues of the holonomy $\Omega(\text{x}) = \text{P} \text{exp}(- i \oint_{S^1} dx_4 \text{A}_4 (\text{x},x_4))$ constitute its gauge invariant content ($P = \text{tr} \Omega$). 
At tree level the eigenvalues of the holonomy can take any value, as there is no potential for $\Omega$ in the classical  Yang-Mills Lagrangian (\ref{eq:2.1}).
To find an effective potential for the eigenvalues of the holonomy  at one-loop, we expand \eqref{eq:2.1} around a constant diagonal $A_4$ field and evaluate the one loop contribution to the effective potential by integrating out the quadratic terms of gauge and ghost fields \cite{01,06}, to find:
\begin{flalign}
\label{eq:2.4}
V_1[\Omega] = -{2\over \pi^2 L^3}\underset{n=1}{\overset{\infty}{\sum}}{1\over n^4}|\textrm{tr}\Omega^n|^2, \ \ \text{where} \ \Omega = \text{exp}(-iLA_4)~.
\end{flalign}
From \eqref{eq:2.4} it can be seen that $V_1[\Omega]$ is minimized when $\Omega$ is an element of the centre of the gauge group, i.e. $\omega_N^k \;I$, with $I$ the unit matrix.\footnote{We defined $\omega_N = e^{ 2 \pi i \over N}$.} This would imply $\langle P(\text{x})\rangle = \langle \text{tr}(\Omega)\rangle = N \omega_N^k \ne 0 $, indicating a deconfined  center-symmetry broken phase of Yang-Mills theory at high temperatures, or small circle sizes $L$ (owing to asymptotic freedom, the small-$L$/high-$T$ regime is the one where the calculation leading to (\ref{eq:2.4}) can be trusted).

In order to change this picture and have a model of confinement at arbitrary small circle sizes $L$ we can add a deformation potential term to Yang-Mills theory \cite{01,Myers:2007vc}:
\begin{equation}
\label{eq:2.5}
S = \int_{{\rm I\!R}^3 \times S^1}{1 \over 2g^2} \text{tr} F^2_{\mu\nu}(x) + \Delta S, \   \Delta S \equiv \int_{{\rm I\!R}^3}{1 \over L^3} P[\Omega(\text{x})], \ P[\Omega] \equiv {2 \over \pi^2} \overset{[N/2]}{\underset{n=1}{\sum}}{b_n \over n^4}|\textrm{tr}(\Omega^n)|^2~,
\end{equation}
with $b_n$---sufficiently large and positive coefficients. The effect of the $\Delta S$ term is to dominate the gluonic and ghost potential \eqref{eq:2.4} in a way that the minimum of $V_1[\Omega] + {1 \over L^3}P[\Omega]$ occurs when $\text{tr}(\Omega^n) = 0$ for $n$ mod $N$ $\ne$ 0. This would imply $\langle P(\text{x})\rangle = \langle \text{tr}(\Omega)\rangle = 0$ and hence a confinement phase for deformed Yang-Mills theory at arbitrarily small circle sizes. 

The $\Delta S$ deformation term in \eqref{eq:2.2} would make the theory non-renormalizable. To have a well-behaved theory at high energies, the deformation can be considered as an effective potential term generated by some renormalizable dynamics, notably  $n_f$ flavors of massive adjoint Dirac fermions with periodic boundary conditions along the $\S^1$. Following~\cite{05} for conventions on Euclidean formulation of Dirac fermions we have:
\begin{equation}
\label{eq:2.6}
S_{dYM} = \int_{{\rm I\!R}^3 \times S^1}\{ {1 \over 2g^2} \text{tr}F^2_{\mu\nu}(x) -i \underset{i=1}{\overset{n_f}{\sum}}\bar{\psi}_i(\slashed{D}+m)\psi_i \}
\end{equation}
The effective potential for the holonomy generated by the $n_f$ massive adjoint Dirac fermions is given by~\cite{07,08}:
\begin{equation}
\label{eq:2.7}
V_2[\Omega] = +{2\over \pi^2 L^3}\underset{n=1}{\overset{\infty}{\sum}}n_f(nLm)^2K_2(nLm){|\textrm{tr}\Omega^n|^2 \over n^4}~,
\end{equation}
where $K_2$ is the modified Bessel function of the second kind. It has to be noted that in the deformed theory \eqref{eq:2.6} the compactified dimension $S^1$ can only be a spatial dimension since the heavy fermions satisfy periodic boundary conditions along this direction.   

There are two free parameters $n_f$ and $NLm$ in the effective potential \eqref{eq:2.7}.\footnote{The $n_f = 1/2$ massless case leads to vanishing potential, as is clear by comparing the massless limit of  (\ref{eq:2.7}) with (\ref{eq:2.4}). This case corresponds to the minimally supersymmetric Yang-Mills theory in four dimensions. } The beta function of $SU(N)$ Yang-Mills theory with $n_f$ flavours of Dirac fermions in the adjoint representation of the gauge group is, at the one loop level, $\beta(g) = - {g^3 \over (4 \pi)^2} ({11 \over 3}N - {4 \over 3}n_fN)$, hence to assure asymptotic freedom $n_f = 1 \ \text{or} \ 2$. If we allow for massive Majorana flavors, $n_f = 5/2$ is the maximum value. 
On the other hand, if we want the effective potential $V_2$ to  dominate the gluonic potential $V_1$, $NLm$ should be of order 1 ($NLm \sim 1$; for larger values of $m$, the fermions decouple and the theory loses confinement at small $L$).
To  gain some intuition on how the coefficients of the potential $V_2$ behave let $c_n \equiv n_f ({n\over N}LmN)^2 K_2({n\over N}LmN)$. Choosing $n_f = 2$ ($n_f = 1$) and $NLm = 4$ ($NLm = 3$) gives $c_n \approx 4$ ($c_n \approx 2$) for $n/N \approx 0$, $c_n \approx 2$ ($c_n \approx 1.3$) for $n/N \approx 0.5$, $c_n \approx 0.56$ ($c_n \approx 0.55$) for $n/N \approx 1$ and $c_n$ approaching zero exponentially for $n/N > 1$. Minimizing\footnote{This has been explicitly performed for the above choices of parameter up to $SU(10)$ and with considering the effective potentials up to $n=20$.} the combined potential  $V_1[\Omega] + V_2[\Omega]$ gives $\langle\Omega\rangle = \text{diag} (\omega_N^{N-1}, ..., \omega_N, 1)$ for odd $N$, and $\langle\Omega\rangle = e^{i {\pi \over N}} \text{diag}  (\omega_N^{N-1}, ..., \omega_N, 1)$ for even $N$, which gives $\text{tr} \langle\Omega\rangle^n  = 0$ for $n$ mod $N \neq 0$, hence, a confined phase for deformed Yang-Mills theory.\footnote{General $SU(N)$ theories with semiclassically calculable dynamics at small-$L$ have been classified in \cite{Anber:2017pak}.}

\subsection{Perturbative and non-perturbative content of dYM}
\label{sec:2.2}

\subsubsection{Perturbative content}
\label{sec:2.2.1}

 The eigenvalues of the holonomy $\Omega (\text{x}) = {\cal{P}}\text{exp}(i\oint d\text{x}_4 A_4(\text{x},\text{x}_4))$, are the only gauge invariant content of the gauge field component in the compact direction $A_4(x)$ and are invariant under any periodic gauge transformation. Working in a gauge  such that the $A_4(x)$ field assumes these eigenvalues ($A_4(x) = -i {\text{ln}(\Lambda(\text{x})) / L}$, with $\Lambda(\text{x})$---the diagonal matrix of eigenvalues of $\Omega(\text{x})$) and expanding around a center-symmetric $vev$ 
 \begin{eqnarray}
 \label{a4vev}
 A^{vev}_4 &=& {1 \over N L} \text{diag}(2\pi (N-1), ..., 2\pi, 0) \; \; \text{for odd}\; N, \nonumber \\
 A^{vev}_4 &=& {1 \over N L} \text{diag}(2\pi (N-1) + \pi, ..., 3\pi, \pi)\;\;\text{for even} \; N, 
 \end{eqnarray}
     the perturbative particle content of  dYM theory with  action \eqref{eq:2.6} can be worked out by writing the second order Lagrangian of the modes expanded around the above center-symmetric $vev$. 
  Clearly, the $vev$ of the ``Higgs field'' $A_4$ breaks the gauge symmetry $SU(N) \rightarrow U(1)^{N-1}$. The gauge fields associated with the non-compact direction can be written as:
\begin{equation}
\label{eq:2.8}
A_i(\text{x},\text{x}_4) = {\sqrt{2}  }\; A_{i,0}(\text{x}) + {}\overset{+\infty}{\underset{k = - \infty} {\sum}^\prime} A_{i,k}(\text{x})\;\text{exp}(ik{2\pi \over L}\text{x}_4) \ \   ({\sum}^\prime  \text{ over}\;    k \neq 0)  ~,
\end{equation}
with $A_{i,k}(\text{x}) = A_{i,k1}(\text{x}) + iA_{i,k2}(\text{x}) = A^{\dagger}_{i,-k}(\text{x})$ in order to ensure reality of the $A_i$ fields. It turns out that the gauge-boson field content  is a non-trivial one. We work out the quadratic Lagrangian  in Appendix \ref{app:wboson}, by substituting (\ref{eq:2.8}) in the action and expanding around  (\ref{a4vev}).

We begin with a discussion of the abelian spectrum.
The diagonal components of the gauge fields $A_i$  commute with the $vev$   $A_4^{vev}$. Hence,  their zeroth Fourier modes along $\S^1$ correspond to massless 3d photons and their higher Fourier modes gain mass of ${2 \pi m \over L}$ where $m=1,2, ...\infty$ is the non-zero momentum in the compact direction. At tree level, the Lagrangian for the $N-1$ photons is simply the reduction of (\ref{eq:2.1}) to the Cartan subalgebra of $SU(N)$. The leading-order coupling of the 3d $U(1)^{N-1}$ gauge theory is 
  given by $g_3^2 = g^2/L$, where $g$ is the four-dimensional gauge coupling at the scale of the lightest $W$-boson mass, $m_W = {2\pi \over NL}$, see below.\footnote{At subleading order, threshold corrections from the $W$-bosons cause the $N-1$ photons (and consequently, the dual photons) to mix. These mixing effects are expected to be similar to the ones in  super-Yang-Mills \cite{08} and QCD(adj) \cite{Anber:2014sda, vito}. They become  important in the abelian large-$N$ limit \cite{Cherman:2016jtu}, where dYM has a curious ``emergent dimension'' representation. The mixing between the $N-1$ photons is also expected to affect the $k$-string tensions in the abelian large-$N$ limit. In this paper, we have not taken these effects into account.}
   
 The physical components of the ``Higgs'' field $A_4(x)$ are diagonal and $x_4$ independent. They are massless at the classical level but gain mass of order at least $g \over  \sqrt{N}  L$ via the one loop effective potential $V_1 + V_2$ generated by quantum corrections.\footnote{The $N$-dependence of the lightest $A_4$  shows that the mass scale of the holonomy fluctuations remains fixed in the abelian large-$N$ limit, where $g^2 N$ and $m_W \gg \Lambda$ remain fixed.}
 
The massive adjoint  Dirac fermions are also  expanded in their Kaluza-Klein modes. Taking into account the effects of $A^{vev}_4$, it can be seen that there are massive Dirac fermions with masses $m + {2 \pi k \over L} + {2 \pi p \over LN}$ for $k = 0, 1, ...$ and $p = 0, 1, ..., N-1$. As we are interested in physics below the scale of the lightest fermion, we shall not present details of the massive fermion spectrum.

Finally, the relation \eqref{eq:2.13}  from Appendix \ref{app:wboson} shows that there are $W$-bosons   with masses $|{2 \pi m \over L} - {2\pi |l-k| \over NL}|$ and $|{2 \pi m \over L} + {2\pi |l-k| \over NL}|$ respectively for $m =0,1,2,...$ and $1\leq l < k \leq N$. The mass of the lightest $W$-boson is $m_W = {2 \pi \over NL}$. Clearly, below that scale, there are no fields charged under the unbroken $U(1)^{N-1}$ gauge group. Thus, the gauge coupling of the $N-1$ photons is frozen at the scale $O(m_W)$. The condition that the theory be weakly coupled, therefore, is that $m_W \gg \Lambda$, or $N L \Lambda \ll 1$. This is the semiclassically calculable regime that we  study in this paper.

 In summary, the perturbative particle content of deformed Yang-Mills theory expanded around the center-symmetric $vev$ consists of $N-1$ photons (the diagonal  Cartan  components of the gauge fields, whose zero Fourier components along the $\S^1$ are massless),  $N^2 - N$ massive gauge fields, charged under the $U(1)^{N-1}$ unbroken gauge symmetry, whose spectrum  is given in (\ref{eq:2.13}), $N-1$ massive eigenvalues of the holonomy, neutral under $U(1)^{N-1}$ and charged and uncharged massive Dirac fermions.


\subsubsection{Non-perturbative content: minimal action instanton solutions}
\label{sec:2.2.2}

Finite action Euclidean configurations of pure Yang-Mills theory   on $\R^3 \times \S^1$ were studied in 
  \cite{06}. It was shown that they are classified by 
their magnetic charge $q_\alpha$, Pontryagin index $p$, and asymptotics of the $\S^1$ holonomy $\Omega$ at  infinity, which are related by
the following formula:
\begin{equation}
\label{eq:2.14}
Q = p + {\text{ln} \mu_{\alpha}  \over 2 \pi i}\; q_{\alpha}~.
\end{equation}
Here, $Q = {1 \over 32 \pi^2}\int_{{  \R}^3 \times \S^1} d^4 x \;F_{\mu\nu}^a \widetilde{F}_{\mu\nu}^a$ is the topological charge  with $\widetilde{F}_{\mu \nu} = {1\over2} \epsilon_{\mu \nu \alpha \beta} F_{\alpha \beta}$. In this  Section, we use 
$\mu_{\alpha}$ with $\alpha = 0, ..., \kappa \leq N-1$ to label the distinct eigenvalues of the holonomy $\Omega (\text{x})$ at spatial infinity. Notice that, for finite action configurations, the eigenvalues of $\Omega$ are independent of the direction that we approach spatial infinity, and that, for the center-symmetric holonomy, $\kappa = N-1$ as all eigenvalues are distinct, given by the $N$ values of $e^{i L A_4^{vev}}$ with $A_4^{vev}$ of (\ref{a4vev}): $\ln \mu_\alpha = {2 \pi i (N-1-\alpha) \over N}$ for odd $N$.
The integer magnetic charges are denoted by $q_{\alpha}$, satisfy $\sum^{\kappa}_{\alpha = 0}q_{\alpha} = 0$, and will be explicitly defined further below, see paragraph after eq.~(\ref{eq:2.21}). The Pontryagin index $p$ is the winding number for mappings of $\S^3$ onto the full group $SU(N)$.\footnote{After any twist of $\Omega$ at infinity associated with the magnetic charges $q_{\alpha}$  is removed by a (singular) gauge transformation, the resulting field may be regarded as a mapping of compactified three space (or $\S^3$) onto the group $SU(N)$, leading to the familiar Pontryagin index. More details regarding the definitions of these quantities can be found in \cite{06}.} 

It is expected that for any value of the quantities $p$, $q_{\alpha}$, and $\mu_{\alpha}$ there is a separate sector of finite  action configurations, with the self-dual ($F_{\mu\nu} = \widetilde{F}_{\mu\nu}$) or anti-self dual ($F_{\mu\nu} = - \widetilde{F}_{\mu\nu}$) solution corresponding to the minimum action configuration in that sector. For self-dual or anti-self-dual solutions the topological charge is proportional to the value of the action. Therefore finding configurations with the minimal non-zero topological charge is equivalent to finding the minimal non-zero action configurations.
Based on the values of $\mu_{\alpha} = \text{exp}(i {2 \pi (N-1 -\alpha) \over N})$ for $\alpha = 0, ..., N-1$ in a center-symmetric vacuum, and the fact that $p$ and $q_{\alpha}$ are integers with $\sum^{N-1}_{\alpha = 0}q_{\alpha} = 0$, it can be clearly seen that the minimal non-zero topological charge is $|Q| = {1 \over N}$. Configurations of minimal $Q = {1 \over N}$ would then correspond  to the following values of $q_{\alpha}$ and $p$:
\begin{flalign}
\label{eq:2.15}
&p = 0, \ q^i_m = (0,..,0,\overset {\text{i-th}}{\widehat{1}},-1,0,...,0), \ \ \ \ \ \  i = 1, ..., N-1, \\
\label{eq:2.16}
&p = 1, \ q^N_m = (-1,0,... ,0, 1),
\end{flalign}
where the $N$ components of the vector $q_m^i$ are the magnetic charges $q_\alpha$  corresponding to the $i$-th minimal action configuration. 
Minimal action configurations with Pontryagin indices and magnetic charges  given in \eqref{eq:2.15} will be referred to as the  $N-1$ $SU(N)$ BPS solutions and the minimal action configurations with Pontryagin number and magnetic charges from  \eqref{eq:2.16}---as the Kaluza-Klein (KK) solution.\footnote{This terminology is adopted for historical reasons. In the limit when the mass of the physical holonomy fluctuations is neglected, both our BPS and KK solutions satisfy a BPS bound and can be found by solving first-order equations.} The $\overline{\text{BPS}}$ (anti-BPS) and $\overline{\text{KK}}$ (anti-KK) configurations have the opposite sign for the magnetic charges and Pontryagin index and thus a negative topological charge $Q = -{1 \over N}$. In total this classification shows that there exist  $2N$ minimum finite action non-trivial configurations. We will refer to these finite action configurations as the ``non-perturbative content'' of deformed Yang-Mills theory---because, as we shall see, it is these Euclidean configurations that lead to confinement of charges, to leading order in $NL\Lambda \ll 1$. 

In order to construct such configurations we start from the $SU(2)$ BPS and Kaluza-Klein monopoles and embed them in $SU(N)$. For the BPS solution, this can be done in $N-1$ different ways leading to the $N-1$ different configurations in \eqref{eq:2.15} and for the Kaluza-Klein monopole this can be done in only one way. The $SU(2)$ BPS monopole solution is given by \cite{23,Diakonov:2009jq}:
\begin{equation}
\label{eq:2.17}
\begin{split}
& A^a_4 = \mp n_a \nu \mathcal{P}(\nu r)\ \ , \ \ \ \ \ \ \ \ \ \ \ \ \  \mathcal{P}(y) = \text{coth}(y) - {1 \over y} \\
& A^a_i = \epsilon_{aij} n_j {1-\mathcal{A}(\nu r) \over r}\ \ , \ \ \ \ \ \ \ \mathcal{A}(y) = {y \over \text{sinh}(y)}~,
\end{split}
\end{equation}
where $n_a$ for $a = 1, 2, 3$ refer to the components of a unit vector in ${\rm I\!R}^3$ and $\nu$ is related to the eigenvalues of the holonomy at infinity. In what follows, as in the above relations, the upper sign always corresponds to the self-dual BPS solution and the lower one to the anti-self-dual $\overline{\text{BPS}}$ solution. The magnetic field strength $B_i ={1 \over 2} \epsilon_{ijk}F_{jk}$ of this solution is:
\begin{equation}
\label{eq:2.18}
B^a_i = (\delta_{i}^a - n^an_i) \nu^2 F_1(\nu r) + n^a n_i \nu^2 F_2(\nu r)
\end{equation}
The functions $F_1$ and $F_2$ are given below in  \eqref{eq:2.21}. In order to embed these solutions in $SU(N)$ and in a center-symmetric vacuum $A^{vev}_4$, we first make a gauge transformation that will make the $A_4$ component diagonal in colour space along $\tau^3 /2$. For this we solve for the equation $S_- \tau^a n_a S_-^{ \dagger} = -\tau^3$ for the BPS and $S_{+} \tau^a n_a S_+^{ \dagger} = \tau^3$ for the $\overline{\text{BPS}}$. This gives:
\begin{equation}
\label{eq:2.19}
\begin{split}
&  S_+(\theta, \phi) = \text{cos} { \theta \over 2} + i \tau^2 \text{cos} \ \phi \ \text{sin} {\theta \over 2} -i \tau^1 \text{sin} \ \phi \ \text{sin}  { \theta \over 2} = e^{-i\phi \tau^3 / 2} e^{i\theta \tau^2 / 2} e^{i\phi \tau^3 / 2} \\
&  S_-(\theta, \phi) = - \text{sin}{ \theta \over 2} \; \text{cos} \ \phi - i\; \text{sin}{ \theta \over 2} \;\text{sin} \ \phi \ \tau^3 + i\; \text{cos}{ \theta \over 2} \tau^2 = e^{i\phi \tau^3 / 2} e^{i (\theta + \pi) \tau^2 /2 } e^{i\phi \tau^3 / 2}~.
\end{split}
\end{equation}
After performing the gauge transformation $A_{\mu} \longrightarrow A^S_{\mu} = S A_{\mu} S^{\dagger} + iS \partial_{\mu} S^{\dagger}$ for $S = S_- \ \text{or} \ S_+$ we get:
\begin{equation}
\label{eq:2.20}
\begin{aligned}
& A^S_4 = \nu \mathcal{P}(\nu r) {\tau^3 \over 2} \\
& A^S_r = 0 \\
& A^S_{\theta} = {\mathcal{A}(\nu r) \over 2r} (\pm \tau^1 \text{sin} \phi + \tau^2 \text{cos} \phi) \\
& A^S_{\phi} = {\mathcal{A}(\nu r) \over 2r} (\pm \tau^1 \text{cos} \phi - \tau^2 \text{sin} \phi) \pm \tau^3 {1 \over 2r} \text{tan} {\theta \over 2} ~,
\end{aligned}
\end{equation}
where $A^S_r = \hat{r}_i A^S_i$, $A^S_{\theta} = \hat{\theta}_i A^S_i$, $A^S_{\phi} = \hat{\phi}_i A^S_i$ are the components of $A^S_i$ along the unit vectors in spherical coordinates.\footnote{Our convention for spherical coordinates is $r(\text{sin} \ \theta \ \text{cos} \ \phi,\text{sin} \ \theta \ \text{sin} \ \phi,\text{cos} \ \theta) = (\text{x}_1,\text{x}_2,\text{x}_3 )$).} It has to be noted that the $A^S_{\phi}$ solution shows a singular string along $\theta = \pi$. This is  a gauge artifact and does not cause any problems for \eqref{eq:2.20} to satisfy the self-duality or anti-self-duality condition. In other words, the magnetic  fields evaluated from \eqref{eq:2.20} are everywhere smooth functions of the spherical coordinates, as can be seen by finding the magnetic field strength in the stringy gauge:
\begin{equation}
\label{eq:2.21}
\begin{aligned}
& B^S_r = \mp {\nu}^2 F_2(\nu r) \tau^3 /2 , \ \ \ \ \ \ \ \ \ \ \ \ \ \ \ \ \ \ \ \ \ \ \ \ \ \ \ \ F_2(y) = {1 \over \text{sinh}^2 y } - {1 \over y^2} \\
& B^S_{\theta} = \nu^2 F_1(\nu r)/2 (\mp \tau^1 \text{cos} \phi + \tau^2 \text{sin} \phi) , \ \ \ \ \ \ \ \ F_1(y) = {1 \over \text{sinh} \ y }( {1 \over y} -  \text{coth} \ y) \\
& B^S_{\phi} = \nu^2 F_1(\nu r)/2 (\pm \tau^1 \text{sin} \phi + \tau^2 \text{cos} \phi) ~.
\end{aligned}
\end{equation}
Using the diagonal components of the field $B^S$ at infinity, diag$(B^S)$,  the magnetic charge vector for the $SU(2)$ BPS monopole solution, in the normalization of (\ref{eq:2.15}) can now be defined\footnote{This definition applies to when the eigenvalues of the holonomy ${\Omega}$ at infinity are distinct. For the general definition of magnetic charges that would also apply to holonomies with degenerate eigenvalues at infinity refer to relation (B.6) in \cite{06}}.by a surface integral at infinity, $(q_1, q_2) = \oint\limits_{S^2_\infty} d^2 \sigma  \;\text{diag} (B_r^S)/(2 \pi) = (1,-1)$.\footnote{A direct calculation of $Q$ for the $SU(2)$ BPS solution yields $Q=1/2$, thus verifying explicitly (\ref{eq:2.14}) with $p=0$ and  the appropriate expression for $\mu_\alpha$. }

There are $N-1$ $SU(2)$ Lie subalgebras, corresponding to the elements $a_{ii}$, $a_{ii+1}$, $a_{i+1i}$, $a_{i+1i+1}$ for $i = 1, ..., N-1$, along the diagonal of an $SU(N)$ Lie algebra matrix  and we can embed an $SU(2)$ BPS monopole in each of them. Only these embedded BPS monopoles will have the lowest topological charge $|Q| = {1 \over N}$. 
We will illustrate the embedding for the top left $SU(2)$ Lie subalgebra. We simply place the $SU(2)$ solution \eqref{eq:2.20},  with $\nu = {2\pi \over NL}$, in the top left $SU(2)$ Lie subalgebra of an $SU(N)$ Lie algebra matrix with all other elements being zero. Next, in order to make the value of $A^{S,SU(N)}_4$ ($\equiv$ $SU(2)$ $A^S_4$ solution of \eqref{eq:2.20} embedded in $SU(N)$) at infinity the same as $A^{vev}_4$ of (\ref{a4vev}), we add the matrix $\bar{A} ={1 \over NL} \text{diag}(2\pi(N-1) -\pi, 2\pi(N-1) -\pi, 2\pi(N-2),...,0)$ for odd $N$ and $\bar{A} ={1 \over NL} \text{diag}(2\pi(N-1) , 2\pi(N-1) , 2\pi(N-2) + \pi,...,\pi)$ for even $N$ to $A^{S,SU(N)}_4$. Similarly,  BPS monopoles can be  embedded in the remaining $N-2$ diagonal $SU(2)$ subalgebras  of $SU(N)$.

For the Kaluza-Klein solution\footnote{More details regarding this solution and its explicit form  can be found in, e.g. \cite{23}. These ``twisted" solutions were first found in \cite{Lee:1997vp,Kraan:1998sn} using different techniques.} we start from the BPS solution of \eqref{eq:2.17} in a vacuum where $\nu$ is replaced by $\nu \rightarrow {2 \pi \over L} - \nu$. To obtain the KK solution ($\overline{\text{KK}}$) we gauge transform the BPS solution of \eqref{eq:2.17} with $S_+$ (with $S_-$) using the upper sign (lower sign). Now the asymptotic behaviour of the $A_4$ field for both solutions is $(\nu - {2\pi \over L}) {\tau^3 \over 2}$. In order to make the asymptotics similar to the BPS $A^S_4$ field in \eqref{eq:2.20}, we perform an $x_4$-dependent gauge transformation $U(\text{x}_4) = \text{exp}(i {2 \pi \over L} \text{x}_4 {\tau^3 \over 2})$, which brings the asymptotics back to $\nu {\tau^3 \over 2}$. This gauge transformation  gives a non-trivial $x_4$-dependence to the cores of the KK and $\overline{\text{KK}}$ solutions. Since the Pontryagin index $p$ of a KK monopole in relation \eqref{eq:2.16} is $p = 1$, in order to obtain the lowest topological non-zero charge (which is $|Q| = {1 \over N}$), the second term in \eqref{eq:2.14} should equal $-{N-1 \over N}$ therefore, as already discussed, there is only one way to embed an $SU(2)$ KK monopole in $SU(N)$ in a centre-symmetric $vev$ that would give the lowest action and that is to choose the $SU(2)$ subalgebra corresponding to the components $a_{11}, a_{1N}, a_{N1}, a_{NN}$ of an $SU(N)$ Lie algebra matrix (i.e.~with $q_1 = - q_N = -1$, as per (\ref{eq:2.16})).

This was a brief summary of the non-perturbative solutions in dYM theory that are responsible for confinement of charges to leading order in the limit $NL \Lambda \rightarrow 0$.

\subsection{Action of a dilute gas of monopoles}
\label{2.3}

The action of the (anti-)self-dual solution \eqref{eq:2.20} embedded in $SU(N)$, with $\nu = 2\pi /NL$, is given by:
\begin{flalign}
\label{eq:2.22}
S_{\text{BPS}} = {2 L \over g^2} \int_{{\rm I\!R}^3} d^3 \text{x} \text{tr}(B_i B_i)  = {8 \pi L \nu \over g^2} \int^{\infty}_0 d\bar{r} \bar{r}^2 \{ {1 \over 2} F^2_2(\bar{r})  + F^2_1(\bar{r})\}= {4 \pi L \nu \over g^2} = {8 \pi^2 \over g^2 N} ~,&&
\end{flalign}
where $\bar{r} = \nu r$, $r$ being the radial coordinate in spherical coordinates.

Next, we calculate the action of two far-separated BPS solutions of \eqref{eq:2.20} embedded in $SU(N)$ and living in a center-symmetric vacuum $A^{vev}_4$. We embed the first monopole (second monopole) in the $i \text{-th}$ ($j \text{-th}$) subalgebra of $SU(N)$ along the diagonal for $1 \leq i, j \leq N-1$. We work in the limit $\nu^{-1} = {NL \over 2\pi} \ll r_0 \ll d$, where $d$ denotes the distance between the centers of the monopoles and $r_0$ is the radius of a two-sphere surrounding each monopole. In constructing far separated monopole solutions, first we need to mention how the monopoles are patched together. To patch the monopoles together, we use  the string gauge 
and  first subtract $A^{vev}_4$ from the $A_4$ component of each monopole solution. The resulting configuration will have an asymptotically vanishing behaviour at infinity for all its gauge field components. Now we simply add up the fields corresponding to the various monopole configurations, with their centres being separated by a large, in the precise sense defined above, distance $d$ from each other. At the end, we add $A^{vev}_4$ to obtain the  final configuration (had we simply added the two monopole configurations at a large separation $d$, asymptotically the $A_4$ component of the two monopole configuration would be $2A^{vev}_4$; this way of construction avoids this double counting of $A^{vev}_4$).

In calculating the action of far separated monopole configurations we only consider gauge invariant leading order terms in the self-energy and interaction energy of the monopoles.\footnote{While we use energetics terminology, motivated by the electro-/magneto-static analogy,  we clearly mean Euclidean action. Also by gauge-invariant terms we refer to any terms in the action of two far separated monopoles that are independent of the Dirac string (singularity of the solutions at $\theta = \pi$ in \eqref{eq:2.20}) or its orientation.} We write the fields as $A_{\mu} = A^{(1)}_{\mu} + A^{(2)}_{\mu}$, for $\mu =1,... , 4$. $A^{(1)}_{\mu}$ and $A^{(2)}_{\mu}$ refer to the contribution of the first and second monopole to the total $A_{\mu}$ field of the monopole configuration respectively. When $A_4$ appears in the commutator term of $F_{k4}$ the overall $A^{vev}_4$ is considered as part of $A^{(i)}_4$ in the two-sphere region of radius $r_0$ surrounding the $i$-th monopole for $i =1, 2$ and otherwise can be distributed in an arbitrary smooth way between $A^{(1)}_{4}$ and $A^{(2)}_{4}$ and for the $\partial_k A_4$ term in $F_{k4}$ the overall $A^{vev}_4$ vanishes and can be neglected. The total action of the far separated two-monopole configuration can be written as:
\begin{equation}
\label{eq:2.23'}
\begin{split}
S^{\prime}_{2 \text{-monopole}} & = {L \over g^2} \int d^3 \text{x} \{ \text{tr} ( B_kB_k ) +  \text{tr}  ( F_{k4}F_{k4}) \} \\ & = \sum^{2}_{i=1}S^{(i)}_{\text{self-energy}} + S_{\text{inter.}, > r_0} + S_{\text{inter.}, < r_0} + S_{\text{non-gauge-invariant}}
\end{split}
\end{equation}
Each of the above terms in \eqref{eq:2.23'} will be explained and evaluated below:
\begin{enumerate}
\item For the self-energy, we calculate the contribution of one monopole to the action neglecting the other monopole. Similar to \eqref{eq:2.22} we find:
\begin{equation}
\label{eq:2.23}
S^{(i)}_{\text{self-energy}} = {L \over g^2} \int d^3 \text{x} \{ \text{tr} ( B^{(i)}_kB^{(i)}_k ) +  \text{tr}  ( F^{(i)}_{k4}F^{(i)}_{k4}) \} = {8 \pi^2 \over g^2 N} + O(\text{exp}(-\nu r_0))  , \ \ \ \  i =1, 2
\end{equation}
where $B_k^{(i)}$ and $F^{(i)}_{k4}$ refer to the magnetic and ``electric'' field\footnote{At the classical level, the $A_4$-field, mediating the so-called ``electric'' interactions,  is massless hence it is of long range. We stress that the term ``scalar interaction'' is the precise one, within the framework of spatial-$\S^1$ compactifications; for brevity, we continue calling these interactions ``electric'' and omit the quotation marks in what follows.
Furthermore, as already explained, at the quantum level the $A_4$ field gains mass hence the electric interaction is short range and not important in the derivation of the string tension action. We only discuss the electric interaction here for the sake of mentioning some points not usually explicitly discussed with regard to the classical interaction of monopole-instantons.} of the $i$-th monopole respectively. The fact that the overall $A^{vev}_4$ is distributed between $A^{(1)}_{4}$ and $A^{(2)}_{4}$ outside their surrounding two-spheres of radius $r_0$ would make the monopole self-energies in \eqref{eq:2.23} to differ from \eqref{eq:2.22} by $O(\text{exp}(-\nu r_0))$. 

\item The first contribution to the interaction between two monopoles at the classical level comes from the long range magnetic and electric fields of each monopole. This  contribution comes from the region outside the two-spheres of radius $r_0$ surrounding each monopole (the second contribution comes from the long range electric influence that each monopole has on the other monopole inside their surrounded sphere of radius $r_0$, see next item).
The magnetic and electric interactions beyond the two surrounding spheres can be evaluated as:
\begin{flalign}
\label{eq:2.24}
S_{\text{inter.}, > r_0} &=   {2L \over g^2} \int d^3 \text{x} \{ \text{tr} ( B^{(1)}_kB^{(2)}_k ) +  \text{tr}  ( F^{(1)}_{k4}F^{(2)}_{k4}) \} \\
&=  {2 L \over g^2} \int d^3\text{x} \{ { \text{x} - \text{x}_1 \over 2|\text{x} - \text{x}_1|^3 } \cdot { \text{x} - \text{x}_2 \over 2|\text{x} - \text{x}_2|^3 }q^{i}_{m1} \cdot q^{j}_{m2}  + { \text{x} - \text{x}_1 \over 2|\text{x} - \text{x}_1|^3 } \cdot {\text{x} - \text{x}_2 \over 2|\text{x} - \text{x}_2|^3 }q^{i}_{e1} \cdot q^{j}_{e2} \} \nonumber \\ & \ \ \ + O({L \nu^{-2} \over g^2 d^3}), \ \ \ \ \ \ \ \ d = |\text{x}_1 - \text{x}_2| \nonumber &&
\end{flalign}
Here, $q^{i}_{m1} = q^{i}_{m}$ refers to the magnetic charge of the first monopole with $q^{i}_{m}$---an $N$-component charge vector given by relation \eqref{eq:2.15}. Similarly $q^{j}_{m2} = q^{j}_{m}$, $1 \leq i, j \leq N-1$. We substituted, from the Cartesian form of the radial terms proportional to ${1 \over r^2}$ in (\ref{eq:2.21}), the magnetic field $B^{(1)}_{k,r^{-2}} \equiv \text{diag}(q^i_{m1} { \text{x}_k - \text{x}_{1k} \over 2|\text{x} - \text{x}_1|^3})$ and $B^{(2)}_{k,r^{-2}} \equiv \text{diag}(q^j_{m2} { \text{x}_k - \text{x}_{2k} \over 2|\text{x} - \text{x}_2|^3})$ (understood as a diagonal matrix with entries determined by the charge vectors $q_{m1}^i$ and $q_{m1}^j$ (\ref{eq:2.15}) of the two monopoles) into the first line in (\ref{eq:2.24}), and similarly for $F_{k4}$. We  then replaced the trace of the product of these abelian matrices with an inner product over the vector of magnetic (or electric, i.e. scalar) charges corresponding to the diagonal elements of these abelian matrices.
For a self-dual BPS solution, the electric charges are  $q^i_{e1} = q^i_{m1}$ and $q^j_{e2} = q^j_{m2}$.
The error term in \eqref{eq:2.24} comes from the inner product of the long range magnetic (or electric) field of one monopole with the term $\sim \hat{r}^{\text{other}} {1 \over \text{sinh}^2 \nu r}$ of magnetic (or electric) field of the other monopole in \eqref{eq:2.21}, integrated over the two-sphere of radius $r_0$ surrounding the other monopole, with $\hat{r}^{\text{other}}$ being the unit vector in the radial direction of the other monopole. After writing ${ \text{x} - \text{x}_i \over 2|\text{x} - \text{x}_i|^3} = -\nabla_{\text{x}} {1 \over 2|\text{x} - \text{x}_i|}$ for $i =1, 2$ in \eqref{eq:2.24} and integrating by parts with $\nabla^2_{\text{x}} {1 \over |\text{x} - \text{x}_i|} = -4 \pi \delta^3 (\text{x} - \text{x}_i)$ we get:
\begin{equation}
\label{eq:2.25}
S_{\text{inter.} > r_0} = {2 \pi L \over g^2 d} \;{q^i_{m1}\cdot q^j_{m2} } + {2 \pi L \over g^2 d}\; {q^i_{e1}\cdot q^j_{e2} } + \text{O}({L \nu^{-2} \over g^2d^3})~.
\end{equation}

\item The final Dirac-string independent contribution to the interaction between the two monopoles is the electric influence of the second monopole on the core of the first monopole\footnote{For another discussion on the core interaction between dyons refer to \cite{60}.}  There is also a similar electric influence from the first monopole on the core of the second one.  It originates from the following cross terms in the action:
\begin{equation}
\label{eq:2.26}
S^{2-1}_{\text{inter.}, < r_0} = -{2 iL \over g^2} \int_{< r_0 , 1} d^3 \text{x} \text{tr} ([A^{(1)}_k,A^{(2)}_4] F^{(1)}_{k4}) - {L \over g^2} \int_{< r_0 , 1} d^3 \text{x} \text{tr} ([A^{(1)}_k,A^{(2)}_4]^2 ) ~.
\end{equation} 
The integration region is within a sphere of radius $r_0$ centered around the first monopole. The main contribution ($\sim 1/d$) in \eqref{eq:2.26} comes from the first integral. Since we excluded the overall $A^{vev}_4$ from $A^{(2)}_4$ within the two-sphere of radius $r_0$ around the first monopole, we have\footnote{$\tau^3_{(j)}$ refers to the $\tau^3$ Pauli matrix placed in the j-th Lie subalgebra of $SU(N)$ along the diagonal.} $A^{(2)}_4 \approx -{1 \over d} {{\tau^3_{(j)} \over 2}}$ ; one can check that there are no other (Dirac-string independent) terms that  can contribute order $1/d$ interaction terms. We can work out the integrand of the first integral, using the $A_4^{(2)}$ asymptotics just given, as:
\begin{equation}
\label{eq:2.27}
\text{tr} ([A^{(1)}_k,A^{(2)}_4] F^{(1)}_{k4}) = \text{tr} ([F^{(1)}_{k 4}, A^{(1)}_k] A^{(2)}_4) = -{i \over 4d}  ( F^{(1),1}_{k4} A^{(1),2}_k - F^{(1),2}_{k4} A^{(1),1}_k  ) \; q^i_{e1} \cdot q^j_{e2} ~.
\end{equation}
Here, $A^{(1),1}_k$ refers to the component of the $A^{(1)}_k$ along the first generator of the $i$-th $SU(2)$ subalgebra along the diagonal of an $SU(N)$ Lie algebra matrix (similar to $\tau^1 /2$ in $SU(2)$) and similarly for the others.
The values of the fields $A^{(1)}_k$ and $F_{k4}^{(1)}$  can be read from relations \eqref{eq:2.20} and \eqref{eq:2.21} for a self-dual ($B_k = E_k$) solution. For the integral of the first term in (\ref{eq:2.26}), we find $ \int_{< r_0, 1} d^3 \text{x} (F^{(1),1}_{k4} A^{(1),2}_k - F^{(1),2}_{k4} A^{(1),1}_k) = - 8 \pi \int\limits_{0}^{r_0 \nu} dy  y {F_1(y) {\cal{A}}(y)} = 4 \pi ({\cal{A}}^2(0) - {\cal{A}}^2(r_0 \nu)) \simeq 4 \pi + \text{O}(e^{- 2\nu r_0})$, where we used $y F_1 = \partial_y {\cal{A}} $. Thus, going back to (\ref{eq:2.26}), one obtains in total:
\begin{equation}
\label{eq:2.28}
S_{\text{inter.}, < r_0} = S^{2-1}_{\text{inter.}, < r_0} + S^{1-2}_{\text{inter.}, < r_0} = - {2\pi L \over g^2 d}\; q^i_{e1}\cdot q^j_{e2} - {2 \pi L \over  g^2 d}\; q^j_{e2}\cdot q^i_{e1} + \text{O}({L \nu^{-1} \over   g^2d^2})~.
\end{equation}
The $\text{O}({L \nu^{-1} \over   g^2d^2})$ error term comes from the evaluation of the second integral in \eqref{eq:2.26} and the error of the first integral in \eqref{eq:2.26} coming from the variation of $A_4^{(2)}$ from its value at the center of the sphere around the first monopole over the region of integration is of order $\text{O}({L \nu^{-2} \over   g^2d^3})$ which we have neglected.

Summing up the electric interactions in \eqref{eq:2.28} and \eqref{eq:2.25} we get $- {2\pi L \over  g^2 d}\; q^i_{e2} \cdot q^j_{e1}$, which shows a negative potential for same-sign  electric charges, hence an attractive electric force between the two monopoles with same electric charges. Since the electric interaction is mediated by the exchange of a massless (at the classical level) scalar field $A_4$, which is attractive for same sign charges, this is expected and was originally observed in \cite{25} using a slightly different approach. Although for simplicity we initially assumed that the solutions are BPS, eq.~\eqref{eq:2.28} is general, meaning that if we had done the same calculation in \eqref{eq:2.26} with two other monopoles (e.g. a KK and a BPS) we would have reached the same relation in \eqref{eq:2.28}, but with their appropriate electric charges replaced.

\item The $S_{\text{non-gauge-invariant}}$ term in \eqref{eq:2.23'} consists of any term in the action that depends on the Dirac-string singularity (in \eqref{eq:2.20} it occurs at $\theta = \pi$) or on its orientation. These non-gauge-invariant terms are unphysical and will be neglected; we were careful to only evaluate contributions that are independent of the Dirac string or its orientation. To be more specific on this matter, we notice that the $A^S_{\phi}$ component in \eqref{eq:2.20} is singular at $\theta = \pi$. Considering the commutator term $[A^{(1)}_i,A^{(2)}_j]$ in $F_{ij}$ for two far separated monopoles at the location of the string of the first monopole, we realize that the ${\text{tan} \ \theta \over 2r} {\tau^3\over 2}$ term in $A^{S, (1)}_{\phi}$ for this monopole does not commute with terms proportional to $\tau^1$ or $\tau^2$ in the components $A^{S, (2)}_{\theta}$ or $A^{S, (2)}_{\phi}$ of the second monopole therefore (even though these terms would be exponentially suppressed outside the sphere of radius $r_0$ of the second monopole) for the action they would give a term proportional to $\int^{\pi}_0 d \theta \text{sin} (\theta) \text{tan}^2 {\theta \over 2}$ which is singular when integrated near $\theta = \pi$. Or, similar to the electric interaction inside the monopole cores as in \eqref{eq:2.26}, another contribution can be evaluated for the magnetic interaction coming from the term $\approx$ ${\text{tan} {\theta \over 2} \over d} {\tau^3 \over 2}$ of $A^{S,(2)}_{\phi}$ near the center of the first monopole which would depend on the orientation of the Dirac string of the second monopole. These contributions are unphysical and a more precise treatment of the interaction of far separated monopoles, as in \cite{25,29,30}, does not involve any orientation dependent contributions, at least in the $d  \nu \gg 1$ limit, but will only involve interactions similar to the gauge-invariant interaction terms $S_{\text{inter.}, > r_0} + S_{\text{inter.}, < r_0}$ evaluated above.\footnote{This also implies that a more precise treatment (as opposed to simply summing far separated monopoles) for the construction of far separated monopole solutions is required, in particular, one that will not involve any non-gauge-invariant contributions. The construction of the monopole gas by summing far separated monopole solutions is appealing due to its simplicity and the fact that its leading gauge-invariant interaction terms reproduce results consistent with the more accurate far separated solutions, as studied in \cite{25,29,30}.}
\end{enumerate}

{\flushleft{To}} summarize, using the relations \eqref{eq:2.28}, \eqref{eq:2.25}, and \eqref{eq:2.23}, the (Dirac-string-independent) action of two far separated monopole solutions in the limit $\nu^{-1} = {NL \over 2\pi}  \ll r_0 \ll d$, with $|\text{x}_1 - \text{x}_2| = d$, can be summarized as:
\begin{equation}
\label{eq:2.29}
S_{2 - \text{monopoles}} = 2 \times {8 \pi^2 \over g^2 N} + {2 \pi L \over  g^2 d}\; {q^i_{m1} \cdot q^j_{m2}}  - {2 \pi L \over  g^2 d} \;{q^i_{e1} \cdot q^j_{e2}} + \text{O}({L \nu^{-1} \over g^2d^2})~.
\end{equation}
 Although  for simplicity we assumed that both solutions are BPS, the  relation \eqref{eq:2.29} is general and applies to two arbitrary monopoles or anti-monopoles. Therefore the action of a dilute gas of $n^{(i)}$ monopoles of type $i$ and $\bar{n}^{(i)}$ anti-monopoles of type $i$ for $i = 1, ..., N$, referring to the KK monopole as the monopole of type $N$,  with $n = \sum^{N}_{i = 1}({n}^{(i)} +\bar{n}^{(i)})$ their total number, is given by:\footnote{The $n^{4/3}$ power in the last term is an attempt at a better than naive estimate of the error. Naively, one could imagine the correction scaling as $n^2$, with $d$ being the typical separation between monopoles, but it is clear that not all monopoles are separated by the same distance. Assuming a uniform distribution of monopoles, with $d$ the closest distance between a given monopole and its neighbors, one can arrive at the estimate given (one expects some power $n^p$ with $1 < p < 2$). Note also the fact that not all $N$ types of monopoles have classical interactions, is not taken into account in writing the last term in (\ref{eq:2.30}).}
\begin{equation}
\label{eq:2.30}
S_{\text{monopole-gas}} = {8 \pi^2 \over g^2 N}n + S_{\text{int}, m} + S_{\text{int}, e} + \text{O}({n^{4/3} L \nu^{-1} \over g^2d^2})~.
\end{equation}
In (\ref{eq:2.30}), $S_{\text{int},m}$ ($S_{\text{int},e}$) is the sum of magnetic (``electric'') interaction terms similar to \eqref{eq:2.29} for every pair of monopoles in the gas:
\begin{equation}
\label{eq:2.31}
S_{\text{int},m} = {2\pi L \over g^2}\big{[} {1 \over 2} \sum_{ \underset{\text{dist. pairs}}{i, j, k_i, k_j} } {q^i_m  \cdot  q^j_m \over |r^{(i)}_{k_i} - r^{(j)}_{k_j}|} + {1 \over 2} \sum_{ \underset{\text{dist. pairs}}{i, j, \bar{k}_i, \bar{k}_j} }  {\bar{q}^i_m  \cdot  \bar{q}^j_m \over |\bar{r}^{(i)}_{\bar{k}_i} - \bar{r}^{(j)}_{\bar{k}_j}|} + \sum_{{i, j, k_i, \bar{k}_j }}  {q^i_m  \cdot  \bar{q}^j_m \over |r^{(i)}_{k_i} - \bar{r}^{(j)}_{\bar{k}_j}|} \big{]}~,
\end{equation}
\begin{equation}
\label{eq:2.32}
S_{\text{int},e} = - {2\pi L \over g^2} \big{[} {1 \over 2} \sum_{ \underset{\text{dist. pairs}}{i, j, k_i, k_j} } {q^i_e  \cdot  q^j_e \over |r^{(i)}_{k_i} - r^{(j)}_{k_j}|} + {1 \over 2} \sum_{ \underset{\text{dist. pairs}}{i, j, \bar{k}_i, \bar{k}_j} }  {\bar{q}^i_e  \cdot  \bar{q}^j_e \over |\bar{r}^{(i)}_{\bar{k}_i} - \bar{r}^{(j)}_{\bar{k}_j}|} + \sum_{{i, j, k_i, \bar{k}_j }}  {q^i_e  \cdot  \bar{q}^j_e \over |r^{(i)}_{k_i} - \bar{r}^{(j)}_{\bar{k}_j}|} \big{]}~,
\end{equation}
with 
$1 \leq i, j \leq N$, $1 \leq k_i \leq n^{(i)}$ and $1 \leq \bar{k}_i \leq \bar{n}^{(i)}$. The summation is being performed over distinct pairs of monopole-monopole and anti-monopole-anti-monopole interactions and a factor of $1 \over 2$ has been included to cancel the double counting of pairs in these summations.
Note that since the notion of anti-monopole is in regard to opposite magnetic charges, although $\bar{q}^j_m = -q^j_m$, this is not true for the electric  charges, which satisfy $\bar{q}^j_e = q^j_e$.

The reader should also be reminded that the electric interaction term will not be important for us in the quantum theory since it is gapped due to the one loop effective potential for the $A_4$ field and hence is of short range ($\sim { \nu^{-1}}$). Therefore in the next  Section we will be only concerned with the magnetic interaction term \eqref{eq:2.31}.

\subsection{Derivation of the string tension action}
\label{sec:2.4.1}
The static quark-antiquark potential in a representation $r$ of the gauge group is determined by evaluating the expectation value of a rectangular Wilson loop of size $R\times T$ in representation $r$, and  considering the leading exponential in the large Euclidean time $T$ limit \cite{04}:
\begin{equation}
\label{eq:2.33}
\underset{T\rightarrow \infty}{\lim} \langle W_r (R,T) \rangle = \underset{T\rightarrow \infty}{\lim} \langle \text{tr}_r ({\cal{P}}\text{exp}(\int_{R\times T}A_{\mu}dx^{\mu}))\rangle \ \sim \ \text{exp}(-V_r(R)T)~.
\end{equation}
In confining gauge theories in the absence of string breaking effects the potential $V_r(R)$ has a linear behaviour $V_r(R) = \sigma_r R$ at large distances,\footnote{At distances $R$ $\gtrapprox \Lambda^{-1}$ with $\Lambda$ being the strong scale of the theory.}  where  $\sigma_r$ is referred to as the string tension for quarks in the representation $r$.  At intermediate distances ($\approx \Lambda^{-1}$), the string tension can have a dependence on the particular representation $r$, it is known that the asymptotic---a few $\Lambda^{-1}$ and more---string tension, because of  colour screening by gluons, depends only on the $N$-ality $k$ of the representation $r$, hence asymptotically $\sigma_r$ is referred to as the $k$-string tension $\sigma_k$.

In this  Section we will be deriving an expression for the $k$-string tensions in dYM theory by evaluating  \eqref{eq:2.33}.
We want to calculate \eqref{eq:2.33} using the low energy degrees of freedom, to leading order in the limit of $NL \Lambda  \rightarrow 0$:
\begin{equation}
\label{eq:2.34}
\langle W_r (R,T)\rangle = {\int [D\psi][D\bar{\psi}][D A] \text{tr}_r {\cal{P}} \text{exp}(i\oint_{R \times T} dx_{\mu} A^{\mu}) \text{exp}(-S_{dYM}) \over \int [D\psi][D\bar{\psi}][A] \text{exp}(-S_{dYM}) }~.
\end{equation}
To evaluate the partition function $Z = \int [D\psi][D\bar{\psi}][DA]\text{exp}(-S_{dYM})$, we expand the action around the perturbative and non-perturbative minimum action configurations (the $2N$ minimum action monopole solutions discussed in  Section \ref{sec:2.2.2}), including the contribution of the approximate saddle points made up of far separated monopole configurations (dilute gas of monopoles), to second order and evaluate  the functional determinant in these backgrounds, using the approximate  factorization of determinants around widely separated monopoles. The result is the grand canonical partition function of a multi-component Coulomb gas \cite{01}\footnote{ {More details regarding the derivation of this partition function can be found in \cite{01}. In this  Section we will use this partition function to derive the Wilson loop inserted dual photon action for the evaluation of the $k$-string tensions. $Z_{\text{pert.}}$ refers to the perturbative contribution of the effective dual photon action: $Z_{\text{pert.}} =  { \int D[ \sigma] \;\text{exp}(-\int_{{\rm I\!R}^3} d^3\text{x} {1 \over 2} {g^2 \over 8 \pi^2 L}( \nabla \sigma )^2 )}$.}}:
\begin{equation}
\label{eq:2.35}
Z = {Z_{\text{pert.}}} \underset{i=1} {\overset{N}{\Pi}} \{ \sum^{\infty}_{n^{(i)} = 0} {\zeta^{n^{(i)}} \over n^{(i)} !} \sum^{\infty}_{\bar{n}^{(i)} = 0} {\zeta^{\bar{n}^{(i)}} \over \bar{n}^{(i)} !} \int_{{\rm I\!R}^3}\underset{k=1} {\overset{n^{(i)}}{\Pi}} d\text{r}^{(i)}_k \int_{{\rm I\!R}^3}\underset{l=1} {\overset{\bar{n}^{(i)}}{\Pi}} d\bar{\text{r}}^{(i)}_l    \} \text{exp}(- S_{\text{int},m})~,
\end{equation}
where the product over $i$ implies the inclusion of   the $N$ types of minimal action BPS and KK monopole-instantons (and anti-monopole-instantons)  and the sum over $n^{(i)}, \bar{n}^{(i)}$ indicates that arbitrary numbers of such configurations with centers at  $r_k^{(i)},\bar r_k^{(i)}$ are allowed. 
For any term in \eqref{eq:2.35} involving $n^{(i)}$ monopoles and $\bar{n}^{(i)}$ anti-monopoles for $i = 1, ..., N$, $S_{\text{int},m}$ is given by \eqref{eq:2.31} and the fugacity is:
\begin{equation}
\label{eq:2.36}
\zeta = C \text{e}^{-S_0} = \bar{A}  {\bar{D}_f} \; m_W^3   (g^2(m_W)  N)^{-2}  \text{e}^{-8 \pi^2 /N g^2(m_W)}, 
\end{equation}
 similar to the expression for the fugacity derived in \cite{01}. The only difference is that now   ${\bar{D}_f}$ the finite part of ${D_f \equiv {\text{det}^{n_f}(\slashed{D} + m) \over \text{det}^{n_f}(\slashed{D} + M )}}$, the Pauli-Villars regulated determinant of massive adjoint fermions, is replacing $\text{e}^{- \Delta S}$ in the expression for fugacity in \cite{01} (instead of the $\Delta S$ term in \eqref{eq:2.5} we now have massive adjoint fermions, of mass $m \sim m_W$, in \eqref{eq:2.6}).  $\bar{A}$ is a dimensionless and $N$-independent coefficient and the finite part of $D_f$ can be absorbed in its redefinition (after taking into account its effect on coupling renormalization; we omit any details on this). 
 
 Consider now the following identity,\footnote{For simplicity it has been written for only two insertions of $e^{i q \cdot \sigma}$.
} where $\sigma$ denotes the $N$-component vector $(\sigma_1,$ $\sigma_2,...\sigma_N)$:
\begin{equation}
\label{eq:2.37}
\begin{split}
& \int D[ \sigma] \;\text{exp}(-\int_{{\rm I\!R}^3} d^3\text{x} {1 \over 2} {g^2 \over 8 \pi^2 L}( \nabla \sigma )^2 ) \; \text{exp}( iq^i_m \cdot \sigma (\text{x}_1)) \;\text{exp}( iq^j_m \cdot \sigma(\text{x}_2)) = \\ & {Z_{\text{pert.}}} \times  \;  \text{exp} (-{2 \pi L \over g^2} {q^i_m\cdot q^j_m \over |\text{x}_1 - \text{x}_2|} ) \; \text{exp} (-2 \times {2 \pi L \over g^2} \lim_{|\text{x}| \rightarrow 0} {1 \over |\text{x}|})~.
\end{split}
\end{equation}
After regularizing the infinite self-energies ($\lim_{|\text{x}| \rightarrow 0} \{{1 \over |\text{x}|} -{\text{exp} (-\mu |\text{x}|) \over |\text{x}|} \} = \mu$) using the Pauli-Villars method, a typical term in \eqref{eq:2.35}, abbreviated t.t. below, using the analog of \eqref{eq:2.37} for $n$ monopoles, with $n^{(i)}$ monopoles and $\bar{n}^{(i)}$ anti-monopoles of each kind, can be written as:
 \begin{equation}
\label{eq:2.38}
\begin{split}
Z_{\text{t.t.}} & = \int D[ \sigma]\; \text{exp}(-\int_{{\rm I\!R}^3} d^3\text{x} {1 \over 2} {g^2 \over 8 \pi^2 L}( \nabla \sigma )^2 ) \; \times \\ & \underset{i=1} {\overset{N}{\Pi}} \{ {( \widetilde{N} \zeta)^{n^{(i)}} \over n^{(i)} !} { (\widetilde{N} \zeta )^{\bar{n}^{(i)}} \over \bar{n}^{(i)} !} \underset{k=1} {\overset{n^{(i)}}{\Pi}} \int_{{\rm I\!R}^3} d\text{r}^{(i)}_k \text{exp}( iq^i_m  \cdot  \sigma (\text{r}^{(i)}_k)) \underset{l=1} {\overset{\bar{n}^{(i)}}{\Pi}} \int_{{\rm I\!R}^3} d\bar{\text{r}}^{(i)}_l \text{exp}( i\bar{q}^i_m  \cdot  \sigma(\bar{\text{r}}^{(i)}_l) ) \}~,
\end{split}
\end{equation}
where $\tilde{N} = \text{exp} ( + {2 \pi L \over g^2} \mu )$. 

 Before we continue, we pause to note that the scalar fields $(\sigma^1,...,\sigma^N)$ are the magnetic duals to the U(N) Cartan-subalgebra electric gauge fields, the so-called ``dual photons.'' For the purpose of the paragraph that follows, in order to elucidate the physical meaning of gradients of the $\sigma$ fields, we revert to Minkowski space. The duality relation, with $(+,-,-)$ metric, is
 \begin{equation}
 \label{dualityrelation}
 F_{kl}^A =  - {g^2 \over 2\sqrt{2} \pi L} \epsilon_{klm} \partial^m \sigma^A~,~~A=1,...,N~.
 \end{equation}
 The kinetic term in the Minkowski space version of (\ref{eq:2.38}) is nothing but a rewriting of the first (``magnetic'') term in Minkowski space version of the action (\ref{eq:2.23'}) restricted to its Cartan subalgebra and considered for a U(N) gauge group via dual variables.\footnote{See  Footnote \ref{modes1} for  the relation between the $U(N)$ and $SU(N)$ Cartan fields and further comments on the duality.} In order to do this in a proper way consider the Minkowski space action of the 3-dimensional low energy theory in perturbation theory with the Bianchi identity imposed as a constraint via the auxiliary field $\sigma$ to eliminate gauge degrees of freedom:
 \begin{equation}
 \label{bianchi}
S = \int_{\mathbb{R}^{1,2}} \{ -{L \over 4 g^2}F^A_{kl}F^{A  kl} +  h   \epsilon_{k l m } \partial^m F^{A  kl}\sigma^A \}~, \;\;\;\;\;\; A =1, ...,N.
 \end{equation}
Integrating by parts the Lagrange multiplier term, completing the square of the $F^a_{kl}$ fields and integrating them out leaves an action only in terms of the dual fields $\sigma$: $S_{\text{dual}} = {2 \over L} h^2 g^2 \int_{\mathbb{R}^{1,2}} (\partial_k \sigma)^2 $. Demanding ${2 \over L} h^2 g^2 = {g^2 \over 2} {1 \over 8 \pi^2 L}$, the coefficient of the gradient of the $\sigma$ fields in \eqref{eq:2.38}, gives $h = 1/(4\sqrt{2}\pi)$. Varying the action \eqref{bianchi} with respect to $F_{kl}$, we obtain the duality relation \eqref{dualityrelation}. An immediate remark, relevant for the discussion in  Section \ref{bagmodelsection}, is   that the duality relation (\ref{dualityrelation}) implies that spatial gradients of $\sigma$ represent perpendicular electric fields $\vec{E}^A$, i.e. 
\begin{equation}
E_i^A  \equiv F_{i0}^A =  {g^2 \over 2 \sqrt{2} \pi L} \epsilon_{ij} \partial_j \sigma^A~. \label{dualityelectric}
\end{equation} 

Returning to our main objective---obtaining the effective theory of the dYM vacuum---we sum  over the contributions of all monopoles and antimonopoles in (\ref{eq:2.38}), and find that  the full partition function becomes:
\begin{equation}
\label{eq:2.39}
\hspace*{-1cm}Z = \int D[ \sigma] \text{exp}(-\int_{{\rm I\!R}^3} d^3\text{x} {1 \over 2} {g^2 \over 8 \pi^2 L}( \nabla \sigma )^2 ) \sum^{\infty}_{n = 0} { (\widetilde{N} \zeta )^{n} \over n!} \{ \sum^{N}_{i = 1} \int_{{\rm I\!R}^3} d^3\text{x}( e^{iq^i_m  \cdot  \sigma (\text{x})} + e^{i\bar{q}^i_m  \cdot  \sigma (\text{x})} ) \}^n~.
\end{equation}
Thus, the final form of the dual photon action reads:
\begin{equation}
\label{eq:2.40}
Z = \int D[ \sigma] \text{exp}(-\int_{{\rm I\!R}^3} d^3\text{x} \{ {1 \over 2} {g^2 \over 8 \pi^2 L}( \nabla \sigma )^2 - \widetilde{\zeta}\; \sum^{N}_{i = 1} \text{cos}(q^i_m  \cdot  \sigma) \} )~,
\end{equation}
where $\widetilde{\zeta} = 2 \tilde{N} \zeta$.\footnote{\label{modes1}The remarks that follow are tangential to our exposition, but serve to convince the reader of the consistency  of our dual action coefficients with charge quantization  (our Eq.~(\ref{eq:2.40}) was derived solely by demanding that the long-distance interactions between monopole-instantons from  Section \ref{2.3} are correctly reproduced)  and to correct minor typos in  expressions that have appeared previously in the literature.  Integrating the duality relation (\ref{dualityelectric}), we obtain  \begin{equation} 
{g^2 \over  \sqrt{2}   L}\;  \; \oint_C { d \sigma^A \over 2 \pi} = \oint_C d \vec{n} \cdot \vec{E}^A,
\label{dual12}
\end{equation} 
 representing the fact that a static electric charge inside $C$ generates flux through $C$ ($\vec{n}$ is an outward unit normal to $C$) which duality relates to the $\sigma$-field monodromy around $C$. Eq.~(\ref{dual12})  implies that the $\sigma^A$ fields have periodicities determined by the fundamental electric charges. 
To find them, we begin with the relation between the $N$ Cartan field strengths $F^A_{kl}$ that first appeared in (\ref{dualityrelation}, \ref{bianchi}) and the original $N-1$ $SU(N)$ fields $F^a_{kl}$ in (\ref{eq:2.1}). The reader can convince themselves that it is given by $F^A_{kl} = {1 \over \sqrt{N}} F_{kl}^0  + \sum\limits_{a=1}^{N-1} F_{kl}^a \lambda^{a A}$, with $\lambda^{aA} \equiv (\theta^{aA} - a \delta^{a+1, A})/\sqrt{a(a+1)}$, where $\theta^{aA} = 1$ for $a \ge A$ and $\theta^{aA}=0$  otherwise. The spectator $U(1)$ field $F^0_{kl}$  is not coupled to dynamical sources. The relations $\sum\limits_{A=1}^N \lambda^{a A} \lambda^{b A} = \delta^{ab}$ and $ \sum\limits_{A=1}^N \lambda^{a A} =0$  help establish that $\sum\limits_{A=1}^N (F^A)^2 = \sum\limits_{a=1}^{N-1} (F^a)^2 + (F^0)^2$. A fundamental static charge is represented by the insertion of a static Wilson loop in the fundamental representation. Early on, see (\ref{eq:2.1}), we stated that our fundamental representation  generators are normalized as  tr($T^a T^b) = \delta^{ab}/2$. Thus, using the definitions just made,   it follows that the fundamental representation Cartan generators are $T^a = {\rm diag}(\lambda^{a 1},..., \lambda^{a N})/\sqrt{2}$. A  fundamental static Wilson loop  is then represented by insertions of  $\int dt A_0^a(\vec{r},t) \lambda^{aA}/\sqrt{2}$  ($A$ labels the Wilson loop eigenvalues) in the path-integral action. Considering one of the eigenvalues of the Wilson loop (one component of the fundamental static quark), the corresponding electric flux is  found by solving the static equation of motion, ${L \over g^2} \nabla^2 A_0^a = {\lambda^{a A}\over \sqrt{2}}\delta(\vec{r})$, thus $ \oint_{C_A} d \vec{n}\cdot \vec{E}^a = -  \lambda^{a A}{g^2 \over   \sqrt{2} L}$. From the earlier relations, we also have that $\vec{E}^a = \sum\limits_{A=1}^N \lambda^{a A} \vec{E}^A$, thus $\sum\limits_{B=1}^N \lambda^{a B} \oint_{C_A}  d \vec{n} \cdot \vec{E}^B =  {g^2 \over   \sqrt{2} L} \lambda^{a A}$. Finally, from (\ref{dual12}), this leads to $\sum\limits_{B=1}^N \lambda^{a B}  \oint_{C_A} { d \sigma^B \over 2 \pi} =  \lambda^{a A}$. It can be already seen, from the explicit form of $\lambda^{a A}$, that this relation implies that the periodicity (monodromies) of differences of $\sigma^A$'s have to be proportional to $2 \pi$. Even more explicitly, from the relation $\sum\limits_{a=1}^{N-1} \lambda^{a A} \lambda^{a B} = \delta^{AB} - {1 \over N}$, one finds that the monodromies of the dual photons are given by $2 \pi$ times the weights of the fundamental representation.
 This  is consistent with the periodicities of the potential terms in (\ref{eq:2.39}) and with the dual photon actions given in   e.g.~\cite{Simic:2010sv, 23, 08}.}

 Before working out the Wilson loop integral, we will derive the $NL \Lambda$ dependence of the fugacity and the dual photon mass, verify the dilute gas  limit conjecture and discuss the hierarchy of scales in this theory. Using the one loop renormalization group invariant scale $\Lambda$ for $n_f = 1, 2$ flavours of Dirac fermions in the adjoint representation of the gauge group:
\begin{equation}
\label{eq:2.41}
\Lambda^{b_0} = \mu^{b_0} \; \text{exp}(- {8 \pi^2 \over N g^2}),  \ \ \ \ \ \ \ \ b_0 = (11 - 4n_f)/3~,
\end{equation}
we can determine the leading dependence of the fugacity on $NL\Lambda$. The Pauli-Villars scale used in (\ref{eq:2.37}) should be thought of as the cutoff of the long-distance theory containing no charged excitations and should be taken below the scale of any charged excitation; for the sake of definiteness, we shall take   $\mu \sim {g \sqrt{N} \over N L}$, the lowest eigenvalue of the holonomy fluctuations. Then, we can neglect $\tilde{N}$ compared to $\text{exp} (- S_0)$ in the small-$NL \Lambda$ limit.\footnote{We note that with this choice the effect of $\mu$ on $\tilde\zeta$ is comparable to the effect of finite $A_4$ mass on the classical monopole action, an effect that we have neglected throughout. Matching between the UV theory, valid at scales $\ge m_W$, and the IR theory, valid at scales $\ll m_W$, to better precision that has been attempted so far is needed to properly account for  these effects.  We also note that in the supersymmetric case, the only case where the determinants in the monopole-instanton backgrounds have actually been computed, where $\sigma$ is replaced by a chiral superfield and the monopole-instantons are ``localized in superspace,'' this ambiguity is absent \cite{08}---in super-Yang-Mills, divergent self energies of monopoles due to electric and magnetic charge cancel out in the analogue of (\ref{eq:2.37}).} The one-loop massive fermion determinant contributes some calculable constant  and renormalizes the coupling of the long-distance theory, as already accounted for in (\ref{eq:2.41}). From \eqref{eq:2.36} and \eqref{eq:2.41}, neglecting any $\text{log}(NL \Lambda)$ (or, equivalently, $g^2 {N}$) dependence, for the leading $NL \Lambda$ dependence  of the fugacity  we obtain:
\begin{equation}
\label{eq:2.42}
\widetilde{\zeta} \sim \zeta \sim ({ 1 \over NL })^3 (NL)^{b_0}  \Lambda^{b_0} = (NL\Lambda)^{b_0 - 3}\Lambda^{3}~.
\end{equation}
The fugacity $\tilde{\zeta}$ or $\zeta$ is proportional to the monopole density $n_d$. For a gas with density $n_d$ the average distance between the particles in the gas is $\sim {1 \over n_d^{1/3}}$ and  in order to verify the dilute gas conjecture we should have that this separation be much larger than the size of the monopoles, of order $NL$:
\begin{equation}
\label{eq:2.43}
d \sim {1 \over \zeta^{1/3} } \gg  NL \longrightarrow (NL \Lambda)^{-{b_0\over 3}} \gg 1 \longrightarrow b_0 > 0 , \ \ n_f \leq 2~ .
\end{equation}
 Since we are working in the limit $NL \Lambda \rightarrow 0$, the condition \eqref{eq:2.43} will be satisfied if $b_0 > 0$, which is the same condition as the asymptotic freedom condition and gives $n_f \leq 2$ (or $n_f \leq 5/2$ if Majorana masses are considered instead).
 
 The mass of the dual photon can be read from \eqref{eq:2.40}. The coefficient of the quadratic term in the dual photon action, after expanding the cosine term and factoring out ${g^2 \over 8 \pi^2 L}$ is ${8 \pi^2 L \over 2g^2}  \tilde{\zeta}$. 
 We define the dual-photon mass scale 
 \begin{equation}
 \label{eq:2.44}
 m^2_{\gamma} = {8 \pi^2 NL \over Ng^2}  \; \widetilde{\zeta} \sim (NL\Lambda)^{b_0-2} \;\Lambda^2 ~, \end{equation}
  and note that it is of order the mass of the heaviest dual photon (the dual photon mass eigenvalues, after diagonalizing the quadratic term via a discrete Fourier transform, are $ m_\gamma \sin {\pi k \over N}$ with $k=1,...,N-1$).

 Thus, the hierarchy of scales in this theory can be summarized as: 
\begin{equation}
\label{eq:2.45}
 {\begin{split}
 m_W \sim {1 \over NL}\; &\gg \; m_H \sim {g \sqrt{N} \over NL} \sim \mu \; \gg {1 \over d} \sim ({1 \over NL \Lambda })^{1 - {b_0 \over 3}} \Lambda \; \gg \; m_{\gamma} \sim ({1 \over NL \Lambda})^{1 - {b_0 \over 2}} \Lambda~.
\end{split}}
\end{equation}
For $n_f = 1$, we have  ${1 \over d} \gg \Lambda \gg m_{\gamma}$ , but for $n_f = 2$: ${1 \over d} \gg m_{\gamma} \gg \Lambda$; we stress again that the scale $\Lambda$ has no physical significance in the small-$L$  theory (except that to ensure weak coupling, we must have $m_W \gg \Lambda$).

Now we will work out the Wilson loop integral. In the dilute gas limit ($NL\Lambda \rightarrow 0$) the leading contribution to the Wilson loop integral comes from the long distance abelian behaviour of the monopole gas far from the cores ($\sim NL$) of the monopoles. Therefore for a Wilson loop in the $\text{x}_1, \text{x}_2$ plane and representation $r$ with $N$-ality $k$ and for a typical monopole gas background involving $n$ monopoles we have:
\begin{equation}
\label{eq:2.46}
\begin{split}
\hspace{-1cm}\{ \text{tr}_r \ \text{exp}( i \oint_{R\times T} A^c_m t_r^c d\text{x}^m) \}_{\text{typ. mon.}} & = \text{tr}_r \ \text{exp} (i \int_{S(R\times T)}\epsilon_{anm} \partial_n A^c_m t_r^c d\text{S}^a ) \\ & = \sum^{d(r)}_{j=1} \text{exp}( i \int_{S(R\times T)} d\text{x}_1d\text{x}_2 \overset{n}{\underset{i=1}{\sum}} \mu^j_r \cdot q^i_m\; {\text{R}^i_3 \over 2 |R^i - \text{x}|^{3}})~,
\end{split}
\end{equation}
where $c=1, ..., N-1$ labels the Cartan generators of $SU(N)$ in the representation $r$ and $\mu^j_r$ is it's $j$-th weight. On the first line above, we used Gauss' law to rewrite the Wilson loop integral as an integral of the magnetic flux through a surface $S$ spanning the loop, and on the second line, we replaced the magnetic field by the field of $n$ monopole-instantons at positions $R^i \in \R^3 $, $i=1,...,n$.\footnote{A more detailed derivation of \eqref{eq:2.46} is done in Appendix \ref{sec:B2}.} Defining  the solid angle $\eta(\text{x})$ that the Wilson loop is seen at from the point $\text{x} \in \R^3$, $\eta(\text{x})$  $\equiv$ $\int_{S(R\times T)} d\text{y}_1d\text{y}_2 {\text{x}_3 \over 2 |y - \text{x}|^{3}}$, we have for the contribution to the Wilson loop expectation value of an $n$-monopole configuration:
\begin{equation}
\label{eq:2.47}
\{ \text{tr}_r \ \text{exp}( i \oint_{R\times T} A_m d\text{x}^m) \}_{\text{typ. mon.}} = \sum^{d(r)}_{j=1} \text{exp}( i \sum^n_{i=1} \mu^j_r \cdot q^i_m \eta (\text{R}^i) )~.
\end{equation}
Comparing with \eqref{eq:2.38}, we see that the effect of the Wilson loop insertion is to shift the $\sigma (\text{r}^{(i)}_k)$ field multiplying the magnetic charges in \eqref{eq:2.38} by $\mu^j_r \eta (\text{r}^{(i)}_k)$ (and similarly for $\sigma (\bar{\text{r}}^{(i)}_k)$ and the $\sigma(\text{x})$ field in \eqref{eq:2.39}). Thus, shifting the $\sigma(\text{x})$ field by $\sigma(\text{x}) \rightarrow \sigma(\text{x}) - \mu^j_r \eta (\text{x})$ gives the final form of the expectation value of the Wilson loop in dYM theory to leading order in $NL\Lambda \rightarrow 0$, calculated using  the low energy effective theory:
\begin{equation}
\label{eq:2.48}
\langle W_r(R,T)\rangle  = \int D[ \sigma] \sum^{d(r)}_{j=1} \text{exp}(-\int_{{\rm I\!R}^3} d^3\text{x} \{ {1 \over 2} {g^2 \over 8 \pi^2 L}( \nabla \sigma - \mu^j_r \nabla \eta )^2 - \widetilde{\zeta} \sum^{N}_{i = 1} \text{cos}(q^i_m \cdot \sigma) \} ) / Z~.
\end{equation}
The string tension action is given by making a saddle point approximation to \eqref{eq:2.48}. The Lagrange equations of motion for the contribution of the $j$-th weight of $r$ to the Wilson loop expectation value are:
\begin{equation}
\label{eq:2.49}
\nabla^2 \sigma_i = 2 \pi (\mu^j_r)_i \theta_A(\text{x}_1, \text{x}_2) \partial_3 \delta (\text{x}_3) + m^2_{\gamma} (\text{sin}(\sigma_i - \sigma_{i+1}) + \text{sin}(\sigma_i - \sigma_{i-1})) , \ \ \ i = 1, ..., N~,
\end{equation}
where $\sigma_{0} \equiv \sigma_N$, $ \sigma_{N+1} \equiv \sigma_{1}$ and $\nabla^2 \eta (\text{x}) = 2 \pi\theta_A(\text{x}_1, \text{x}_2) \partial_3 \delta (\text{x}_3)$. $\theta_{A}(\text{x}_1, \text{x}_2)$ is one for $(\text{x}_1, \text{x}_2) \in A$ and zero for $(\text{x}_1, \text{x}_2) \notin A$, with $A$ being the area of the Wilson loop $R \times T$ in the $\text{x}_1, \text{x}_2$ plane. For large Wilson loops ($R, T \rightarrow \infty$) the saddle point configuration of \eqref{eq:2.48} is zero for regions outside the Wilson loop. The solution near the boundaries would be more complicated and gives a contribution proportional to the perimeter of the Wilson loop. The solution interior to the Wilson loop far from the boundaries would depend on $\text{x}_3$ only. The corresponding one-dimensional equation is:
\begin{equation}
\label{eq:2.50}
{\partial^2 \sigma_i \over \partial \text{x}_3 ^2} = 2 \pi (\mu^j_r)_i \theta_A(\text{x}_1, \text{x}_2) \partial_3 \delta (\text{x}_3) + m^2_{\gamma} (\text{sin}(\sigma_i - \sigma_{i+1}) + \text{sin}(\sigma_i - \sigma_{i-1})) , (\text{x}_1 , \text{x}_2) \in A~.
\end{equation}
Eq.~\eqref{eq:2.50} represents a boundary value problem showing a discontinuity of $2 \pi (\mu^j_r)_i$ for the $\sigma_i$ $(i=1,...,N)$ fields at $\text{x}_3 = 0$. Therefore \eqref{eq:2.48} to leading order in $NL \Lambda \rightarrow 0$ (the saddle point approximation is valid in this limit) and $R, T \rightarrow \infty$ is
\begin{equation}
\label{eq:2.51}
\langle W_r(R,T)\rangle  \sim \sum^{d(r)}_{j=1} \text{exp}(- T^j_r RT)~,
\end{equation}
given by a sum over the exponential contributions of the different weights of the representation $r$. 
Sources in every weight have their own string tension, 
  $T^j_r$, given by:
\begin{equation}
\label{eq:52}
T^j_r = \underset{\sigma(\text{x}_3)}{\text{min}}\int^{+\infty}_{-\infty} d\text{x}_3 \{{1\over {2L}}{g^2 \over {8\pi^2}}({\partial \sigma \over \partial \text{x}_3})^2+ \tilde{\zeta} \overset{N}{\underset{i=1}{\sum}}[1- \text{cos}(\sigma_i-\sigma_{i+1})]\} \Big |_{\Delta \sigma(0) =2 \pi \mu^j_r}~,
\end{equation}
with $\Delta \sigma(0) \equiv \sigma(0^+) - \sigma(0^-)$. 

Notice that because the long-distance theory is abelian,  within the abelian theory, we can insert static quark sources with charges given by any $\mu_r^j$, $j=1, ... , d(r)$. Clearly, this is not the case in the full theory, where the entire representation appears and color screening from gluons is operative. In our $N L \Lambda \rightarrow 0$ limit, the gauge group is broken,  $SU(N) \rightarrow U(1)^{N-1}$, and the screening is due to the heavy off-diagonal $W$-bosons, which were integrated out to arrive at (\ref{eq:2.51}).  Thus, we expect that  at distances $R$ such that  $T_r^j R > O(m_W)$, $W$-bosons can be produced (as in the Schwinger pair-creation mechanism)  causing the strings in representation $r$ with higher string tensions to decay to the string with lowest tension in $r$. Hence, we shall not study all $T^j_r$ tensions, but will focus only on the strings of lowest tension confining quarks in representation $r$.\footnote{The order of magnitude of the string tension is $g^2 N m_\gamma m_W$. Thus  $W$-boson production takes place once $R \sim {\rm O}(1/(g^2 N m_\gamma))$ and  Higgs production (recall $m_H \sim g \sqrt{N} m_W$)  when $R \sim {\rm O}(1/(g \sqrt{N} m_\gamma))$. Notice that the values on the r.h.s., owing to small coupling, are much larger than the Debye screening length $1/m_\gamma$. }

It is shown, in Appendix \ref{sec:B1}, that any representation of $SU(N)$ with $N$-ality $k$ ($= 1, ..., N-1$) contains the $k$-th fundamental weight, $\mu_k$ (given by \eqref{eq:3.2} below) as one of its weights. Furthermore, in  Section \ref{sec:5.1.1} it is shown that this weight would give the lowest string tension action among the other weights of that representation. Therefore \eqref{eq:2.51} reduces to:
\begin{equation}
\label{eq:2.53}
\langle W_r (R,T)\rangle \sim \text{exp}(- T_k RT)~,
\end{equation}
up to pre-exponential factors and subdominant terms corresponding to the higher string tensions.
The $k$-string tension $T_k$, defined by:
\begin{equation}
\label{eq:54}
T_k=\underset{\sigma(\text{x}_3)}{\text{min}}\int^{+\infty}_{-\infty} d\text{x}_3 \{{1\over {2L}}{g^2 \over {8\pi^2}}({\partial \sigma \over \partial \text{x}_3})^2+ \tilde{\zeta} \overset{N}{\underset{i=1}{\sum}}[1- \text{cos}(\sigma_i-\sigma_{i+1})]\} \Big |_{\Delta \sigma(0) =2 \pi \mu_k}
\end{equation}
will be the object of our numerical and analytical studies in the rest of this paper.

\section{String tensions in dYM: a numerical study}
\label{numericsection}

\subsection{String tension action}
\label{sec:3.1}

As derived in  Section \ref{sec:2.4.1} the ``$k$-string tension action'' is given by:
\begin{equation}
\label{eq:3.1}
T_k=\underset{\sigma(\text{z})}{\text{min}}\int^{+\infty}_{-\infty} d\text{z}\{{1\over {2L}}{g^2 \over 8\pi^2}({\partial \sigma \over \partial \text{z}})^2+ \tilde{\zeta}\; \overset{N}{\underset{j=1}{\sum}}[1- \text{cos}(\sigma_j-\sigma_{j+1})]\} \Big |_{\Delta \sigma(0) =2 \pi \mu_k}, \ \ \sigma_{N+1}\equiv \sigma_1~,
\end{equation}
 {where $\mu_k$, the fundamental weights of $SU(N)$, are obtained by solving the equation \;\;\;\;\;     ${2\alpha_i \cdot \mu_k / \left | \alpha_i \right |^2}=\delta_{ik}$~\cite{09}, and are given by:}
\begin{equation}
\label{eq:3.2}
\mu_k=({N - k \over N}, ..., \overset {\text{k-th}} {\widehat{{N-k \over N}}}, {-k \over N}, ..., {-k \over N}),\ \ \ \ \ \ 1\leq k \leq N-1,
\end{equation}
Here $\alpha_i$ are the simple roots of $SU(N)$,  given in their $N$-dimensional representation by:
\begin{equation}
\label{eq:3.3}
\alpha_i=(0,..,0,\overset {\text{i-th}}{\widehat{1}},-1,0,...,0), \ \ \ \ \ \ 1\leq i \leq N-1.
\end{equation}
In deriving the relation for the $\mu_k$'s, it is assumed that the $\mu_k$'s and $\alpha_i$'s span the same subspace in ${\rm I\!R}^N$, orthogonal to the vector $(1,1,1,...,1,1)$. The string tension action \eqref{eq:3.1} will be minimized when the discontinuity $\Delta \sigma(0) = 2 \pi \mu_k$ is equally split between $\sigma (0^+)$ and $\sigma (0^-)$.\footnote {This can be seen as follows. Consider the boundary conditions $\sigma (0^+) = \pi \mu_k + \bar{\text{z}}$ and $\sigma (0^-) = - \pi \mu_k + \bar{\text{z}}$. We will show that the minimum of \eqref{eq:3.1} is when $\bar{\text{z}} = 0$. If $\sigma(\text{z})$ is an extremum solution of \eqref{eq:3.1} so is $\sigma(-\text{z})$ and $- \sigma(\text{z})$, therefore it can be seen that $\bar{\text{z}} = 0$ is an extremum point. It is a minimum since otherwise the kinetic term will increase if we make the magnitude of the boundaries larger than $\pi (\mu_k)_j$ on either side of $\text{z} = 0^+$ or $\text{z} = 0^-$.}  Therefore the value of the action would also be equally split between the positive and negative $z$-axis and we can consider half the $k$-string action. Defining $m_{\gamma}^2 \equiv {8\pi^2 \over g^2} L \tilde{\zeta}$, the parameter-free form of half of \eqref{eq:3.1} is given by making the change of variable $z ={z' {1 \over \sqrt {2} m_{\gamma}}}$:
\begin{flalign}
\label{eq:3.4}
& {\sqrt {2} m_{\gamma} \over \tilde{\zeta}} {T_k \over 2} \equiv {\bar{T_k}}=\underset{f(\text{z}')} {\text{min}} \int_{0}^{+\infty} d\text{z}'\{({\partial f \over \partial \text{z}'})^2+\underset{j}{\sum}[1-\text{cos}(f_j-f_{j+1})]\} ~,\\ & f(+ \infty) = 0, f(0) = \pi \mu_k \ \text{and} \ f(\text{z}')=\sigma( {\text{z}' \over \sqrt {2} m_{\gamma}}) ~.\nonumber 
\end{flalign}
The equations of motion for $f$ are given by:
\begin{equation}
\label{eq:3.5}
{d^2f_j \over d\text{z}^2}={1\over 2} (\text{sin}(f_j-f_{j+1})+ \text{sin}(f_j-f_{j-1})),\ \  1\leq j \leq N, \  f_0\equiv f_{N}, \ f_{N+1} \equiv f_{1}.
\end{equation}
It is possible to solve the equations of motion \eqref{eq:3.5} directly for $SU(2)$ and $SU(3)$ and derive the exact value of \eqref{eq:3.4}.\footnote{That an ansatz with a single exponential works for $SU(3)$ is a consequence of the existence of only a single mass scale in the dual-photon theory, a fact that only holds for $N=2,3$.}  The solution for $SU(2)$ and its corresponding $\bar{T}_1$ value is:
\begin{equation}
\label{eq:3.6}
-f_2(\text{z}) = f_1(\text{z}) = 2 \text{ arctan}(\text{exp}(-\sqrt{2}\;\text{z}))\xrightarrow{\text{after inserting in \eqref{eq:3.4}}} \bar{T}_1 = 8/\sqrt{2}~.
\end{equation}
Due to charge conjugation symmetry $\bar{T}_{k} = \bar{T}_{N-k}$ \cite{01} hence for $SU(3)$ $\bar{T}_1 = \bar{T}_2$ and therefore it suffices to solve the equations of motion and find the action for the first fundamental weight $\mu_1$ of $SU(3)$:
\begin{equation}
\label{eq:3.7}
-2f_2(\text{z}) = -2f_3(\text{z}) = f_1(\text{z}) = {8 \over 3} \text{ arctan}(\text{exp}(-\sqrt{3/2}\;\text{z}))\xrightarrow{\text{after inserting in \eqref{eq:3.4}}} \bar{T}_1 = 16/\sqrt{6}~>
\end{equation}
In the following  Sections, these exact values will be used as a check on our numerical methods. 
\subsection{Discretization of the string tension action}
\label{sec:3.3}

We will obtain the numerical value of the string tensions in deformed Yang-Mills theory by discretizing \eqref{eq:3.4} and minimizing the multivariable function obtained upon discretization.
We set the boundary conditions at $f_j(0)=\pi (\mu_k)_j$ and $f_j(J)=0$ for $J > 0$.

To discretize \eqref{eq:3.4} divide the interval $[0,J]$ into $m$ partitions and consider the following array of discretized variables:
\begin{equation}
\label{eq:3.8}
\{f_{jl}\,|\,\,\, 1\leq j \leq N\,,\,\,\,0\leq l \leq m\,,\,\,\, f_{j0}=\pi (\mu_k)_j\,,\, f_{jm}=0\}~.
\end{equation}
We denote $\delta \text{z} = J/m$, and introduce the  discretized functions $f_j^{(m)}$, linearly interpolated in every interval of width $\delta \text{z}$:\footnote{The superscript $(m)$ indicates that this is the discretization with $m$ partitions of the interval.}
\begin{equation}
\label{eq:3.9}
f^{(m)}_j(\text{z})=f_{jh}+{f_{jh+1}-f_{jh}\over \delta \text{z}}( \text{z}-h\delta \text{z})\,, \,\,\, \text{z} \in [h\delta \text{z},(h+1)\delta \text{z}]\,,\, h = 0, ..., m-1~.
\end{equation}
Inserting \eqref{eq:3.9} in \eqref{eq:3.4} and performing the $\text{z}$ integration, the linearly discretized action is:
\begin{equation}
\label{eq:3.10}
\begin{split}
\bar{T}^{m,J}_k=\bar{T}^{m,J}_{k1}+\bar{T}^{m,J}_{k2}, \
&\bar{T}^{m,J}_{k1} = \underset{j,h}{\sum}{(f_{jh+1}-f_{jh})^2\over \delta \text{z}},
\\
&\bar{T}^{m,J}_{k2} = NJ-\delta \text{z} \underset{j,h}{\sum}{\text{sin}(f_{jh+1}-f_{j+1h+1})- \text{sin}(f_{jh}-f_{j+1h})\over{(f_{jh+1}-f_{j+1h+1})-(f_{jh}-f_{j+1h})}}~,
\end{split}
\end{equation}
where we split the discretized action into a kinetic ($\bar{T}^{m,J}_{k1}$) and potential ($\bar{T}^{m,J}_{k2}$)parts. Notice their different scalings with the width of partition $\delta \text{z}$ (we shall make use of this fact in  Section \ref{bagmodelsection} when discussing similarities with the MIT Bag Model). 

\subsection{Minimization of the string tension action and error analysis}
\label{sec:minimize}
In order to obtain more accurate numerical results and have control over the minimization process, a systematic method is utilized for minimizing the multivariable function \eqref{eq:3.10}. For sufficiently small $\delta \text{z}$, $\bar{T}^{m,J}_k$ has a parabolic structure along the direction of any variable $f_{lp}$ (i.e.${\partial^2\bar{T}^{m,J}_k\over \partial f_{lp}^2}>0$). The second derivative of the 1st term in \eqref{eq:3.10} with respect to $f_{lp}$ is ${4\over \delta \text{z}}$ and the second derivative with respect to the 2nd term is at least\footnote{Minimizing the second derivative of $\bar{T}^{m,J}_{k,2}$ with respect to $f_{lp}$, gives $-{\delta \text{z} \over 3}$ for each time the variable $f_{lp}$ appears in the sum over $j$ and $h$. Since it appears 4 times when replacing each variable $f_{jh+1}$, $f_{j+1h+1}$, $f_{jh}$ and $f_{j+1h}$ in the expression and the minimum value is the same for all 4 cases, this gives $-{4\over 3} \delta \text{z}$ for a lower bound on the 2nd derivative of the second term.}  $-{4\over 3}\delta \text{z}$, hence:
\begin{equation}
\label{eq:3.11}
{\partial^2\bar{T}^{m,J}_k\over \partial f_{jh}^2}> 0 \Longrightarrow {4\over \delta \text{z}}-{4\over 3}\delta \text{z} > 0 \Longrightarrow \delta \text{z} < \sqrt{3}~.
\end{equation}
To minimize \eqref{eq:3.10} we assume $\delta z$ small enough in order to satisfy \eqref{eq:3.11} and have a parabolic structure along the direction of any variable. The string tension action \eqref{eq:3.4} and its discretized form \eqref{eq:3.10} are positive quantities with the extremum solution of \eqref{eq:3.5} being the minimum point of the action, therefore following the parabolas along the direction of any variable downward should lead us to this minimum point. In order to do this in a systematic way $R$ random points are generated in the $N\times (m-1)$ dimensional space of discretized variables and starting from each random point the multivariable function \eqref{eq:3.10} is \textit{minimized to width} $w$ (i.e.~it is minimized to a point such that moving $w$ in either direction along any variable $f_{lp}$ and keeping other variables fixed gives a higher value for the action). This process is continued by dividing $w$ in half and \textit{minimizing to width} ${w\over 2}$ and further continued to \textit{minimization to width} ${w\over 2^n}$ at the $n$-th step until the difference between the string tension value at step $n$ and $n-1$ is sufficiently small. Let $X_n$ denote the random variable for the value of the string tension at step $n$ obtained by this minimization process. The \textit{minimization error} (i.e.$| \text{min}\bar{T}^{m,J}_k - \langle X_n\rangle |$) in minimizing the multivariable function is reduced to the desired accuracy if the following quantities are sufficiently small:
\begin{flalign}
\label{eq:3.12}
i) ~{\sigma_n \over \sqrt{R}}~, \ \ \ \ \ \  \ \    ii)~ |\langle X_n\rangle -\langle X_{n-1}\rangle |~,&& 
\end{flalign}
where $\sigma_n$ is the standard deviation of $X_n$ and $\langle X_n\rangle$ denotes the average of $X_n$.\par
The same analysis described above is done for $2m$ number of partitions and the \textit{discretization error} (i.e. $| \text{min}\bar{T}^{m,J}_k - \text{min} \bar{T}^{\infty,J}_k|$) of the string tensions is reduced to the desired accuracy if the difference between the string tension values obtained for $m$ number of partitions and $2m$ number of partitions is small enough. We consider the difference $|\text{min}\bar{T}^{m,J}_k - \text{min} \bar{T}^{2m,J}_k|$ as an estimate for the \textit{discretization error} of the string tension values obtained for $2m$ number of partitions.

The boundary value number $\text{z} = J$ is assumed large enough to ensure the \textit{truncation error} (i.e. $|\text{min} \bar{T}^{\infty,J}_k - \text{min} \bar{T}^{\infty,\infty}_k(=\bar{T}_k)|$) is small enough. An upper bound estimate for the truncation error is given by \ref{sec:A1}:
\begin{equation}
\label{eq:3.13}
|\text{min} \bar{T}^{\infty,J}_k-\bar{T}_k|<2| \text{min} \bar{T}^{\infty,J}_{k1} - \text{min}\bar{T}^{\infty,J}_{k2}|.
\end{equation}
The total error estimate in minimizing \eqref{eq:3.4} is given by:
\begin{equation}
\label{eq:3.14}
\begin{split}
\text{Total Error} & = \text{Min. E.} + \text{Dis. E.} + \text{Trunc. E.}
\\
& = |\langle X_n\rangle -\langle X_{n-1}\rangle | +{\sigma_n \over \sqrt{R}}+|\text{min}\bar{T}^{m,J}_k- \text{min}\bar{T}^{2m,J}_k|
\\
&+ 2|\text{min} \bar{T}^{\infty,J}_{k1} - \text{min}\bar{T}^{\infty,J}_{k2}|~.
\end{split}
\end{equation}
The  analysis of the errors defined above is discussed at length  in Appendix \ref{errorappendix}.

\subsection{Numerical value of ${k}$-string tensions in dYM}
\label{sec:3.4}

The numerical values of \eqref{eq:3.4} obtained by the minimization method above with their corresponding errors are listed in Table \ref{table:1} below.\footnote{Numerical computations of the string tensions were performed on the gpc supercomputer at the SciNet HPC Consortium \cite{22}. Due to a high number of $k$-string calculations ($>1000$) with most of them involving minimization of multivariable functions with more than 500 variables, using a cluster that could perform many $k$-string computations at the same time in parallel was necessary.}
Since the minimum value in \eqref{eq:3.4} always lies below the numerical values obtained in a numerical minimization procedure the upper bound estimate for the error has been indicated with a minus sign only.
\\

\begin{table}[h]
\centering
\begin{tabular}{|c|c|c|c|c|c|c|c|c|c|}
\hline
\begin{tabular}{c|c}
$SU(N)$& k
\end{tabular}
&1&2 &3&4&5&6&7&8 &9\\
\hline 
2 & 5.6576&-&-&-&-&-&-&-&-\\
\hline 
3 & 6.5326&6.5326&-&-&-&-&-&-&-\\
\hline 
4&6.8583&8.0006&6.8583&-&-&-&-&-&-\\
\hline
5&7.0140&8.6602&8.6602&7.0140&-&-&-&-&-\\
\hline
6&7.1001&9.0168&9.5547&9.0168&7.1001&-&-&-&-\\
\hline
7&7.1526&9.2318&10.0744&10.0744&9.2318&7.1526&-&-&-\\
\hline
8&7.1868&9.3713&10.4051&10.7192&10.4051&9.3713&7.1868&-&-\\
\hline
9&7.2104&9.4670&10.6292&11.1455&11.1455&10.6292&9.4670&7.2104&-\\
\hline
10&7.2273&9.5355&10.7882&11.4434&11.6491&11.4434&10.7882&9.5355&7.2273\\
\hline
\end{tabular}

\caption{The numerical values, \eqref{eq:3.4}, of half $k$-string tensions for gauge groups ranging from $SU(2)$ to $SU(10)$. The upper bound estimate for error is $-0.006$.}
\label{table:1}
\end{table}

The minimization method was carried out for $J=14.0$, $m=100$ and $m=200$ number of partitions and the initial width was $w=1$. The number of random points $R$ generated initially is $R=24$. The multivariable function \eqref{eq:3.10} for $m=100$ was minimized to width $w \over 2^n$ for $n=20$. For $m=200$, \eqref{eq:3.10} was minimized to step $n=20$ for $SU(2 \leq N \leq 7)$ and to step $n=22$ for $SU(8 \leq N \leq 10)$. The numerical values listed in Table \ref{table:1} refer to the numbers obtained with $m=200$ number of partitions rounded to the fourth decimal.
A comparison of the known analytical results for $SU(2)$ (\eqref{eq:3.7}) and $SU(3)$ (\eqref{eq:3.8}) half $k$-string tensions with the numerical results is made in Table \ref{table:2}.
\begin{table}[h]
\centering
\begin{tabular}{ |c|c|c|}
\hline
&Numerical value&Analytical (exact) value\\
\hline
$SU(2)$ & $5.6576_{- 0.006}$ & ${8/\sqrt{2}}\approx 5.6569$\\
\hline
$SU(3)$ & $6.5326_{-0.006}$& ${16/ \sqrt{6}}\approx 6.5320$\\
\hline
\end{tabular}
\caption{$SU(2)$ \& $SU(3)$ numerical and analytical half $k$-string tensions.}
\label{table:2}
\end{table}

{\flushleft{T}}he same minimization process and error analysis used to derive the $SU(2)$ and $SU(3)$ half $k$-string tensions was utilized for the higher gauge groups.

A discussion of the results shown in Table \ref{table:1}, especially regarding the $k$-scaling of string tensions and the large-$N$ limit will be given in  Section \ref{sec:5}.

\section{String tensions in dYM:~perturbative evaluation}
\label{sec:4}

Here, we will rederive the half $k$-string tensions in Table \ref{table:1} by a perturbative evaluation of the saddle point. We stress that this is not an oxymoron and  that, indeed, we will be using (resummed) expansions and only Gaussian integrals to compute a nonperturbative effect. 

In order to explain the main ideas, we briefly summarize them now, in an attempt to divorce them from the many technical details given later.
Our starting point is the partition function of the Wilson loop inserted dual photon action for a fundamental weight $\mu_k$ from  Section \ref{sec:2.4.1}:
\begin{equation}
\label{eq:4.1}
Z^{\eta} =\int [D\sigma] \text{exp}(-\int_{{\rm I\!R}^3} d^3\text{x}\{{1\over {2L}}{g^2 \over 8\pi^2}({\nabla \sigma - \nabla \eta \mu_k })^2- \tilde{\zeta} \;\overset{N}{\underset{j=1}{\sum}}\text{cos}(\sigma_j-\sigma_{j+1})\})~.
\end{equation}
For a Wilson loop in the $\text{y}_1, \text{y}_2$ plane, $\eta(\text{x}) = \underset{\text{A}}{\int} \text{dy}_1\text{dy}_2 {\text{x}_3 \over 2|\text{x} -\text{y}|^3}$ with $\text{y} = (\text{y}_1,\text{y}_2, 0)$ and ``A'' stands for the area of the rectangular Wilson loop $R \times T$ where the integral is being evaluated.
We will rewrite \eqref{eq:4.1} in a form appropriate for a perturbative evaluation. Defining ${1\over \beta} \equiv {\tilde{\zeta} \over m_{\gamma}^3}$, rescaling $\text{x}_l \rightarrow {1 \over m_{\gamma}} {\hat{\text{x}}_l}$, $\text{y}_l \rightarrow {1 \over m_{\gamma}} \hat{\text{y}}_l$ ($l = 1,2,3$) with $m^2_{\gamma} = {8 \pi^2 \over g^2} L \tilde{\zeta}$ and expanding the cosine term (neglecting the leading constant term) we have:
\begin{equation}
\label{eq:4.222}
Z^{\eta} =\int [D\sigma] \text{exp}(-{1 \over \beta}\int_{{\rm I\!R}^3} d^3\hat{\text{x}}\{{1\over 2}(\partial_l \sigma)^2 - (\mu_k)_j\partial_l \sigma_j \partial_l \eta + {1 \over 2} (\partial_l \eta)^2 {\mu_k}^2 + {1 \over 2} (\sigma_j-\sigma_{j+1})^2 -{1 \over 4!} (\sigma_j-\sigma_{j+1})^4 + ... \})~,
\end{equation}
with $\sigma_{N+1} \equiv \sigma_1$ and an implicit sum over $j=1,..,N$ and $l = 1,2,3$. We note that based on \eqref{eq:2.42} and  \eqref{eq:2.44}  for the leading $NL \Lambda$ dependence of $\beta$ we have $\beta \sim (NL \Lambda)^{b_0 \over 2}$ with  $b_0>0$ given by \eqref{eq:2.41}. 
Thus, in the regime of validity $NL \Lambda \rightarrow 0$ of the semiclassical expansion $\beta \rightarrow 0$ and the partition function (\ref{eq:4.222}) can be evaluated using the saddle point approximation, which was done numerically in  Section \ref{numericsection} and analytically in this  Section.

 We shall present details for both $SU(2)$, in  Section \ref{su2analyticsection}, and $SU(N)$, in  Section \ref{suNanalyticsection} below, but begin by explaining the salient points of the analytic method here.
For this purpose consider \eqref{eq:4.222} for an SU(2) gauge group. Following steps from equations \eqref{eq:4.3} to \eqref{eq:4.7} we obtain:
\begin{equation}
\label{eq:4.71}
Z^{\eta}_{g^4} =\int [Dg] \text{exp}(-\int_{{\rm I\!R}^3} d^3\hat{\text{x}}\{{1\over 2}(\partial_l g)^2 + {m^2 \over 2} g^2 + \beta \lambda g^4 + {1 \over 2}({b \over 2\pi})^2 (\partial_l \eta)^2  \}) \text{exp}(+ {b \over \sqrt{\beta} } \underset{\text{A}}{\int} d\hat{\text{x}}_1 d\hat{\text{x}}_2  \partial_3g)~.
\end{equation}
The differences between (\ref{eq:4.71}) and (\ref{eq:4.222}) are that: {\it i.}) the integration variable, the single dual photon of $SU(2)$, is now called $g$, {\it ii.})  nonlinear terms higher than quartic are discarded in order to simply illustrate the procedure, {\it iii.}) arbitrary dimensionless constants are introduced: mass parameter $m$, quartic coupling $\lambda$ and the boundary coefficient $b$ in (\ref{eq:4.71}). Naturally the values of $m, \lambda$ and $b$ are determined by the original action in (\ref{eq:4.222}) (or see below equation \eqref{eq:4.4}), but it is convenient to keep them general in order to organize the expansion. As we explain below, we perform  a combined expansion in  $\beta$ to all orders, and in  $\lambda$ to any desired order.\footnote {As is evident from equation \eqref{eq:4.30}, the expansion parameter is $ {\lambda b^2\over 4 m^2}$; as discussed there, convergence of the perturbative expansion of the saddle point for a $g^4$ interaction term only requires that this parameter be less than $ {1/2}$. This condition is met in dYM theory, but not in QCD(adj) \cite{Anber:2015kea}, for the choices of parameters following from the underlying action (In dYM from below equation \eqref{eq:4.4}), although not strictly required since the full potential in both theories includes higher non-linearities and in taking these into account the perturbative series evaluation of the saddle point would be a convergent one.}

To explain the procedure, from \eqref{eq:4.222} it can be seen that the $\beta$ parameter is similar to an $\hbar$ parameter and in (\ref{eq:4.71}) the fields have been rescaled by this parameter for a perturbative evaluation of the saddle point. A rescaling of fields by a parameter will not change how an expansion in that parameter behaves therefore the expansion in $\beta$ is similar to an $\hbar$ expansion. In the limit of an infinite Wilson loop the $\beta$ expansion will be organized in the following way:
\begin{equation}
\label{eq:betaexpansion}
e^{- \hat{R} \hat{T}\{ {1 \over \beta}S_0(\lambda) + f_{1\text{-loop}}(\lambda) + \beta f_{2\text{-loop}}(\lambda) + ... \} }~ ,
\end{equation}
$S_0$ is the one-dimensional saddle-point action, $f_{1\text{-loop}}$, $f_{2\text{-loop}}$, etc correspond to the summation of the one loop, two loop, etc diagrams. For brevity we have only included a $\lambda$ dependence although generally they depend on $\lambda$, $m$ and $b$. In this Section we will be only concerned with the leading saddle point result of order $1 \over \beta$ in the exponent of \eqref{eq:betaexpansion}. To carry this out we expand the Wilson loop exponent and the $g^4$ term in \eqref{eq:4.3} and look for connected terms of order $1 \over \beta$. These will be the terms that exponentiate to produce the saddle point action in \eqref{eq:betaexpansion}. As an example consider the Wilson loop exponent expanded to second order: 
 ${1 \over 2!}({b \over \sqrt{\beta} } \underset{\text{A}}{\int} d\hat{\text{x}}_1 d\hat{\text{x}}_2  \partial_3g)^2$. When evaluated using the free massive propagator it is a connected diagram of order $1 \over \beta$, combined with the evaluation of the non-connected terms ${1 \over 4!}({b \over \sqrt{\beta} } \underset{\text{A}}{\int} d\hat{\text{x}}_1 d\hat{\text{x}}_2  \partial_3g)^4$, etc it would exponentiate to produce the term $- {\hat{R} \hat{T} \over \beta}S_0(\lambda = 0)$ in the exponent of \eqref{eq:betaexpansion}. The odd terms in the expansion of the Wilson loop exponent vanish due to an odd functional integral. Similarly higher order contributions in $\lambda$ to the saddle point action can be evaluated. The order $\lambda$ contribution comes from the exponentiation of the connected diagram involving one $g^4$ term and four Wilson loop terms which is of order $1 \over \beta$: $ - \beta \lambda \int_{{\rm I\!R}^3} d^3{\text{x}} g^4$ ${1 \over 4!}({b \over \sqrt{\beta}} \int d\text{x}_1 d\text{x}_2 \partial_3 g)^4$. Terms of order $\lambda^2$, etc in the expansion of the saddle point action in \eqref{eq:betaexpansion} can also be evaluated perturbatively which would result in the perturbative expansion of the saddle point in $\lambda$ (or more precisely ${ \lambda b^2 \over m^2}$) as in \eqref{eq:4.30}. Clearly the large value of $1 \over \beta$ causes no problem for these exponentiations since the radius of convergence of an exponential function is infinite. At every order in this combined expansion, we are faced with the calculation of Gaussian integrals only---hence the  ``perturbative evaluation'' in the title of this  Section. 

In this paper, we compute the leading-order contributions to the Wilson loop expectation value that behave as 
\begin{equation}
\label{wilsonperturbative}
e^{- {\hat{R} \hat{T}\over \beta}( a_1 + a_2 \lambda)}~,
\end{equation} where ${\hat{R} \hat{T}}$ is the dimensionless area of the Wilson loop, defined in  Section \ref{su2analyticsection}, and $a_1$( $= S_0(\lambda = 0)$), $a_2$ are numerical coefficients that we compute (for $SU(2)$ we will evaluate a few higher order corrections as well).

Setting $\lambda =0$ in (\ref{wilsonperturbative}) corresponds to ignoring non-linearities and is equivalent to a calculation of the saddle point action using the Gaussian approximation for the dual photon action. This was previously done in the 3d Polyakov model in \cite{Antonov:2003tz, Anber:2013xfa}. However,  as noted in  \cite{Anber:2013xfa} and also follows from our results, the neglect of nonlinearities introduces an order unity error in the string tension. On the other hand, incorporating even only the leading quartic nonlinearity and setting $\lambda$ equal to the value that follows from (\ref{eq:4.222}) at the end of the calculation leads to a significantly better agreement with the exact analytic or numerical data. One explanation for this is that the value of the saddle point functions approach zero quickly from its boundary value at $\text{x}_3 = 0$ which for $SU(2)$ is $\pi$ and for higher gauge groups is less than $\pi$ therefore the non-linearities will be suppressed. We show this in great detail in Appendix \ref{sec:C} for a wide range of $N$ and $k$. Here, to  illustrate the utility of the method, in  Table \ref{table:71}, we only list the results for the $k=1$-string tension for gauge groups $SU(2)$---$SU(10)$, obtained via the method explained above and keeping the quartic nonlinearity only. A look at Table \ref{table:71} shows that   the convergence to the numerical (or exact analytic, when available) result is evident.\footnote{This agreement can be further improved, as we have verified for  $SU(2)$. Summing only contributions to (\ref{wilsonperturbative}) to order $\lambda$ we obtain the value $7.84$  shown in Table \ref{table:71}. Including the higher-order correction terms show an oscillatory convergence: Including the first order correction due to the $g_1^6$ term in \eqref{eq:4.5} gives $8.285$ and including order $\lambda^2$ of the quartic term expansion, we obtain $8.007$, to be compared with the exact value $8$.}

\begin{table}[ht]
\centering
\begin{tabular}{|c|c|c|c|c|}
\hline
$SU(N)$& $a_1$ & $a_2 \lambda$ & $a_1 + a_2 \lambda$  & Num. value \\
\hline 
2 & 9.870 & -2.029 & 7.841 & 8.000 {\tiny (exact)}  \\
\hline 
3 & 11.396 & -2.343 & 9.053 & 9.238 {\tiny (exact)} \\
\hline 
4 & 11.913 & -2.396 & 9.517 & 9.699  \\
\hline
5 & 12.150 & -2.410 & 9.740 & 9.919  \\
\hline
6 & 12.277 & -2.415 & 9.862 & 10.041  \\
\hline
7 & 12.355 & -2.417 & 9.938 & 10.114 \\
\hline
8 & 12.405 & -2.418 & 9.987 & 10.163  \\
\hline
9 & 12.439 & -2.418 & 10.021 & 10.196  \\
\hline
10 & 12.463 & -2.418 & 10.045 & 10.221  \\
\hline
\end{tabular}
\caption{Comparison of $N$-ality $1$ $k$-string tensions for $SU(2 \leq N \leq 10)$, obtained using the perturbative method explained here---leading contribution $a_1$ plus first subleading $a_2 \lambda$, from  eq.~(\ref{wilsonperturbative})---with the results of the numerical study. To avoid confusion, we  note that the exact analytic values for $SU(2)$ and $SU(3)$ in the dimensionless units used here are $8$ and $9.238$, respectively which agree with the numerical values listed in Appendix \ref{sec:C}; see also the end of  Section \ref{su2analyticsection}.}
\label{table:71}
\end{table}

Our final comment is that, in principle, this approach would also allow one to compute corrections to the leading semiclassical result. In the case at hand, this would necessitate a more precise matching of the long-distance theory to the underlying gauge theory; needless to say, any detailed study of such corrections is left for future work.

\subsection{Evaluation for $\text{SU(2)}$}
\label{su2analyticsection}

We will first demonstrate the basic ideas of the method in the simpler case of $SU(2)$ which an analytic solution to the saddle point is available, hence a direct comparison can be made with the perturbative evaluation. Defining $g_1 \equiv (\sigma_1 - \sigma_2)/\sqrt{2}$ and $g_2 \equiv (\sigma_1 + \sigma_2)/\sqrt{2}$ with $\mu_1 = (0.5,-0.5)$, \eqref{eq:4.222} for $SU(2)$ reduces to:
\begin{equation}
\label{eq:4.3}
Z^{\eta} =\int [Dg] \text{exp}(-{1 \over \beta}\int_{{\rm I\!R}^3} d^3\hat{\text{x}}\{{1\over 2}(\partial_l g_1)^2 +{1\over 2}(\partial_l g_2)^2 - {1\over \sqrt{2}}\partial_l g_1 \partial_l \eta + {1 \over 4} (\partial_l \eta)^2 + {4 \over 2} (g_1)^2 -{8 \over 4!} (g_1)^4 + ... \})~.
\end{equation}
From \eqref{eq:4.3} it is clear that $g_2$ only appears in the kinetic term hence can be neglected. In what follows we will neglect the higher order interactions and demonstrate how the method works for a $g_1^4$ interaction term only. We replace $g_1$ with $g$ and use general dimensionless parameters for the mass, the $g^4$ coupling constant and the coefficient of the Wilson loop terms:
\begin{equation}
\label{eq:4.4}
Z^{\eta}_{g^4} =\int [Dg] \text{exp}(-{1 \over \beta}\int_{{\rm I\!R}^3} d^3\hat{\text{x}}\{{1\over 2}(\partial_l g)^2 + {m^2 \over 2} g^2 +\lambda g^4 - {b \over 2 \pi}\partial_l g \partial_l \eta + {1 \over 2}({b \over 2\pi})^2 (\partial_l \eta)^2  \})~.
\end{equation}
For later use we note that the corresponding values of $m$, $\lambda$ and $b$ in \eqref{eq:4.3} are $m=2$, $\lambda = -{8 \over 4!}$ and $b = \sqrt{2} \pi$. Integrating by parts the Wilson loop term (linear term in $\partial_l g$) with:
\begin{equation}
\label{eq:4.5}
 \partial_l \partial_l \eta(\text{x}) = -2\pi \underset{\text{A}}{\int} \text{dy}_1\text{dy}_2 \partial_3 \partial^2 {1 \over 4 \pi |\text{x} -\text{y}|} = 2\pi \theta_A(\text{x}_1,\text{x}_2) \partial_3 \delta (\text{x}_3)
\end{equation}
gives:
\begin{equation}
\label{eq:4.6}
Z^{\eta}_{g^4} =\int [Dg] \text{exp}(-{1 \over \beta}\int_{{\rm I\!R}^3} d^3\hat{\text{x}}\{{1\over 2}(\partial_l g)^2 + {m^2 \over 2} g^2 +\lambda g^4 + b \partial_3 \delta (\hat{\text{x}}_3) \theta_A(\hat{\text{x}}_1, \hat{\text{x}}_2) g  + {1 \over 2}({b \over 2\pi})^2 (\partial_l \eta)^2  \})~,
\end{equation}
where $\theta_A(\hat{\text{x}}_1,\hat{\text{x}}_2)$ is $1$ on the Wilson loop area and zero otherwise. To evaluate \eqref{eq:4.4} perturbatively rescale $g \rightarrow \sqrt{\beta} g$:
\begin{equation}
\label{eq:4.7}
Z^{\eta}_{g^4} =\int [Dg] \text{exp}(-\int_{{\rm I\!R}^3} d^3\hat{\text{x}}\{{1\over 2}(\partial_l g)^2 + {m^2 \over 2} g^2 + \beta \lambda g^4 + {1 \over 2}({b \over 2\pi})^2 (\partial_l \eta)^2  \}) \text{exp}(+ {b \over \sqrt{\beta} } \underset{\text{A}}{\int} d\hat{\text{x}}_1 d\hat{\text{x}}_2  \partial_3g)~.
\end{equation}
In what follows, we will drop the $hat$ on $\text{x}$; due to the rescaling made earlier it should be remembered that we are working with dimensionless variables.

 We will first calculate the Wilson loop exponent using the quadratic terms (kinetic term + mass term) in \eqref{eq:4.7}. In expanding the exponential $\text{exp}(+ {b \over \sqrt{\beta} } \underset{\text{A}}{\int} \text{dx}_1\text{dx}_2  \partial_3g)$ the odd terms vanish due to an odd functional integral and the even terms will be organized in the form of an expansion of an exponent (hence they would exponentiate) therefore it would be sufficient to only evaluate the second order term:\footnote{We have to note that since we are summing to all orders such a perturbative expansion is justified although ${1 \over \sqrt{\beta}}$ becomes large as $\beta \rightarrow 0$.} 
\begin{equation}
\label{eq:4.8}
\big\langle {b^2 \over 2 \beta } \underset{\text{A A}}{\iint}' \partial_3g  \partial'_3g \big\rangle_{0} = {b^2 \over 2 \beta } \underset{\text{A A}}{\iint}'  \partial_3 \partial'_3 P(\text{x}-\text{x}') = {b^2 \over 2 \beta } \underset{\text{A A}}{\iint}'  \partial_3 \partial'_3 {\text{exp}(-m|\text{x}-\text{x}'|) \over 4 \pi |\text{x}-\text{x}'|}, \ \text{at} \ \text{x}_3= \text{x}'_3 = 0~.
\end{equation}
The last expression can be evaluated as:
\begin{equation}
\label{eq:4.9}
\underset{\text{A  A}}{\iint}' \partial_3 \partial'_3 P(\text{x}-\text{x}') = \underset{\text{A  A}}{\iint'} \{ (-\partial^2 + m^2)P(\text{x}-\text{x}') + (\partial^2_1+\partial^2_2)P(\text{x}-\text{x}') - m^2P(\text{x}-\text{x}') \}~.
\end{equation}
For a Wilson loop in the $\text{x}_1, \text{x}_2$ plane we can bring the second term on the right hand side of \eqref{eq:4.9} on the boundaries of the Wilson loop using the identity:
\begin{equation}
\label{eq:4.10}
\begin{split}
\underset{b(\text{A}) \  b(\text{A})}{\iint'} d\text{x}^l d\text{x}'^k \delta^{l k} P(\text{x}-\text{x}') & = \{ \underset{\text{A A}}{\iint'} dS^l dS'^l \partial_n\partial'_n - \underset{\text{A A}}{\iint'} dS^l dS'^k \partial_k\partial'_l \} P(\text{x}-\text{x}') \\ & = - \underset{\text{A A}}{\iint'} d^2\text{x} d^2\text{x}' \{ \partial_1^2 + \partial_2^2 \} P(\text{x}-\text{x}')~.
\end{split}
\end{equation}
Here, $b(\text{A})$ stands for the boundary of the Wilson loop area.

Using relations \eqref{eq:4.8}, \eqref{eq:4.9}, \eqref{eq:4.10} and noting that $P(\text{x}-\text{x}')$ is the Greens function of the operator $-\partial^2 + m^2$ we have:
\begin{equation}
\label{eq:4.12}
\big\langle {b^2 \over 2 \beta } \underset{\text{A A}}{\iint}' \partial_3g  \partial'_3g \big\rangle_{0} ={b^2 \over 2 \beta } \{ \delta(0) \hat{R}\hat{T} - \underset{b(\text{A}) \  b(\text{A})}{\iint'} d\text{x}^l d\text{x}'^k \delta^{l k} P(\text{x}-\text{x}') - {m^2} \underset{\text{A  A}}{\iint'} P(\text{x} - \text{x}') \}~.
\end{equation}
The subscript of zero of the expectation value refers to it being evaluated using the free theory Lagrangian (i.e. at $\lambda =0$). The first term and the infinite part of the second term on the right hand side of \eqref{eq:4.12} would cancel with the infinite parts of the ${1 \over 2} ({b \over 2 \pi})^2(\partial_l \eta)^2$ term in \eqref{eq:4.7} to give a finite perimeter law for the Wilson loop and the third term on the right hand side of \eqref{eq:4.12} would give rise to an area law in the large area limit.
Evaluating the ${1 \over 2} ({b \over 2 \pi})^2(\partial_l \eta)^2$ term in \eqref{eq:4.7} using \eqref{eq:4.5} and similar methods used to evaluate \eqref{eq:4.12} gives:
\begin{flalign}
\label{eq:4.13}
\int_{{\rm I\!R}^3} d^3{\text{x}}{1 \over 2}({b \over 2\pi})^2 (\partial_l \eta)^2 & = - {1 \over 2}({b \over 2\pi})^2   \int_{{\rm I\!R}^3} d^3{\text{x}} \eta(\text{x}) 2\pi \theta_A(\text{x}_1,\text{x}_2) \partial_3 \delta (\text{x}_3) = -{b^2 \over 2} \int_A d^2{\text{x}}\int_A d^2{\text{y}} \partial^2_3 {1 \over 4 \pi |\text{x} -\text{y}|} \nonumber \\ & = {b^2 \over 2} \{ \delta(0) \hat{R}\hat{T} - \underset{b(\text{A}) \  b(\text{A})}{\iint'} { d\text{x}^l d\text{y}^k} \delta^{l k}{1 \over 4 \pi |\text{x} -\text{y}|}    \} ~.&&
\end{flalign}
Further, 
\eqref{eq:4.7}, \eqref{eq:4.12} and \eqref{eq:4.13} give:

%
%

\begin{equation}
\label{eq:4.14'}
\hspace{-0.7cm} {\{ Z^{\eta}_{g^4} \}_{\lambda =0} \over \{ Z^{\eta}_{g^4} \}_{b=\lambda =0} } = \text{exp}(- {1 \over \beta} \{ {m^2 b^2 \over 2} \underset{\text{A  A}}{\iint'} P(\text{x} - \text{x}') + {b^2 \over 2} \underset{b(\text{A}) \  b(\text{A})}{\iint'} d\text{x}^l d\text{x}'^k \delta^{l k} ( P(\text{x}-\text{x}') - {1 \over 4 \pi |\text{x} -\text{x}'|}) \} )~.
\end{equation}

%
%

In the limit that the area of the Wilson loop goes to infinity the first term on the right hand side of  \eqref{eq:4.14'} can be evaluated explicitly. Consider a Wilson loop with $\hat{R} = \hat{T} \equiv a$. Rescaling $\text{x}_k \rightarrow a \text{x}_k$, $\text{x}^{\prime}_k \rightarrow a \text{x}^{\prime}_k$ for $k = 1, 2$ and considering the limit $a \rightarrow \infty$ we have:
\begin{flalign}
\label{eq:4.11}
\underset{\text{A  A}}{\iint}'P(\text{x}-\text{x}') & = a^3 \underset{\text{A}_{1 \times 1}}{\int}d^2\text{x} \underset{\text{A}_{1 \times 1}}{\int'}d^2\text{x}' {\text{exp}(-am|\text{x}-\text{x}'|) \over 4 \pi |\text{x}-\text{x}'|} \nonumber\\& = a^3 \underset{\text{A}_{1 \times 1}}{\int}d^2\text{x} 2\pi \int^{\infty}_0 rdr {\text{exp}(-am\sqrt{(\text{x}_3 - \text{x}'_3)^2/a^2 + r^2}) \over 4 \pi \sqrt{(\text{x}_3 - \text{x}'_3)^2/a^2 + r^2}}  \\ &= {1\over 2m}\text{exp}(-m|\text{x}_3 - \text{x}'_3|) a^2 \rightarrow {a^2 \over 2 m}= {\hat R \hat T \over 2 m} .\nonumber
\end{flalign}
To arrive at the result above, we noted that in the limit of a large Wilson loop area ($a \rightarrow \infty$) due to the exponential suppression the main contribution to the $d^2\text{x}'$ integral comes from a small circle with radius $r \sim {1 \over a}$ centered at the point $\text{x} = (\text{x}_1, \text{x}_2 , 0)$ in the Wilson loop. Therefore it can be seen that the exact value of the $d^2\text{x}'$ integral in this limit would be given when the $dr$ integral is evaluated from zero to infinity. This would imply that the $d^2\text{x}'$ integral is independent of $\text{x}$, hence the $d^2\text{x}$ integral would be a trivial one over a unit square.

Using \eqref{eq:4.11}, \eqref{eq:4.14'} in the limit of a large Wilson loop becomes:
\begin{equation}
\label{eq:4.14}
{\{ Z^{\eta}_{g^4} \}_{\lambda =0} \over \{ Z^{\eta}_{g^4} \}_{b=\lambda =0} } = \text{exp}(- {1 \over \beta} \{ {mb^2 \over 4} \hat{R}\hat{T} + {b^2 \over 2} \underset{b(\text{A}) \  b(\text{A})}{\iint'} d\text{x}^l d\text{x}'^k \delta^{l k} ( P(\text{x}-\text{x}') - {1 \over 4 \pi |\text{x} -\text{x}'|}) \} )~.
\end{equation}
\eqref{eq:4.14} contains an area law term (first term in the exponent) and a perimeter law term (last two terms in the exponent). We note that without the use of the perturbative saddle point method the evaluation of the perimeter law term, due to the complicated behaviour of the saddle point solution near the boundaries of the Wilson loop, would have been a difficult task. The perimeter law term in \eqref{eq:4.14} is a finite quantity and proportional to $\text{Log}(a)a$ (for $a = \hat{R} = \hat{T}$) hence negligible compared to the area law term in the limit $a \rightarrow \infty$. Due to this and the fact that our main focus in this  Section is the area law term we will drop this term in what follows. Another point worth mentioning, as will be seen in what follows, is that in the limit of $a \rightarrow \infty$ only the area law term in \eqref{eq:4.14} will receive $\lambda$ corrections.

The saddle point equation  of motion of \eqref{eq:4.6} is given by:
\begin{equation}
\label{eq:4.15}
\partial^2 g = m^2 g + 4 \lambda g^3 + b \partial_3 \delta (\text{x}_3) \theta_A(\text{x}_1, \text{x}_2)~.
\end{equation}
For large Wilson loops, far from the boundaries of the Wilson loop the saddle point solution to \eqref{eq:4.6} obeys \eqref{eq:4.15}. For regions outside the Wilson loop the solution is zero. For regions close to the boundaries the saddle point solution will be more complicated and gives the perimeter law contribution in \eqref{eq:4.14}. For regions interior to the Wilson loop far from the boundaries the solution only depends on $\text{x}_3$ with a discontinuity of $b$ at $\text{x}_3 = 0$ and gives the area law contribution in \eqref{eq:4.14}. The corresponding one dimensional problem is given by:
\begin{equation}
\label{eq:4.16}
{d^2 \over d\text{x}^2_3} h = m^2 h + 4 \lambda h^3  \ \ \ \ \ \ h(0^+) = {b / 2} , \ \ \ \ h(0^-) = -{b / 2}~.
\end{equation}
The discontinuity should be equally split above and below the Wilson loop in order to give the lowest action.
The solution to \eqref{eq:4.16} for $\lambda = 0$ is given by:
\begin{equation}
\label{eq:4.17}
h(\text{x}_3) = \left \{
  \begin{array}{c}
  {b \over 2} \text{exp}(-m\text{x}_3) \ \ \ \text{for} \ \ \ \text{x}_3 > 0 \\
  -{b \over 2} \text{exp}(m\text{x}_3) \ \ \ \text{for} \ \ \ \text{x}_3 < 0 \\
  \end{array}
\right .
\end{equation}
The action of this solution is:
\begin{equation}
\label{eq:4.18}
S[h] = \int^{+\infty}_{-\infty} dx_3 \{ {1 \over 2} ({dh \over dx_3})^2 + {1 \over 2} m h^2 \} = {m \over 4}b^2 , \ \ \ \ \text{at} \ \  \lambda = 0~,
\end{equation}
which is the same as the coefficient of the area law term in \eqref{eq:4.14}. This demonstrates the validity of the perturbative saddle point method in producing the corresponding action of the saddle point boundary value problem. In order to further verify this method we will evaluate the saddle point action for a nonzero $\lambda$ and compare it with the corresponding analytic solution. We expand the exponential of the $g^4$ term in \eqref{eq:4.7}. The order $\lambda$ term contracts with the fourth order term in the expansion of the Wilson loop exponent:
\begin{equation}
\label{eq:4.19}
\big\langle - \beta \lambda \int d^3\text{x} g^4 ({b \over \sqrt{\beta}})^4{1\over 4!} (\int_A d^2\text{y} \partial_{\text{y}_3} g)^4 \big\rangle_{0,C} = -{4! \over 4!} {\lambda \over \beta} b^4 \int d^3\text{x} \prod^{4}_{i=1} \int_A d^2\text{y}^i-\partial_{\text{x}_3} P(\text{x}-\text{y}^i)~.
\end{equation}
The subscript $C$ refers to the connected contribution. It has to be reminded  that we will only be interested in connected terms of order $1 \over \beta$ since these terms would exponentiate to produce the series expansion of the saddle point action in \eqref{eq:4.30}. In the limit of $a \rightarrow \infty$ (after rescaling the variables $\text{x}_k \rightarrow a \text{x}_k, \text{y}^i_k \rightarrow a \text{y}^i_k$ for $k =1, 2$) due to the exponential suppression the $\text{x}_1$ and $\text{x}_2$ components of $\text{x}$ will be restricted to the Wilson loop and the main contribution to the integrals would come from a small circle of radius $\sim {1 \over a}$ centred at $(\text{x}_1, \text{x}_2,0)$. Following similar steps as \eqref{eq:4.11} we have:
\begin{flalign}
\label{eq:4.20}
 \int_A d^2\text{y} (-\partial_{\text{x}_3} P(\text{x}-\text{y})) & = -\partial_{\text{x}_3} \int_A d^2\text{y} P(\text{x}-\text{y}) \overset{\text{x}_k \rightarrow a \text{x}_k} {\overset{\text{y}^i_k \rightarrow a \text{y}^i_k}{\overset{a \rightarrow \infty} {=}}} -\partial_{\text{x}_3} {1 \over 2m}\text{exp}(-m|\text{x}_3|) \nonumber \\ & = {\text{sign}(\text{x}_3) \over 2} \text{exp}(-m |\text{x}_3|) ~.&&
\end{flalign}
Then \eqref{eq:4.19} becomes:
\begin{equation}
\label{eq:4.211}
\big\langle - {\lambda \over \beta} {b^4 \over 4!}  \int d^3\text{x} g^4 (\int_A d^2\text{y} \partial_{\text{y}_3} g)^4 \big\rangle_{0,C} = - {1 \over \beta}{\lambda b^4 \over 16m} \Bbbk_1 a^2 = - {1 \over \beta}{\lambda b^4 \over 32m} \hat{R}\hat{T}~,
\end{equation}
with $\Bbbk_1$   given by:
\begin{equation}
\label{eq:4.22}
\Bbbk_1 \equiv \int^{+\infty}_{-\infty} d\text{x}_3 \text{E}(\text{x}_3)^4 = {1 \over 2}, \ \ \ \text{E}(\text{x}_3) \equiv  \text{sign}(\text{x}_3) \text{exp}(- |\text{x}_3|)~.
\end{equation}
We note that the $d \text{x}_1d\text{x}_2$ integral would be a trivial one over a unit square and hence only an integral over $d\text{x}_3$ would remain. In a similar way the $\lambda^2$ term can be evaluated. This term would contract with the sixth order term in the expansion of the Wilson loop exponent:
\begin{flalign}
\label{eq:4.23}
\big\langle {\beta^2\lambda^2 \over 2} (\int d^3\text{x} g^4)^2 ({b \over \sqrt{\beta}})^6{1\over 6!} (\int_A d^2\text{y} \partial_{\text{y}_3} g)^6 \big\rangle_{0,C} = &{16 \times 6! \over 2 \times 6!} {\lambda^2 \over \beta} b^6 \iint d^3\text{x}d^3\text{x}' \prod^{3}_{i=1} \int_A d^2\text{y}^i\partial_{\text{x}_3} P(\text{x}-\text{y}^i) \times \nonumber \\ & P(\text{x}-\text{x}')\prod^{3}_{i=1} \int_A d^2\text{y}'^i\partial_{\text{x}'_3} P(\text{x}'-\text{y}'^i) ~.&
\end{flalign}
Rescaling the variables ($\text{x}_k \rightarrow a \text{x}_k , ... $ for $k = 1, 2$), considering the limit $a \rightarrow \infty$, using \eqref{eq:4.20} and \eqref{eq:4.11} we have:
\begin{equation}
\label{eq:4.24}
\big\langle {\lambda^2 b^6 \over 2 \beta 6!} (\int d^3\text{x} g^4)^2(\int_A d^2\text{y} \partial_{\text{y}_3} g)^6 \big\rangle_{0,C} = {1 \over \beta}{8 \lambda^2 b^6 \over 128} {\Bbbk_2 \over m^3} a^2 = {1 \over \beta}{\lambda^2 b^6 \over 16} {1 \over 24 m^3} \hat{R}\hat{T}~, 
\end{equation}
with $\Bbbk_2$  given by:
\begin{flalign}
\label{eq:4.25}
\Bbbk_2 \equiv \overset{+\infty}{\underset{-\infty}{\iint}} d\text{x}_3d\text{x}'_3  \text{E}(\text{x}_3)^3 \text{exp}(-|\text{x}_3-\text{x}'_3|) \text{E}(\text{x}'_3)^3 = {1 \over 24}~.
\end{flalign}
 Higher order terms in $\lambda$ can be calculated similarly. The $\lambda^n$ term contracts with the $2n + 2$ order term in the expansion of the Wilson loop exponent. For $n > 3$ there would be more than one way of contracting the connected diagrams hence the evaluation would be more complicated but possible in principle 
 
Now we will directly solve for the saddle point and compare the result with the above expressions. \eqref{eq:4.16} is the one dimensional problem of interest. This is the motion of a particle moving in a potential $V(h) = -({m^2 \over 2}h^2 +\lambda h^4)$. Therefore the total energy is a constant of motion ${1\over 2}({dh \over d\text{x}_3})^2 + V(h) = C$. The minimum action corresponds to when $C=0$ therefore ${1\over 2}({dh \over d\text{x}_3})^2 =-V(h)$:
\begin{flalign}
\label{eq:4.29}
S[h] & = \int^{+\infty}_{-\infty} d\text{x}_3 \{ {1 \over 2} ({dh \over d\text{x}_3})^2 - V(h) \} = 2\int^{0}_{b \over 2} {d\text{x}_3 \over dh} dh  \{ - 2V(h) \} = 2m \int_{0}^{b \over 2} dh h \sqrt{ 1 + {2\lambda h^2 \over m^2}}  ~.&&
\end{flalign}
We expand the square root\footnote{\label{footnotetaylor}The Taylor series expansion of $\sqrt{1+ \text{x}}$ converges for $|\text{x}| < r = 1$. Evaluating $\text{x} = {2\lambda h^2 / m^2}$ for $h = {b / 2}$ with values of parameters from below equation \eqref{eq:4.4} gives $|\text{x}| = {\pi^2 / 12} < 1$ which lies within the radius of convergence. {As a reminder we mention that the condition $|\text{x}| = |{2\lambda h^2 / m^2}| < 1$ is not strictly required in dYM since the full potential is cosine which would allow for a wider range of these parameters.}} and evaluate the integral term by term. We also multiply the action by a negative sign to take into account the negative in the exponent. 
We find:
\begin{flalign}
\label{eq:4.30}
-S[h] & = \int_{0}^{b \over 2} dh \{ -2mh + -{2\lambda \over m}h^3 + \overset{\infty}{\underset{n=2}\sum} (-1)^n {2 \lambda^n \over m^{2n-1}} {(2n-3)!! \over n!}h^{2n+1} \} \nonumber \\ & = -m {b^2 \over 4} - {\lambda b^4 \over 32 m} + \overset{\infty}{\underset{n=2}\sum} { (-\lambda)^n b^{2n+2} \over m^{2n-1}} {(2n-3)!! \over 2^{2n+2} (n+1)!} \\ & {= -{mb^2 \over 4} \{ 1 + {1 \over 2} {\lambda b^2 \over 4 m^2} - \overset{\infty}{\underset{n=2}\sum} {(-1)^{n}  {(2n-3)!! \over (n+1)!} ({\lambda b^2 \over 4 m^2})^n} \} }\nonumber~. &&
\end{flalign}
The order $\lambda$ and $\lambda^2$ terms in \eqref{eq:4.30} match with the coefficients of ${\hat{R}\hat{T} \over \beta}$ in \eqref{eq:4.211} and \eqref{eq:4.24} respectively. This further demonstrates the validity of the perturbative evaluation of the saddle point. {The verification to higher orders in $\lambda$ ($n \geq 3$) can be made if the corresponding diagrams are evaluated.}

The higher order terms in \eqref{eq:4.3} ($g^6_1, g^8_1, ...$) and their cross terms with each other can also be evaluated similarly.

Next we will compare the $SU(2)$ $k$-string result of the perturbative evaluation of the saddle point to next to leading term, with the exact result of $SU(2)$ $k$-string in Table \ref{table:2} of  Section \ref{sec:3.4}. The exact $SU(2)$ saddle point area law is given by \eqref{eq:3.1}, \eqref{eq:3.4}:
\begin{flalign}
\label{eq:4.32}
{ Z^{\eta} \over  Z^{0} } = \text{exp}(- T_1 RT) = \text{exp}(-{2 \bar{T}_1 \over \sqrt{2} {\beta}} \hat{R}\hat{T}) = \text{exp}(- {8 \over {\beta}} \hat{R}\hat{T}), \ \hat{R}, \hat{T} \rightarrow \infty , \ \beta \rightarrow 0~,
\end{flalign}
where $R = {1 \over m_{\gamma}} \hat{R}$, $T = {1 \over m_{\gamma}} \hat{T}$, $\beta = {m^3_{\gamma} / \tilde{\zeta}} $ and, from Table \ref{table:2} of  Section \ref{sec:3.4}, $\bar{T}_1 = {8 \over \sqrt{2}}$.

Using \eqref{eq:4.14} and  \eqref{eq:4.211} the perturbative saddle point method gives:
\begin{flalign}
\label{eq:4.33}
{ Z^{\eta} \over  Z^{0} } = \text{exp}(- {1 \over \beta}({mb^2 \over 4} + {\lambda b^4 \over 32m} + ... ) \hat{R}\hat{T}) = \text{exp}(- {1 \over \beta}( 7.84 + ... \ ) \hat{R}\hat{T}), \hat{R}, \hat{T} \rightarrow \infty , \ \beta \rightarrow 0  ~.&&
\end{flalign}
From the comment below equation \eqref{eq:4.4}, the values of $m=2$, $\lambda = - {8 \over 4!}$ and $b = \sqrt{2} \pi$ have been replaced. This shows the convergence of the $SU(2)$ perturbative saddle point method result to the exact value obtained by a direct calculation of the saddle point.

\subsection{Evaluation for $\text{SU(N)}$}

\label{suNanalyticsection}

Having shown how the method works for $SU(2)$, in this  Section we will evaluate the $k$-string tensions perturbatively to next to leading order for $SU(N)$. We start from (recall \eqref{eq:4.222}):
\begin{equation}
\label{eq:4.2333}
Z^{\eta} =\int [D\sigma] \text{exp}(-{1 \over \beta}\int_{{\rm I\!R}^3} d^3\hat{\text{x}}\{{1\over 2}(\partial_l \sigma)^2 - (\mu_k)_j\partial_l \sigma_j \partial_l \eta + {1 \over 2} (\partial_l \eta)^2 {\mu_k}^2 + {1 \over 2} (\sigma_j-\sigma_{j+1})^2 -{1 \over 4!} (\sigma_j-\sigma_{j+1})^4 + ... \})~.  \end{equation}
  The mass term in \eqref{eq:4.2333} can be diagonalized. Let $(\sigma_j-\sigma_{j+1})^2 = \sigma^{T} A \sigma$ (for $j = 1, ...,N$) where $A$ is the following $N \times N$ matrix for $N \geq 3$:
\begin{equation}
\label{eq:4.34}
A_{ij} = 2 \ \text{for} \ i=j \ , \ A_{ij} = -1 \ \text{for} \ |i - j| = 1 \ , \ A_{1N} = A_{N1} = -1 \ , \ A_{ij} = 0 \ \text{otherwise}~, \end{equation}
while for ${SU(2)}$:
\begin{equation}
\label{eq:4.35}
A_{11} = A_{22} = 2 \ \text{and} \ A_{12} = A_{21} = -2~.
\end{equation}
The matrix $A$ is  symmetric   and can be diagonalized by an orthogonal transformation $D$, explicitly $D^TD = I$, $A = D \Lambda D^T$. $D = (v_1 \ v_2 \ ... \ v_N)$ with $v_q$ the eigenvectors of $A$. This gives $(\sigma_j-\sigma_{j+1})^2 = g^{T} \Lambda g$ with $g = D^T \sigma$. $A$ has an eigenvalue\footnote{The diagonalization matrix $D$ has the effect of an $\Z_N$ Fourier transform and the eigenvalues of $A$ are $\Lambda_q = 4 \sin^2{ \pi q\over N}$, $q = 1, ..., N-1$ and $\Lambda_N = 0$. As discussed in \cite{Cherman:2016jtu}, this is the spectrum of a latticized emergent dimension of $N$ sites.}  of zero corresponding to the eigenvector $v^{T}_N \equiv ({1 \over \sqrt{N}}, ... , {1 \over \sqrt{N}})$. The corresponding field $g_N = (\sigma_1 + ... + \sigma_N)/ \sqrt{N}$ would be a massless component which decouples from the rest of the fields and hence can be neglected. We will now express the higher order interaction terms in a form convenient for a perturbative expansion. For this define the matrix $B$ as follows:
\begin{equation}
\label{eq:4.36}
B_{ij} = 1 \ \text{for} \ i=j \ , \ B_{ij} = -1 \ \text{for} \  j = i + 1 \ , \ B_{ij} = 0 \ \text{otherwise}~.
\end{equation}
Defining $h_q \equiv \sigma_q - \sigma_{q+1}$ for $q = 1, ..., N-1$ we have $BDg = B\sigma = \big( \begin{array}{c} h \\ \sigma_N \end{array} \big )$. Also define the top left $(N-1) \times (N-1)$ block of the matrix $BD$ as $K \equiv (BD)_{N-1 \times N-1}$. Since the first $N-1$ elements of the last column of $BD$ are zero\footnote{\label{eigenvectorfootnote}Because $D= (v_1,...,v_N)$, where $v_q$, $q=1,...N-1$ are the eigenvectors of $A$ with eigenvalues $\Lambda_q = 4 \sin^2{ \pi q \over N}$ and  $v_N ={1\over \sqrt{N}} (1,1,1...,1)^T$  is the zero eigenvector. For use below,   the other $N-1$ eigenvectors, for brevity shown for odd $N$ only, with components $v_q^l$ are: $v_{q < {N\over 2}}^l = \sqrt{2\over N} \sin{2 \pi q l\over N}$ and $v_{  {N\over 2}<q<N}^l = \sqrt{2\over N} \cos{2 \pi q l\over N}$.} this gives: $K_{pq} g_q = h_p \equiv \sigma_p - \sigma_{p+1}$ for $p = 1, ..., N-1$. Using the previous notations and definitions, \eqref{eq:4.2333} can now be rewritten as:
\begin{equation}
\label{eq:4.37}
\begin{split}
Z^{\eta} =\int [Dg] \text{exp}( &-{1 \over \beta}\int_{{\rm I\!R}^3} d^3{\text{x}}\{ {1\over 2}(\partial_l g_N)^2+{1\over 2}(\partial_l g_q)^2 + {1 \over 2} \Lambda_q g^2_q - (\mu_k)_j D_{jq}\partial_l g_q \partial_l \eta \\ & + {1 \over 2} (\partial_l \eta)^2 {\mu_k}^2 -{1 \over 4!}\underset{p} {\sum}(K_{pq}g_q)^4 -{1 \over 4!}(\underset{p} {\sum}K_{pq}g_q)^4 + ... \})~,
\end{split}
\end{equation}
where a summation over $p,q$ and $l$ is implicit. Note that $(\mu_k)_j D_{jN} = 0$, hence the massless mode $g_N$ completely decouples from the rest of the modes and interactions. Integrating by parts the Wilson loop term and rescaling $g_q \rightarrow \sqrt{\beta} g_q$, we cast it, similar to (\ref{eq:4.7}), into a form appropriate for a perturbative evaluation of the saddle point:
\begin{flalign}
\label{eq:4.38}
Z^{\eta} =\int [Dg] \text{exp}( &-\int_{{\rm I\!R}^3} d^3{\text{x}}\{ {1\over 2}(\partial_l g_N)^2+{1\over 2}(\partial_l g_q)^2 + {1 \over 2} \Lambda_q g^2_q + {1 \over 2} ({b_q \over 2\pi})^2 (\partial_l \eta)^2 \\ & +\beta {\lambda \over 8}\underset{p} {\sum}(K_{pq}g_q)^4 + \beta {\lambda \over 8}(\underset{p} {\sum}K_{pq}g_q)^4 + ... \})\text{exp}(+ {b_q \over \sqrt{\beta} } \int_A d\text{x}_1d\text{x}_2 \partial_3 g_q)~. \nonumber
\end{flalign}
Here, we defined $b_q = 2\pi (\mu_k)_j D_{jq}$ and $\lambda = - {8 \over 4!}$. To evaluate the Wilson loop exponent using the quadratic terms, we follow steps similar to the ones leading from \eqref{eq:4.8} to \eqref{eq:4.14'}. We obtain the analogue of \eqref{eq:4.14'} for $SU(N)$:
\begin{equation}
\label{eq:4.39'}
{\{ Z^{\eta}_{g^4} \}_{\lambda =0} \over \{ Z^{\eta}_{g^4} \}_{b_q=\lambda =0} } = \text{exp}(- {1 \over \beta} \{ {\Lambda_q b_q^2 \over 2} \underset{\text{A  A}}{\iint'} P_q(\text{x} - \text{x}') + {b_q^2 \over 2} \underset{b(\text{A}) \  b(\text{A})}{\iint'} d\text{x}^l d\text{x}'^k \delta^{l k} ( P_q(\text{x}-\text{x}') - {1 \over 4 \pi |\text{x} -\text{x}'|}) \} )~,
\end{equation}
where $P_q(\text{x}-\text{x}') = {\text{exp}(-\sqrt{\Lambda_q} |\text{x}-\text{x}'|) / (4 \pi |\text{x}-\text{x}'|} )$ and with an implicit summation over $q$. In the limit of a large Wilson loop area ($\hat{R}, \hat{T} \rightarrow \infty$), eq.~\eqref{eq:4.39'}, following a similar calculation as in \eqref{eq:4.11}, reduces to:
\begin{equation}
\label{eq:4.39}
{\{ Z^{\eta}_{g^4} \}_{\lambda =0} \over \{ Z^{\eta}_{g^4} \}_{b_q=\lambda =0} } = \text{exp}(- {1 \over \beta} \{ {\sqrt{\Lambda_q}b^2_q \over 4} \hat{R}\hat{T} + {b_q^2 \over 2} \underset{b(\text{A}) \  b(\text{A})}{\iint'} d\text{x}^l d\text{x}'^k \delta^{l k} ( P_q(\text{x}-\text{x}') - {1 \over 4 \pi |\text{x} -\text{x}'|}) \} )~,
\end{equation}
Using the explicit form of the eigenvectors given in Footnote \ref{eigenvectorfootnote}, we can analytically show that confining strings have finite tension in the large-N limit, despite the vanishing mass gap.\footnote{Note that the string tension remains fixed at large-N, despite the vanishing mass gap, as  there is a number of dual photons of nonzero mass ($\sim m_\gamma$) whose flux is confined, as well as a number of dual photons approaching zero mass ($\sim m_\gamma/N$) whose flux spreads out. Thus the finite tension confining string in the gapless abelian large-N limit is a rather fuzzy object. We defer a further study until the large-N corrections, discussed in \cite{Cherman:2016jtu} for sYM,  are better understood in the dYM case.} In particular, for $k=1$ strings we find the infinite-N limit $a_1 = \lim\limits_{N \rightarrow \infty} \sum\limits_{q=1}^{N-1}{1\over 4} \sqrt{\Lambda_q}b^2_q   = 4 \pi$, consistent with the results from Table~\ref{table:71}. 
For further comments on the large-N limit, see  Sections \ref{sec:compare} and \ref{sec:largeN}. Appendix \ref{sec:appxproduct} discusses the large-N limit of the string tensions in product representations and gives more details on the analytic calculations using the leading-order saddle point method of this  Section.

Next, we evaluate the leading corrections to this result for $SU(N)$ and compare the values obtained with the numerical results in Table \ref{table:1}. The integrals we need to do are the generalization of (\ref{eq:4.19}) from the $SU(2)$ calculation:
\begin{flalign}
\label{eq:4.40}
&\text{I}_{\lambda} \equiv  \big\langle - \int d^3 \text{x} \{ \beta {\lambda \over 8}\underset{p} {\sum}(K_{pq}g_q)^4 + \beta {\lambda \over 8}(\underset{p} {\sum}K_{pq}g_q)^4 \} {1\over 4!}({b_q \over \sqrt{\beta}} \int_A d^2\text{y} \partial_{\text{y}_3} g_q)^4 \big\rangle_{0,C} \\ & = -{1 \over \beta}{4! \over 4!} {\lambda \over 8}P_{q_1q_2q_3q_4} \{ \prod^4_{i=1} K_{pq_i}b_{q_i} + \prod^4_{i=1} K_{p_iq_i}b_{q_i} \} \nonumber~, &&
\end{flalign}
and a summation over $p$, $q_i$, $p_i$ ($i =1,2,3,4$) from $1, ..., N-1$ is implicit. The quantities $P_{q_1q_2q_3q_4}$ are given by:
\begin{equation}
\label{eq:4.41}
P_{q_1q_2q_3q_4} \equiv \int d^3 \text{x} \prod^4_{i=1} \int_A d^2\text{y}^i \partial_{\text{y}^i_3}P_{q_i}(\text{x} -\text{y}^i)~.
\end{equation}
In the limit that $a \rightarrow \infty$ ($a = \hat{R} = \hat{T}$) using \eqref{eq:4.20} we have:
\begin{equation}
\label{eq:4.42}
P_{q_1q_2q_3q_4} \overset{a \rightarrow \infty}{=} {a^2 \over 16} \bar{P}_{q_1q_2q_3q_4} \ \ \text{with} \ \  \bar{P}_{q_1q_2q_3q_4} = \int^{+\infty}_{-\infty} d\text{x}_3 \prod^4_{i=1} \text{E}(\sqrt{\Lambda_{q_i}} \text{x}_3)~,
\end{equation}
hence $\text{I}_{\lambda}$ becomes:
\begin{equation}
\label{eq:4.43}
\text{I}_{\lambda} \overset{a \rightarrow \infty} {=} -{1 \over \beta}{\lambda \over 128}\bar{P}_{q_1q_2q_3q_4} \{ \prod^4_{i=1} K_{pq_i}b_{q_i} + \prod^4_{i=1} K_{p_iq_i}b_{q_i} \} \hat{R}\hat{T}~,
\end{equation}
where $K_{pq} = B_{pj}D_{jq}$ and $b_q = 2\pi (\mu_k)_j D_{jq}$ for $q, p = 1, ..., N-1$. The leading area law term in \eqref{eq:4.39} and its leading correction in \eqref{eq:4.43} need to be evaluated numerically with mathematical software. The results are summarized in Appendix \ref{sec:C}; results for $k=1$ strings for $SU(3)-SU(10)$ were already shown in Table \ref{table:71}; the inclusion of the leading order correction brings the numerical value closer to the numerical or analytic (for $SU(3)$) value.

\section{$\mathbf{N}$-ality dependence and large-$\mathbf{N}$ behaviour of $\mathbf{k}$-strings in dYM  }

\label{sec:5}

In this  Section we give a discussion on two main questions regarding the properties of $k$-string tensions:  their $N$-ality dependence and large-$N$ behaviour in dYM theory. The $N$-ality of an irreducible representation of a gauge group $SU(N)$ refers to the number of boxes in the Young tableaux of the representation mod $N$  \cite{09} or the charge of the representation under the action of the center element $\text{exp}(-i{2 \pi \over N})\text{I}$ of the gauge group. It is believed that asymptotically the string tensions in a gauge theory depend only on the $N$-ality of that representation, see  \cite{04}. This is due to the screening effect by gluons. A cloud of gluons would transform any charge in a representation with $N$-ality $k$ to its $k$-antisymmetric representation which carries the stable lowest energy k-string among different representations with the same $N$-ality $k$. 

We will argue that this is also true in dYM theory and show that the asymptotic string tensions will only depend on the $N$-ality $k$ of the representation $k$. The screening by gluons, in the framework of dYM theory, is due to the pair production of $W$-bosons, an effect (in principle) calculable using weak coupling semiclassical methods.

We discuss qualitatively the role of the unbroken $\Z_N$ center symmetry in dYM for the confining string properties and contrast them to those in another theory with abelian confinement---Seiberg-Witten theory.
We also derive an approximate analytic formula for k-string ratios in dYM theory for $N \sim 10$ and smaller and have a comparison with known scaling laws of k-string ratios.

In regards to their large-$N$  behaviour we show that dYM $k$-string ratios favour even power corrections similar to the sine law scaling and derive the leading terms in the $1/N$ expansion of $k$-string ratios in dYM theory. At the end we will argue that at large $N$  $k$-strings are not necessarily free in gauge theories; in other words, $T$$_{\text{k}}$ can remain smaller than $kT$$_{\text{1}}$ in the large-$N$ limit.
 
\subsection{$N$-ality dependence}

\subsubsection{Asymptotic string tensions depend only on the $N$-ality of the representation}
\label{sec:5.1.1}

The expectation value of the Wilson loop for charges in a representation $r$ with $N$-ality $k$ of $SU(N)$ evaluated using the low energy effective theory in dYM theory in the limit of $\hat{R}, \hat{T} \rightarrow \infty$ and $\beta (= {m^3_{\gamma} \over \tilde{\zeta}})\rightarrow 0$ using \eqref{eq:2.51} and \eqref{eq:3.4} is given by:
\begin{equation}
\label{eq:5.1}
\langle W_r(R,T) \rangle = \sum^{d(r)}_{i = 1} \text{exp}(- T^i_rRT) = \sum^{d(r)}_{i = 1} \text{exp}(- {2 \bar{T}^i_{r} \over \sqrt{2} \beta}\hat{R}\hat{T})~,
\end{equation}
where $d(r)$ refers to the dimension of the representation $r$ and $\bar{T}^i_{r}$ is given by a similar expression as \eqref{eq:3.4} but with $\mu_k$ replaced by the weight $\mu^i_r$ of representation $r$ with $N$-ality $k$:
\begin{flalign}
\label{eq:5.2}
{\bar{T}^i_r}=\underset{f(\text{z})} {\text{min}} \int_{0}^{+\infty} d\text{z}\{({\partial f \over \partial \text{z}})^2+\underset{j}{\sum}[1-\text{cos}(f_j-f_{j+1})]\}, \ \ \ \  f(+ \infty) = 0, f(0) = \pi \mu_r^i  ~.&&
\end{flalign}
Expression \eqref{eq:5.1} is the sum of exponential of area laws. The leading exponential in the limit of large $\hat{T}$ and $\hat{R}$ would give the string tension for charges in representation $r$ with $N$-ality $k$. Any representation of $SU(N)$ with $N$-ality $k$ contains the fundamental weight $\mu_k$ as one of its weights (Appendix \ref{sec:B1}). Therefore in order to show that string tensions would only depend on the $N$-ality of the representation $r$ of the group $SU(N)$ we have to show that the lowest string tension action in \eqref{eq:5.2} corresponds to boundary conditions dictated by the fundamental weight $\mu_k$ among all the weights $\mu_r^i$ ($i=1, ..., d(r)$) of the representation $r$.
Any weight of a representation of $SU(N)$ can be obtained from the highest weight by lowering with the simple roots \cite{09} and as noted above any representation with $N$-ality $k$ contains $\mu_k$ as one of its weights. Therefore any weight of a representation $r$ with $N$-ality $k$ can be obtained from the fundamental weight $\mu_k$ by adding or subtracting the simple roots.

We will now qualitatively (but convincingly) argue that adding or subtracting any simple root from $\mu_k$ would result in boundary conditions that would give a value for the minimum of the action \eqref{eq:5.2} which is equal to\footnote{Degenerate string tensions will  occur when the corresponding weights are related by the unbroken $\Z_N$ center symmetry. For example,  for the fundamental representation all weights have the same string tension, see  Section \ref{sec:compare}. For higher $N$-ality representations, the dim($r$) weights fall into distinct $\Z_N$ orbits, each of which has degenerate string tensions.} or more than the value obtained by boundary conditions of $\mu_k$.

The saddle point solutions $f_j$ of \eqref{eq:5.2} for $\mu_r^i = \mu_k$ start from $\pi (\mu_k)_j$ at $z = 0$ and decrease or increase monotonically to zero at $z = + \infty$. From the form of $\mu_k$ given in (\ref{eq:3.2}), one sees that there are two discontinuities, as a function of $j$, in the boundary conditions for $f_j$. These occur between $j=N$ and $j=1$, since $(\mu_k)_N = -k/N$ and $(\mu_k)_1 =1 - k/N$, and between $j=k$ and $j=k+1$, as $(\mu_k)_k = 1- k/N$ and $(\mu_k)_{k+1} = -k/N$. These two discontinuities in the boundary conditions  make the corresponding terms in the cosine potential $1 - \text{cos}(f_N - f_1)$ and $1 - \text{cos}(f_k - f_{k+1})$ to start from 2 at $z = 0$ and reach $0$ at $z = + \infty$ (this is in contrast to all the other terms, which  start from $0$ at $z=0$ and reach $0$ again at $z = \infty$).
Thus,  for boundary conditions given by $\mu_k$ we would summarize the behaviour of $f_j$'s as follows.
For the kinetic term in \eqref{eq:5.2} we would have $k$ functions $f_1, ..., f_k$ that start from $\pi (1 - k/N)$ at $z = 0$ and reach $0$ at $z = + \infty$ and $N - k$ functions $f_{k+1}, ..., f_N$ that start from $- \pi k/N$ at $z = 0$ and reach $0$ at $z = + \infty$. For the potential term, since only the difference between the $f_j$'s enters the cosine,  the $1 - \text{cos}(f_N - f_1)$ and $1 - \text{cos}(f_k - f_{k+1})$ terms  start at 2 at $z = 0$ and reach 0 at $z= + \infty$, while the rest of the terms  start from 0 at $z = 0$ and reach 0 at $z= + \infty$. 

Now let us ask how this picture would change if simple roots are added or subtracted from $\mu_k$. The picture of the kinetic term will either remain the same (k functions start from $\pi (1-k/N)$ and $N - k$ functions starting from $- \pi k/N$) or it would become worse (and thus increase the value of action) in a way that the boundaries values at $z=0$ become higher and  result in an increase in the kinetic term (since it is the square of the derivative of the functions). The same is true for the cosine term---any addition or subtraction of the simple roots from the weight $\mu_k$ would either not change the picture of the cosine term or it would make it worse (increase the value of action) in a way that we would have more than 2 discontinuities in the boundaries that would result in more than two terms of the potential term having to start from 2 at $z = 0$, or we would still have two discontinuities but the boundary conditions would have become larger and the cosine terms corresponding to these discontinuities would oscillate between 2 and zero more than once. Both our numerical results and the simple variational ansatz of  Section \ref{bagmodelsection}  confirm  this picture.

\subsubsection{Comparing different abelian confinements: strings in dYM vs.  softly-broken Seiberg-Witten theory}
\label{sec:compare}

The two most-studied examples where confinement of quarks becomes analytically calculable within quantum field theory are softly-broken Seiberg-Witten theory on $\R^4$ and QCD(adj) with massive or massless adjoint fermions on $\R^3 \times \S^1$. This paper is devoted to the study of dYM theory, which belongs to the second class, QCD(adj) with massive adjoint fermions. Semiclassical calculability in dYM is achieved, as  mentioned many times, by taking the $N L \Lambda \ll 1$ limit. 

In both dYM and Seiberg-Witten theory confinement is ``abelian:''\footnote{This should not be taken to mean that the nonabelian nature of the theory is not relevant: on the contrary, it is crucial in both examples.} the confining strings form in a regime where $W$-bosons  are not relevant and the dynamics of confinement is described by a weakly-coupled abelian gauge theory. In Seiberg-Witten theory, this is the dual magnetic gauge theory on $\R^4$, while in dYM it is the long-distance theory on $\R^3\times \S^1$---the theory of the dual photons discussed at length in earlier  Sections. In both cases, the confining dynamics involves  magnetically charged---and thus nonperturbative from the point of view of the electric gauge theory---objects: the magnetic monopoles or dyons in Seiberg-Witten theory condense to break the magnetic gauge symmetry, while in dYM, the proliferation of monopole-instantons in the vacuum (which should not really be called ``condensation,'' the title of \cite{Unsal:2007jx} nothwithstanding) leads to the expulsion of electric flux.\footnote{There are hints that the two confinement mechanisms are related, see \cite{Poppitz:2011wy}.}  We shall see, in the next  Section, that the physics of confinement in dYM has a flavor very similar to the picture of the QCD vacuum underlying the MIT Bag Model.

Here, we want to stress two aspects in which dYM confinement  is distinct from Seiberg-Witten theory that have not been much discussed in the literature:
\begin{enumerate}
\item The presence of a global unbroken $\Z_N$ (zero-form\footnote{In the terminology of \cite{Gaiotto:2014kfa}.}) center symmetry in dYM vs. the fact that the Weyl group in Seiberg-Witten theory is broken \cite{15}. The unbroken $\Z_N$ symmetry has implications for the ``meson'' and ``baryon'' spectra of the theory, as we explain further in this  Section.
\item The abelian large-$N$ behaviour: confining string tensions remain finite in dYM in the large-$N$, fixed $\Lambda N L \ll 1$ limit. This is different from their behaviour in the analogous limit of Seiberg-Witten theory, where the string tensions vanish along with the mass gap \cite{15}. For further discussion, see  Section \ref{sec:largeN}.
\end{enumerate} 
Here we concentrate on the first  point above: the  unbroken $\Z_N$ center symmetry in dYM. In the long-distance theory, in the  $N$-dimensional basis of dual photons we are using, this symmetry appears as a clock symmetry, taking $\sigma_i \rightarrow \sigma_{i+1}$, with $N+1 \equiv 1$. In gauge-variant terms, the action of the $\Z_N$ center symmetry resembles that of an unbroken  cyclic subgroup of the Weyl symmetry of $SU(N)$,  as can be seen by noting that it cyclically interchanges the $N$ monopole-instantons associated with the simple and affine root of the Lie algebra.\footnote{In terms independent of the choice of basis vectors of the root lattice, the $\Z_N$ center acts on the dual photons ${\sigma}$ as the ordered product of Weyl reflections with respect to all simple roots, see \cite{Anber:2015wha}.} On the other hand, in Seiberg-Witten theory, the Weyl group is spontaneously broken, as pointed out long ago \cite{15}. 

\begin{figure}
\centering
\begin{minipage}{0.7\textwidth}
	\includegraphics[width=\textwidth]{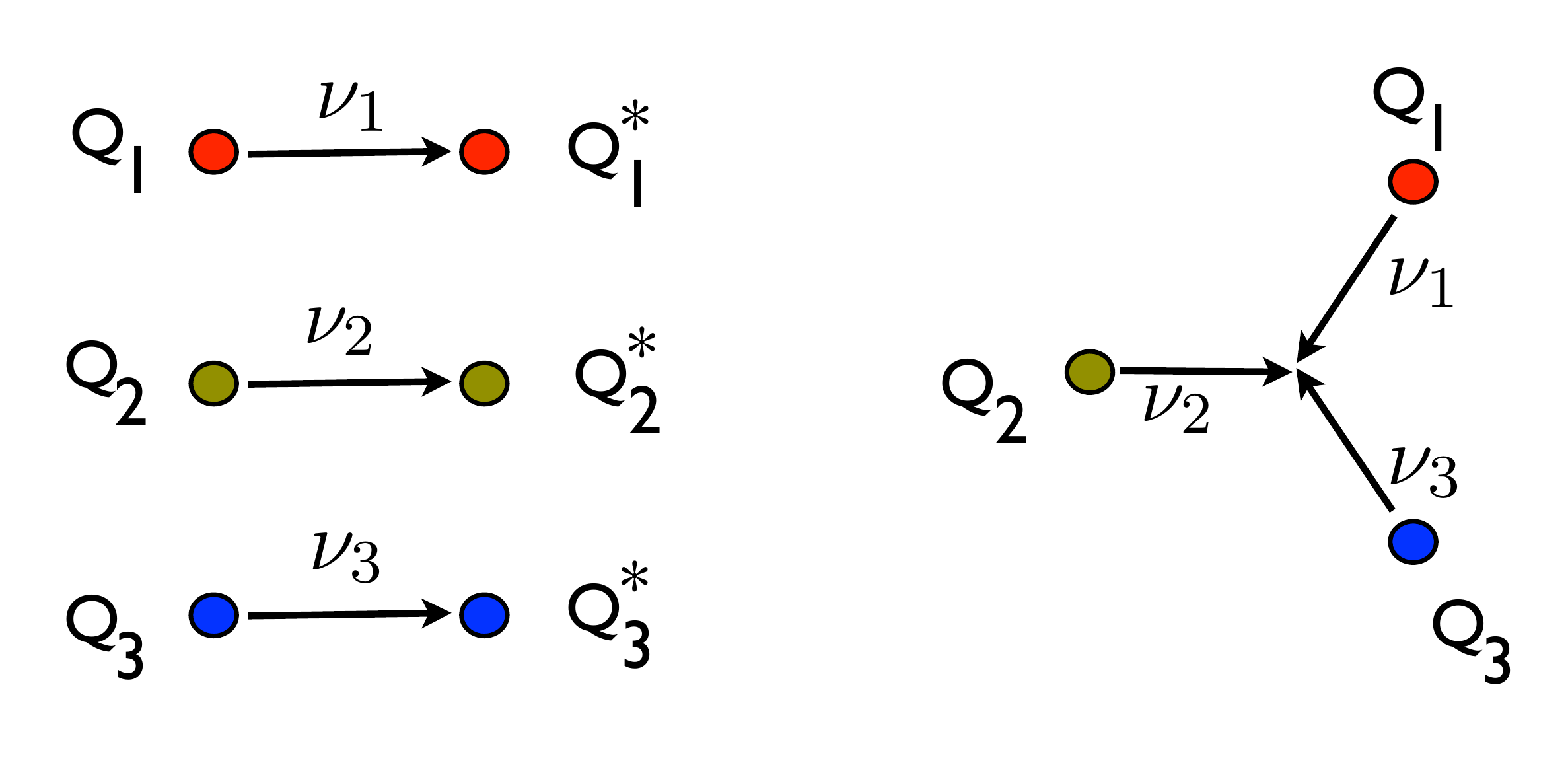}
	\caption{Strings between static quarks of different colors (denoted by color circles) in $SU(3)$ dYM theory. {\it Left panel:} $Q_i \overline{Q}_i$ mesons in $SU(3)$ dYM are degenerate, due to the unbroken $Z_3$ center symmetry. There are three flux tubes carrying fluxes $\nu_i$ ($i=1,2,3$), one for fundamental quarks of each weight (color). {\it Right panel:} A ``baryon vertex'' in dYM is a 3-domain wall junction, which exists due to the vanishing total flux $\nu_1+\nu_2+\nu_3=0$. Similar structures persist for arbitrary number of colors in dYM theory.}{\label{fig:dymstring}}
\end{minipage}
\end{figure}

The different global symmetry realization has interesting implications for the nature of confining strings in the two theories. 
To illustrate the differences it suffices to consider the confinement of fundamental quarks in $SU(3)$. In dYM theory, there are degenerate   ``mesons'' composed of quarks (introduced as static sources) of the three different colors, of weights $\nu_1 = \mu_1$, $\nu_2 = \nu_1 - \alpha_1$, and $\nu_3=\nu_2 - \alpha_2$, respectively. These mesons are confined by distinct flux tubes related by the $\Z_N$ global symmetry action (the action of $\Z_N$ on the weights of the $SU(N)$ fundamental representation is to cyclically permute them).  Furthermore, the fluxes carried by these three strings add up to zero, so one can form a ``baryon,'' where the ``baryon vertex''  is a junction of three strings (a domain wall junction), as illustrated on Figure \ref{fig:dymstring}.

\begin{figure}
\centering
\begin{minipage}{0.8\textwidth}
	\includegraphics[width=\textwidth]{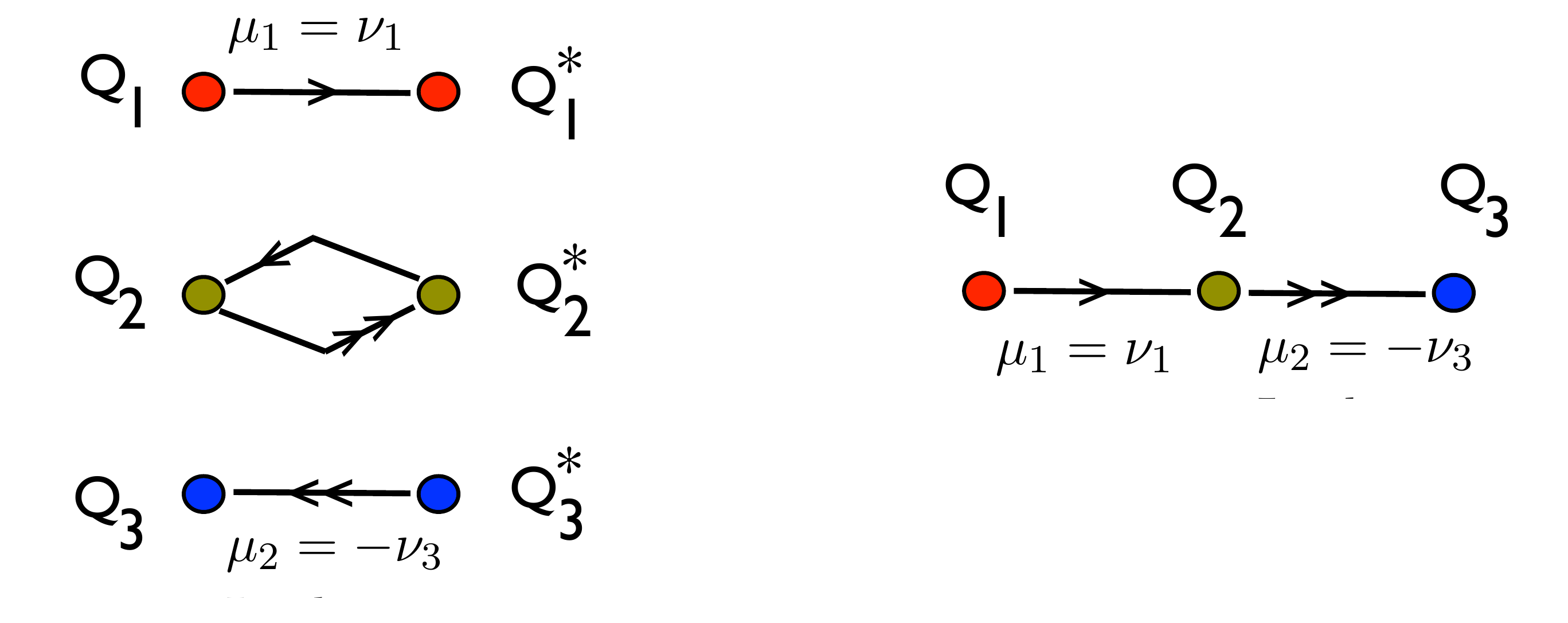}
	\caption{Strings in $SU(3)$ Seiberg-Witten theory. {\it Left panel:} $Q_i \overline{Q}_i$ mesons for different color quarks are non degenerate, due to existence of only two ANO flux tubes (denoted by  lines with a single or double arrow) carrying electric fluxes $\mu_1$ and $\mu_2$ (notice that $\nu_2 = \mu_2-\mu_1$), respectively. {\it Right panel:} Only linear baryons exist in Seiberg-Witten theory. Similar pictures hold for any number of colors.}{\label{fig:SWstrings}}
\end{minipage}
\end{figure}

In contrast, in $SU(3)$ Seiberg-Witten theory, there are two $U(1)$ magnetic gauge groups broken by the monopole condensate, giving rise to two Abrikosov-Nielsen-Olesen (ANO) vortices. The flux of one vortex is proportional to $\mu_1$ and confines quarks in the highest weight of the fundamental representation. The other flux tube carries electric flux proportional to $\mu_2$ (the second fundamental weight of $SU(3)$) and confines quarks in the highest weight of the two-index antisymmetric representation (anti-quarks, for $SU(3)$). There is no third flux tube. The picture of ``mesons'' in $SU(3)$ Seiberg-Witten theory  that results is shown on Figure 
\ref{fig:SWstrings}: the lowest and highest weights of the fundamental quarks are confined by the two ANO flux tubes, while the middle-weight quark is confined by two flux tubes: one of flux $\mu_2$ and an anti-flux tube of flux $\mu_1$. The lack of a third flux tube becomes especially noticeable when baryons are considered: baryons in Seiberg-Witten theory are ``linear molecules'' only, as shown on Figure \ref{fig:SWstrings}. This difference persists and becomes more pronounced for higher rank $SU(N)$ gauge groups.\footnote{See \cite{16} for a description of confining strings in softly-broken Seiberg-Witten theory within its $M$-theory embedding.}

As the $SU(3)$ example illustrates,  the different  symmetry realizations in dYM and Seiberg-Witten theory have implications for the spectrum of mesons and baryons. We shall not pursue this further here, but only note that  in dYM one can add dynamical massive quarks   and the meson, baryon (as well as glueball) spectra can be studied within weakly-coupled field theory, revealing many unusual and surprising features discussed in  \cite{Aitken:2017ayq}.

\subsubsection{An approximate form of $
\text{k}$-string ratios and the MIT Bag Model}
\label{bagmodelsection}

Here, we shall derive a naive analytic upper bound for the half $k$-string tensions in Table \ref{table:1} by approximating the integral in \eqref{eq:3.4} in a simple manner. 
We shall arrive at a simple $k$-string tension scaling law, which is in good agreement with the available data, as described further below. We shall also elaborate on the similarity between confinement in dYM and the MIT Bag Model of the Yang-Mills vacuum. 

We begin by  repeating the dimensionless half $k$-string tension action (recall e.g. eq.~(\ref{eq:5.2})):
\begin{flalign}
\label{eq:3.41}
  {\bar{T_k}}=\underset{f(\text{z}')} {\text{min}} \int_{0}^{+\infty} d\text{z}'\{({\partial 
  \vec{f} \over \partial \text{z}'})^2+ \sum\limits_{j=1}^N[1-\text{cos}(f_j-f_{j+1})]\}, ~~ \vec{f}(+ \infty) = 0, \vec{f}(0) = \pi \vec\mu_k  ~.
   ~ \end{flalign}
Here $\vec{f}$ represents the $N$-dimensional vector of dual photon fields (whose components are summed explicitly in the second term; we omit the arrows in what follows) and the boundary conditions at the origin and at infinity are the ones appropriate for static sources in the highest weight of the $k$-index antisymmetric tensor representation. 

A simple variational ansatz for the half domain wall extremizing (\ref{eq:3.41}) can be obtained by 
 approximating the first term in the action as a linear function connecting the boundary value $\pi \mu_k$ at $z = 0$ to zero at a finite positive $z = J$. The second term is approximated by simply taking its value at $z=0$ (i.e. with $f=\pi \mu_k$) multiplied by $J$; in other words, the fields $f_i$ are taken in the vacuum (where the potential term in (\ref{eq:3.41}) vanishes) outside a region of width $J$ which represents the thickness of the flux tube in our variational ansatz. As the form of $\mu_k$ and the potential term imply, for  $f= \pi \mu_k$ only two terms in the sum of  $N$ cosines contribute a factor of $2$ each, giving rise to second term in (\ref{eq:5.3}), while the remaining $N-2$ terms do not contribute.\footnote{As discussed in  Section \ref{sec:5.1.1},  one of the qualitative reasons why charges $\mu_k$ are confined by strings of the lowest tension (for every representation) is that adding or subtracting any root from $\mu_k$ leads to higher ``vacuum energy'' cost.} 
 Collecting everything, 
using the explicit form of the fundamental weight $\mu_k$ from (\ref{eq:3.2}), we obtain the string tension as a function of the one variational parameter $J$, the flux tube thickness:
\begin{equation}
\label{eq:5.3}
\bar{T_k}^{naive}(x) = J\{ ({\pi {N-k \over N}\over J})^2k + ({\pi {k \over N}\over J})^2(N-k) \} + 4J = {\beta_k \over J} + 4J~,
\end{equation}
where the parameter  
\begin{equation}
\beta_k \equiv (\pi {N-k \over N})^2k + (\pi {k \over N})^2(N-k) = \pi^2 {(N-k)k \over N} ,
\label{betak}
\end{equation}
is proportional to the quadratic Casimir of the $k$-index antisymmetric tensor.
Extremizing \eqref{eq:5.3} with respect to $J$ gives $J_{k, \text{ext}} = {\sqrt{\beta_k} \over 2}$. The value of the string tension at the extremum point is:
\begin{equation}
\label{eq:5.4}
\bar{T}_{k}^{\text{naive}} = 4\pi \sqrt{(N-k)k \over N}~.
\end{equation}
Although the relation \eqref{eq:5.4} is only a naive upper bound estimate for the $k$-strings in Table \ref{table:1}, its ratio with the fundamental ($k=1$) $k$-string  gives a good fit to the ratios of $k$-strings of Table \ref{table:1}:
\begin{equation}
\label{eq:5.5}
{\bar{T}_{k}^{\text{naive}} \over \bar{T}_{1}^{\text{naive}}} = \sqrt{(N-k)k \over N-1}~.
\end{equation}

The relation \eqref{eq:5.5} is, in fact, known as the ``square root of the Casimir'' scaling law for $k$-string ratios. It was first seen to arise  in the MIT Bag Model of the QCD vacuum a long time ago \cite{13}.\footnote{Ref.~\cite{13} studied a rotating string solution, but a simpler static one exists, see discussion below and ref.~\cite{Hasenfratz:1977dt}, which also contains a review of the physical  picture underlying the  MIT Bag Model of the Yang-Mills vacuum.}  
As far as we are aware, dYM theory is the only known example where this ``square root of Casimir'' k-string scaling has been seen to arise within a controlled approximation in quantum field theory. 

We shall now discuss the physics behind (\ref{eq:5.3}) and (\ref{eq:5.5}) and will argue that the similarity of strings in dYM to those in the MIT Bag Model is not an accident.
The first term on the r.h.s. of (\ref{eq:5.3}) represents the gradient energy of the $\sigma$-field. Recall that the duality relation (\ref{dualityrelation}) maps spatial gradients of the dual photon field to electric fields in the perpendicular direction (i.e.~to electric flux going from the quark to the antiquark, which are here taken at infinite separation). Thus, the $\beta_k \over J$ term represents the electric field energy cost (per unit length) for a flux tube of thickness $J$. The coefficient $\beta_k$, the total electric flux, is determined by the sources---quarks in the $k$-index antisymmetric tensor representation---and is proportional to the quadratic Casimir of that representation, as in the classical MIT Bag Model  of the confining string.\footnote{The classical chromoelectric flux of static sources in a given representation is proportional to the quadratic Casimir, see  Section 3.3 in \cite{Hasenfratz:1977dt}. Also note that the ``square root of Casimir'' scaling is obtained in the Bag Model without surface tension and that introducing additional Bag Model parameters, e.g. bag surface tension, modifies the scaling with the Casimir of the representation.}
Naturally,  in order to minimize its energy, the electric flux tube wants to expand, i.e. maximize $J$---in a perturbative vacuum, the chromoelectric field would relax to the dipole field of the quarks. The second term on the r.h.s.~of (\ref{eq:5.3}), equal to $4 J$, represents the energy  cost per unit length  to ``expelling the vacuum'' and replacing it with electric  flux in a region of width $J$. This term represents the ``volume energy cost,'' proportional to the bag constant parameter of the MIT Bag Model. 
In dYM, the vacuum is a monopole-antimonopole medium which abhors electric flux and wants to minimize $J$;   the ``bag constant'' in dYM is not a model parameter, but is determined by the fugacity of monopole-instantons, ultimately fixed by the  underlying gauge theory. The compromise between the two contributions to the energy results in $k$-strings of width $J_{k, \text{ext}} =\sqrt\beta_k/2$ and tensions given in (\ref{eq:5.4}). 

As we already alluded to, the agreement between the dYM and MIT Bag Model $k$-string tensions is not accidental. In the MIT Bag Model, the major assumption is that the chromoelectric fields within the confines of the (presumably small) bag can be treated classically, owing to asymptotic freedom. The ``bag constant" of the YM vacuum, characterizing its abhorrence of electric flux, is introduced as a model parameter. In dYM, both the classical treatment of the Cartan electric fields and the expulsion of electric flux are dynamical features  arising from the judiciously chosen deformation of YM theory and are justified in the $N \Lambda L \ll 1$ limit. 

Finally, we note that while the physical picture in dYM is similar to that in the bag model, the ``square root of Casimir'' scaling of $k$-string tensions discussed here is not exact in dYM, as it results from a simple variational estimate. It is only an upper bound on the string tensions in dYM, see the following  Section and, in particular, Figure \ref{fig:1}.

\subsubsection{Comparison with known scaling laws}

\label{sec:5.1.4}

It is known that the asymptotic string tensions depend only on the $N$-ality $k$ of the representation of the confined charges, hence they are often referred to as the $k$-strings. Different models of confinement make different predictions for the ratios of $k$-string tensions. The main ones are the sine law and Casimir scaling. We  also include the square root of Casimir scaling in the list below, due to its similarity with the $k$-string ratios in dYM theory for N $\sim$ 10 and smaller:
\begin{flalign}
\label{eq:5.6}
& \text{Sine law}: {T_k\over T_1} = {\text{sin}(\pi {k\over N})\over {\text{sin}({\pi \over N})}},  \nonumber \\ & \text{Casimir scaling}: {T_k\over T_1} = {k (N-k) \over N-1},\\ & \text{Square root of Casimir scaling}: {T_k\over T_1} = \sqrt{k (N-k) \over N-1}  ~.\nonumber &&
\end{flalign}
In field theory calculations, usually the corresponding $k$-string tension is calculated to leading order in a small parameter expansion. It has to be noted that the above relations correspond to the leading order result in that expansion and, in each case, are subject to corrections.
\begin{figure}
\centering
\begin{minipage}{0.9\textwidth}
	\includegraphics[width=\textwidth]{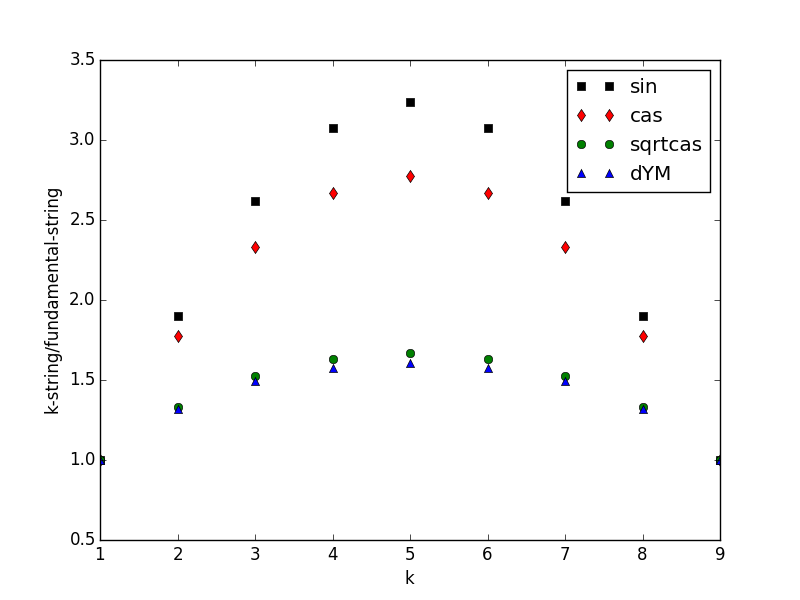}
	\caption{Comparison of $SU(10)$ $k$-string ratios of \eqref{eq:5.4} with dYM $k$-string ratios, labeled by ``dYM'', to other $k$-string tension laws.  
The Sine law  labeled by ``sin'',    the Casimir scaling by ``cas'',  and  scaling with the Square root of the Casimir scaling by  ``sqrtcas''. From the known theoretical models predicting different scalings of $k$-string tensions, the ones in dYM are closest to the MIT Bag Model ``square root of Casimir'' $k$-string tension law. There is a clear physical reason behind this similarity, explained in  Section \ref{bagmodelsection}.}{\label{fig:1}}
\end{minipage}
\end{figure}

The Sine law is found in Seiberg-Witten theory \cite{15}, in MQCD \cite{16},  in three-dimensional SU(N) gauge theories with massless Dirac or Majorana fermions \cite{61}, and in some AdS/CFT-inspired models \cite{17}. Casimir scaling of string tensions refers to the relation between string tensions $T_r/T_F = C_2(r)/C_2(F)$, where $C_2(r)$ and $C_2(F)$ are the quadratic Casimir of representation $r$ and the fundamental representation,  respectively ($T_r$ denotes the string tension for charges in representation $r$). This relation can be derived from the ``dimensional reduction'' form of the Yang-Mills vacuum wave functional \cite{18}, from the stochastic vacuum picture \cite{19}, and from certain supersymmetric dual models \cite{20}. 

$SU(3)$ lattice simulations have shown scaling with the Casimir of the representation $C_2(r)$ with a good accuracy \cite{21}: it holds at intermediate distances ($\lessapprox$ 1 fm) but at larger distances (asymptotically) gluons screen the charges down to the representation of the same non-zero $N$-ality with the lowest dimensionality which carries the most stable lowest string tension---then   $C_2(r)$ is replaced by the Casimir of the $k$-antisymmetric representation which leads to the Casimir scaling relation shown in \eqref{eq:5.6}; notice however, that for $N=3$ $T_1 = T_2$. Lattice studies of $3$-dimensional YM theory seem to also favor Casimir scaling of $k$-string tensions ratios for gauge groups up to $SU(8)$ \cite{Bringoltz:2008nd}, while studies of $4$-dimensional YM theory (for similar number of colors) appear to favor scaling in-between the sine and Casimir laws, see  \cite{Lucini:2012gg} for references and discussion.

 The various $k$-string ratios shown in \eqref{eq:5.6} are compared with dYM $k$-string ratios for $SU(10)$ in Figure \ref{fig:1}. It is clear from the figure that the square root of Casimir scaling shows most similarity with the dYM $k$-string ratios. This scaling arises in the MIT Bag Model of QCD \cite{13} and the reasons for the similarity was discussed in   Section \ref{bagmodelsection}.

\subsection{Large-$\text{N}$ behaviour}
\label{sec:largeN}

One feature of the abelian large-N limit in dYM was already mentioned: in the $N\rightarrow \infty$, fixed-$NL \Lambda \ll 1$ double scaling limit, the mass gap vanishes, but the string tensions stay finite. This large-N behaviour is quite different from a similar abelian large-N limit of Seiberg-Witten theory, where both the string tensions and mass gap vanish \cite{15}.
Furthermore, 
as observed in \cite{Cherman:2016jtu}, in the above double-scaling  limit on $\R^3\times \S^1$, where   the size of the dimension $L\rightarrow 0$ and the number of colors $N\rightarrow \infty$, with $NL$-fixed,  in both super Yang-Mills and dYM, the infrared theory can be viewed as a theory ``living'' in an emergent latticized dimension, in a manner reminiscent of T-duality in string theory. This is a behaviour not quite expected of  quantum field theory and clearly deserves a better understanding. 

The results of this paper  show  the nonvanishing of the string tension in dYM at large N. In the remainder of this  Section, we study the leading large-N corrections to k-string ratios and their large-N behaviour, for a range of N that includes exponentially large values, but does not strictly extent to infinity.
 
The reason for this restriction, already mentioned in   Section \ref{summary} of the Introduction, is that our analysis has neglected the fact that at large values of N, the virtual effects of the W-bosons become  important, as there is a large number of them. In particular,  W-boson loops induce mixing between the Cartan algebra photons (and hence between dual photons), which were not incorporated in our effective Lagrangian. Similar to the discussion of ref.~\cite{Cherman:2016jtu} for sYM (using the calculations of refs.~\cite{08,Anber:2014sda}), we estimate that these mixing terms become important when $N$ becomes comparable to $N^* = 2 \pi e^{+ {24 \pi^2 \over (11 - 4 n_f)(N g^2)}}$. This exponentially large value of $N^*$ is the one that applies to  massless adjoint QCD, and uses the computations in \cite{vito}. The corresponding calculation for dYM (adjoint QCD with massive adjoint flavors) has not yet been performed, but we expect the appearance of  a similar exponentially large  $N^*$. Studying the role of these corrections in dYM is an interesting task for future work, which will allow to further study the intriguing features of the abelian large-N limit.

\subsubsection{Leading large-$\text{N}$ terms}
\label{sec:5.2.1}

\begin{figure}
\centering
\begin{minipage}{.65\textwidth}
	\includegraphics[width=\textwidth]{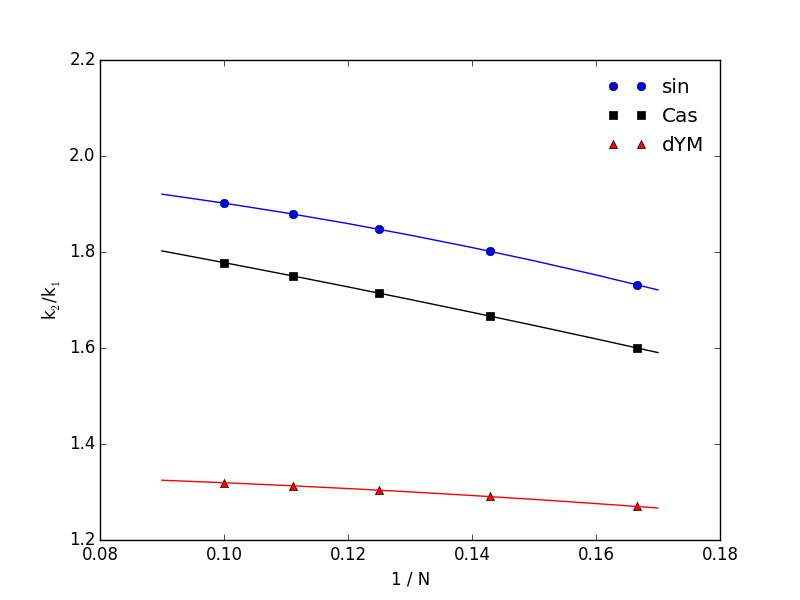}
	\caption{Polynomial fits for $T_2/T_1$ $k$-string ratios in dYM and noised ($\sim$ 0.0005) data in sine, Casimir for $\text{SU}(6 \leq \text{N} \leq 10)$. Sine: $2.007   -0.150\text{x} -9.013\text{x}^2$, Casimir: $1.995 -1.873\text{x} -2.981\text{x}^2$, dYM: $1.345 + 0.035\text{x} -2.917\text{x}^2$.}{\label{fig:2}}
\end{minipage}
\hfill
\begin{minipage}{.65\textwidth}
	\includegraphics[width=\textwidth]{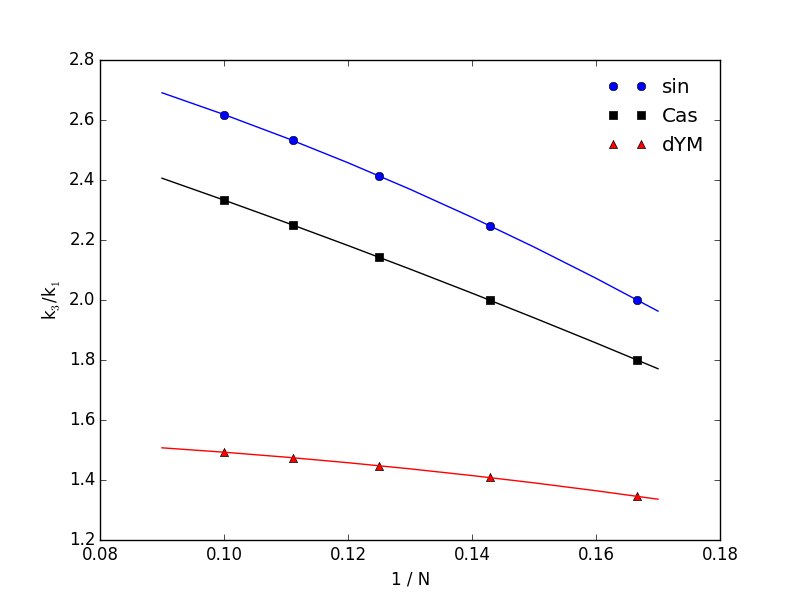}
	\caption{Polynomial fits for $T_3/T_1$ $k$-string ratios in dYM and noised ($\sim$ 0.0005) data in sine, Casimir for $\text{SU}(6 \leq \text{N} \leq 10)$. Sine: $3.105   -2.222\text{x} -26.447\text{x}^2$, Casimir: $2.980 -5.552\text{x} -9.178\text{x}^2$, dYM: $1.551 + 0.390\text{x} -9.734\text{x}^2$.}{\label{fig:3}}
\end{minipage}
\end{figure}

  In this  Section, we derive the leading large-N corrections to k-string ratios in dYM theory for  T$_2$/T$_1$ and  T$_3$/T$_1$. We will show that the k-string ratios in dYM theory favour even  power corrections in $1\over N$. For this we add noise\footnote{Noise of order $\epsilon$ refers to a random fluctuation of order $\epsilon$ imposed on the data. The fluctuation can be a Gaussian, uniform, etc., distribution of width $\epsilon$ centred on the data point. We have used a uniform distribution.} of order $0.0005$, the typical value of error of dYM k-string ratios\footnote{We consider half of the upper bound estimate of the error in Table \ref{table:1} ($-0.006/2 = -0.003$) as the value of error for k-strings. Hence for k-string ratios as a typical example we get: $T_2/T_1 = 8.0006_{-0.003}/6.8583_{-0.003} \approx 1.1666^{+ 0.0005}_{- 0.0004} $. The reader has to be reminded that an error of $-0.003$ is still a high confidence interval for the true value of k-strings.}  to the exact k-string ratios of the Casimir scaling and sine law, whose scaling behaviour is known, and analyze them along with the k-string ratios in dYM theory. From Figures \ref{fig:2} and \ref{fig:3} it can clearly be seen that the coefficient of the linear correction term in dYM k-string ratios similar to the sine law is suppressed (whereas for the Casimir scaling law it is of same order) compared to the constant or the coefficient of the second order term therefore it can be concluded that dYM $k$-string ratios similar to the sine law disfavour a linear correction term and favour even power corrections.

To find the leading term and leading correction term, we add noise of order $0.0005$ to the exact k-string ratios of the sine law for $\text{SU}(5\leq N \leq 10)$ to generate data with errors of order of the errors of the dYM data. Next we generate n = 1000 noised data for dYM and the sine law data with noise and make even power polynomial fits: $c_0 + c_2\text{x}^2 + ... + c_p\text{x}^{p}$ for $p = 2k, k \geq 0$, with $\text{x}=1/N$. The average and standard deviation of $c_0$ and $c_2$ give estimates for the values of these coefficients and their errors. We increase $p$ and make higher order polynomial fits until consistent results are reached. Tables \ref{table:3} and \ref{table:4} summarize the values obtained by this analysis. It can be seen that consistent results are obtained for $p = 6$ and $p = 8$ polynomial fits. In fact the values of the $p = 6$ column for the noised sine law data are in agreement with the exact coefficients of the $1 \over N$ expansion of the sine law k-string ratios as can be seen from \eqref{eq:5.7}. This is not limited to the sine law. Any other function with even power corrections shows a similar behaviour and the results for $c_0$ and $c_2$ coefficients for an even polynomial fit with $p=6$ would agree with the true values of the coefficients in its $1/N$ expansion (e.g. doing the same analysis for a cosine(x) function with $\text{x} = {1 \over N}$). So assuming dYM k-string ratios have only even power corrections, the values of the coefficients in the $p = 6$ column would be in agreement with the true values in dYM theory. 

 The following relations summarize the leading large $N$ corrections in sine, Casimir, square root of Casimir and dYM k-string ratios: 
\begin{flalign}
\label{eq:5.7}
&\text{Sine}: {\text{sin}(k{\pi \over N}) / \text{sin}({\pi \over N}) } = k + (k/6 - k^3/6)\pi^2 ({1 \over N})^2 + ... ~,\nonumber \\
&\text{Casimir}: {k(N -k) / (N-1)} = k + (k - k^2) ({1 \over N}) + ... ~,\nonumber \\
&\text{Sqrt of Casimir}: \sqrt{k(N -k) / (N-1)} = k^{1 \over 2} + {1 \over 2}(k^{1 \over 2} - k^{3 \over 2}) ({1 \over N}) + ...  ~,\\ 
&\text{dYM}: \text{T}_2 / \text{T}_1 = 1.347 \pm 0.001 + (-2.7 \pm 0.2) ({1 \over N})^2 + ... ~,\nonumber \\
& \phantom{{ } = \text{abC}} \text{T}_3 / \text{T}_1 = 1.570 \pm 0.001 + (-7.5 \pm 0.2) ({1 \over N})^2 + ... ~.\nonumber  &&
\end{flalign}

\begin{table}[h]
\centering
\begin{tabular}{|c|c|c || c | | c|}
\hline
 & p=2 & p=4 & p=6 & p=8 \\
\hline
\hline
$\text{c}_0$(sin) & 1.9962 $\pm$ 0.0001 & 2.001 $\pm$ 0.0004  & 2.001 $\pm$ 0.001 & 1.998 $\pm$ 0.005 \\
\hline
$\text{c}_2$(sin) & -9.458 $\pm$ 0.006 & -9.91 $\pm$ 0.03 & -9.9 $\pm$ 0.2 & -9 $\pm$ 1  \\
\hline
\hline
$\text{c}_0$(dYM) & 1.3482 $\pm$ 0.0001 & 1.3465 $\pm$ 0.0004 & 1.347 $\pm$ 0.001 & 1.347 $\pm$ 0.005  \\
\hline
$\text{c}_2$(dYM) & -2.822 $\pm$ 0.006 & -2.65 $\pm$ 0.03 & -2.7 $\pm$ 0.2 & -3 $\pm$ 1 \\
\hline

\end{tabular}
\caption{$c_0$ and $c_2$ for even power polynomial fits of order p for $\text{T}_2 / \text{T}_1$ for noised ($\sim$ 0.0005) sine law data and dYM}
\label{table:3}
\end{table}

\begin{table}[h]
\centering
\begin{tabular}{|c|c|c||c||c|}
\hline
 & p=2 & p=4 & p=6 & p=8 \\
\hline
\hline
$\text{c}_0$(sin) & 2.9397 $\pm$ 0.0001 & 2.9984 $\pm$ 0.0004  & 3.000 $\pm$ 0.001 & 2.999 $\pm$ 0.005 \\
\hline
$\text{c}_2$(sin) & -33.334 $\pm$ 0.006 & -39.17 $\pm$ 0.03 & -39.4 $\pm$ 0.2 & -39 $\pm$ 1  \\
\hline
\hline
$\text{c}_0$(dYM) & 1.5815 $\pm$ 0.0001 & 1.5682 $\pm$ 0.0004 & 1.570 $\pm$ 0.001 & 1.569 $\pm$ 0.005  \\
\hline
$\text{c}_2$(dYM) & -8.594 $\pm$ 0.006 & -7.27 $\pm$ 0.03 & -7.5 $\pm$ 0.2 & -7 $\pm$ 1 \\
\hline

\end{tabular}
\caption{$c_0$ and $c_2$ for even power polynomial fits of order p for $\text{T}_3 / \text{T}_1$ for noised ($\sim$ 0.0005) sine law data and dYM}
\label{table:4}
\end{table}

As a short summary of this  Section, we argued that k-strings in dYM are not free at large N, i.e. $T_k/T_1 \ne k$, and 
  leading corrections to  $T_k/T_1$  are of order $1/N^2$. In the next  Section, we discuss some theoretical questions behind these findings.

\subsubsection{Comments on free $\text{k}$-strings and large-$\text{N}$ factorization}
  \label{sec:5.2.2}
  
An often-discussed expected behaviour of $k$-strings at large $N$ is that they become free, meaning that the string tension with $N$-ality $k$ becomes $k$ times the tension of the fundamental $k=1$-string at large $N$ \cite{14,10}. From the previous  Section, in particular \eqref{eq:5.7}, it can be clearly seen that the k-string tensions in dYM theory show a different behaviour: $\underset{N \rightarrow \infty}{\text{lim}}\text{T}_2 = (1.347 \pm 0.001) \text{T}_1 < 2 \text{T}_1 $ and $\underset{N \rightarrow \infty}{\text{lim}}\text{T}_3 = (1.570 \pm 0.001) \text{T}_1 < 3 \text{T}_1$. The usual line of reasoning that leads to the conclusion  that k-strings become free at large $N$  is based on large-$N$ factorization and assumes the commutativity of the large-$N$ and large Euclidean time $T$ limits. We will show that factorization and commutativity of limits should be treated more carefully.\footnote{For another discussion on the non-commutativity of the large-$N$ and large-$T$ limits refer to \cite{62}.}

{\flushleft{W}}e  first briefly review the usual arguments that lead to free k-strings at large $N$:
\\
\hspace*{5mm} A correlator of two gauge invariant operators A and B can always be written as a factorized expectation value plus a connected expectation value. In the lattice strong coupling expansion and in perturbation theory in gauge theories it is known that the leading term in the large $N$ limit is the factorized one \cite{11}. Assuming a normalization $\langle \text{AB} \rangle\sim O(1)$ we have:
\begin{flalign}
\label{eq:5.8}
\langle\text{A} \text{B}\rangle = \langle\text{A}\rangle\langle\text{B}\rangle + \langle\text{A} \text{B}\rangle_C       ,   \ \ \ \      \langle\text{A}\rangle\langle\text{B}\rangle \sim O(1), \ \ \ \  \langle\text{A} \text{B}\rangle_C \sim O({1 \over N^2})~. &&
\end{flalign}
In particular,  we will apply this formula to the expectation value of a Wilson loop in the product representation:
\begin{flalign}
\label{eq:5.9}
& \langle W_{\square \otimes \square}\rangle \equiv \langle\text{tr} (U_{\square \otimes \square} ... U_{\square \otimes \square})\rangle = \langle\text{tr} \{ (U_{\square} ... U_{\square}) \otimes (U_{\square} ... U_{\square}) \}\rangle = \langle W_{\square} W_{\square}\rangle \nonumber \\ & \Longrightarrow \langle W_{\square \otimes \square}\rangle = \langle W_{\square} W_{\square}\rangle = \langle W_{\square}\rangle\langle W_{\square}\rangle + \langle W_{\square}W_{\square}\rangle_C ~.&&
\end{flalign}
A subscript of a ``square'' (as in $W_{\square}$) refers to the fundamental representation. The product of the link matrices $U$ is being taken along a rectangular Wilson loop $R \times T$.
To find the k-string tensions we take the large $T$ and $R$ limit and consider the leading exponential on the right hand side of \eqref{eq:5.9}. To consider the properties of the k-strings at large $N$ we also take the large $N$ limit. If the large $T$ and $R$ and large $N$ limits commute, then we can reverse the order of limits. Taking the large $N$ limit first  makes the connected term vanish, then taking the large $T$ and $R$ limit we would find: 
\begin{equation}
\label{eq:5.10}\nonumber
\langle W_{\square \otimes \square}\rangle   \sim   \langle W_{\square}\rangle\langle W_{\square} \rangle, ~ \langle W_{\square \otimes \square}\rangle  \sim \text{exp}(- T_2 RT), ~\langle W_{\square}\rangle   \sim \text{exp}(- T_1 RT) ~~ \Longrightarrow~ ~T_2 = 2 T_1,
\end{equation}
i.e. the result that the $k=2$-string tension is twice the fundamental string tension.

 The line of reasoning represented above leads to the result that $k$ strings are ``non-interacting'' and would be correct if the large $T$ and $R$ and large $N$ limits commuted, which is not always true, as we discuss at length below (see  \cite{62} for a discussion in a similar framework and \cite{Witten:1978qu} for a reminder that large-distance and large-$N$ limits' non-commutativity has a long history). An important difference between the large area limit and large-$N$ limit relevant to their non-commutativity is the fact that the large area limit is taken in the same quantum field theory where as the large-$N$ limit is taken in different quantum field theories.
 
To study the general properties of field theories at large $N$, e.g. large $N$ factorization, one takes the large $N$ limit first but to study the asymptotic k-string tensions at large $N$, due to the non-commutativity of the large area and large N limits, one should not take the large $N$ limit first. For any $SU(N)$ theory, the proper way to find asymptotic $k$-string tensions at large $N$ is to first solve for the $k$-string tensions at fixed $N$, this is done by taking the large area $(RT)$ limit and considering the leading exponential in this limit. Then, once the asymptotic $k$-string tensions are determined for each $N$ from the coefficient of the area term of the leading exponential, the large-$N$ limit of $k$-string tensions can be taken. 
 
 The limits cannot be taken in reverse order as the leading exponential in the large area ($RT$) limit, which gives the $k$-string tension for the given value of $N$, can be suppressed in the large-$N$ limit compared to exponentials sub-leading in the large area limit. 
 Let us first illustrate this important point in a toy example, similar to the way the non-commutativity of limits is realized in dYM. As the discussion of dYM is  somewhat lengthy and slightly technical\footnote{See further below the discussion of this  Section (between eqs.~\eqref{eq:5.13'} and \eqref{eq:abd}) as well as the explicit calculations in Appendix \ref{sec:appxproduct}.} we prefer to first illustrate the result by the following example. Consider the function
 \begin{equation} \label{acd}
 g(A,N) = \text{exp}(-T_{N} A) + {1 \over N^p}~ \text{exp}(-T^{\prime}_{N} A)~, ~{\rm with} ~ p>0, 
 \end{equation} where $A$ stands in for the area of the Wilson loop.  Let the large-$N$ limits of $T_N$ and $T'_N$, $\underset{N \rightarrow \infty}{\text{lim}}T^{\prime}_N = T^{\prime}$, $\underset{N \rightarrow \infty}{\text{lim}}T_N = T$, be such that $T^{\prime} < T$. For any large but fixed $N$ the leading term in the large-$A$ limit is $g(A,N) \sim \text{exp}(-T^{\prime} A)$ and for any large but fixed $A$ the leading term in the large-$N$ limit is $g(A,N) \sim \text{exp}(-TA)$. Therefore, if one is interested in the leading exponential in the large-$A$ limit, one should not take the large-$N$ limit first, as this will make the second term in (\ref{acd}), which is leading in the large area limit, vanish. One would then find $g(A,N) \sim \text{exp}(-T A)$, which is an incorrect result for the leading exponential in the large-$A$ limit. A similar behaviour happens in dYM as we discuss in detail further below, see discussion after (\ref{eq:5.13'}).
 
 In what follows, we shall see that in the regime of parameters studied in this work, in particular in the framework of a bounded large-$N$ (see the comments in the beginning of  Section \ref{sec:largeN}), the leading exponential in the large $T$ and $R$ limits, which determines the k-string tensions, comes from the connected term, although it can be shown that for fixed $R$ and $T$ this term will be sub-leading in $N$ compared to the factorized term, similar to the toy example of eqn.~(\ref{acd}).

First, we argue how this can be seen more explicitly from the results of 
   Section \ref{sec:5.1.1}. There, it was argued that the lowest string tension action in the product representation of $N$-ality $2$ corresponds to boundary conditions on the dual photon fields determined by the fundamental weight $\mu_2$ ($\bar{T}^i_{\square \otimes \square}$ in relation \eqref{eq:5.2} for $\mu^i_{\square \otimes \square} = \mu_2$) and for a fundamental representation of unit $N$-ality it corresponds to $\mu_1$ ($\bar{T}^i_{\square}$ in relation \eqref{eq:5.2} for $\mu^i_{\square} = \mu_1$). Hence the leading exponential of the factorized term in the limit of large $\hat{T}$ and $\hat{R}$ is $\text{exp}(-2\times{2\bar{T}_1 \over \sqrt{2} \beta } \hat{R}\hat{T}) $ and the leading exponential for the Wilson loop in the product representation is $\text{exp}(-{2 \bar{T}_2 \over \sqrt{2} \beta } \hat{R}\hat{T})$.  For large values of $N$  from equation \eqref{eq:5.7} we have $\underset{N \rightarrow \infty}{\text{lim}}2\bar{\text{T}}_2 = (1.347 \pm 0.001) 2\bar{\text{T}}_1 <2 \times 2 \bar{\text{T}}_1 $. Clearly,  the factorized term can never produce this leading exponential which should, therefore, come from the connected term. 
  
  This result quoted above can also be obtained without referring to numerics, via the perturbative saddle point method developed in  Section \ref{sec:4}, as shown in Appendix \ref{sec:appxproduct}.     
\\
\par Next, we wish to verify the large $N$ factorization result in dYM and directly argue that, for a Wilson loop in the product representation the factorized term is leading and the connected term is sub-leading for large $N$. 
The discussion that we begin now  becomes more transparent and explicit after reviewing the calculations of Appendix \ref{sec:appxproduct}. 

Consider the expectation value of a  Wilson loop in the product representation ${\square \otimes \square}$. Based on \eqref{eq:5.1}, for fixed but large $R$ and $T$ we have: 
\begin{equation}
\label{eq:5.13'}
\langle W_{{\square \otimes \square}}(R,T) \rangle \sim \sum^{d({\square \otimes \square})}_{h = 1} \text{exp}(- T^h_{{\square \otimes \square}}RT)~,
\end{equation}
where $d({\square \otimes \square}) = N^2$ refers to the dimension of the product representation. In words, the expectation value of the Wilson loop in the product representation is given, in the abelianized dYM theory, by a sum of decaying exponentials, one for each weight $h$ of the product representation, with string tension $T^h_{{\square \otimes \square}}$ corresponding to each weight.

On the other hand, for a Wilson loop in the fundamental representation $\square$ we have:
\begin{equation}
\label{eq:5.14'}
\langle W_{\square}(R,T)\rangle \sim \sum^{d(\square)}_{i = 1} \text{exp}(- T^i_{\square}RT) = N \text{exp}(- T_1RT)~.
\end{equation}
Similar to (\ref{eq:5.13'}), this is a sum of decaying exponentials, one for every weight  of the fundamental representation, with the only simplification ocurring because of the unbroken $\Z_N$ center symmetry, ensuring that 
  the string tensions for all  weights of the fundamental representation have the same value $T_1$. 
 
 Now, let us study (\ref{eq:5.13'}) in more detail. Our considerations from this point to eqn.~\eqref{eq:abd} are more qualitative than quantitatively rigorous (although, as already mentioned, they can be justified in the leading order perturbative evaluation of the saddle point, see Appendix \ref{sec:appxproduct}). They carry similar flavor to our argument of  Section \ref{sec:5.1.1} that strings sourced by quarks with charges in the highest weight of the $k$-index antisymmetric representation have the smallest string tension. However, we find the considerations below quite suggestive and intuitive, supporting the large-$N$ vs. large-$RT$ limit subtlety.
 
  The weights of the product representation are given by the sum of the weights of the fundamental representation in \eqref{eq:B12}: $\mu^{h}_{{\square \otimes \square}} \equiv \mu^{(ij)}_{{\square \otimes \square}} = \mu^{i}_{\square} + \mu^{j}_{\square}$ for $1 \leq h \leq N^2$ and $1 \leq i, j \leq N$. These weights enter the boundary conditions of the string tension action \eqref{eq:52}.  In what follows, we  show that for large $N$ and for $|i - j| \gg 1$ and $|i - j| \ll N$ the string tensions of the product representation become approximately equal to two times the string tension of the fundamental representation at large $N$, i.e. $T^{h}_{\square \otimes \square} \equiv T^{(ij)}_{\square \otimes \square} \approx 2 T_1 $. As there are $O(N^2)$ such tensions, it will be concluded, after considering eq.~\eqref{eq:abd}, that $\langle W_{{\square \otimes \square}}(R,T) \rangle - \langle W_{\square}(R,T) \rangle  \langle W_{\square}(R,T) \rangle < O(N^2)$ and therefore the connected term would be sub-leading in $N$.

 Consider the fundamental string tension $T^q_\square$ given by \eqref{eq:52} with $r = \square$ and $q$ denoting one of the weights of the fundamental representation. The weight $\mu_\square^q$ is given in \eqref{eq:B12}. Recall, from  Section \ref{sec:3.3}, that the dual photon configuration extremizing the action of a given string interpolates between a value at the origin given by $\pi \mu^q_\square$ and zero at infinity. Since the $p$-th component of $\mu_\square^q$ is $(\mu_{\square }^q)_p = -{1 \over N} + \delta^{pq}$, for large values of $N$ all the components of $\sigma_a$, $1 \leq a \leq N$ at $z = 0$ approach zero except for the $q$'th component which approaches $\pi$. The fact that one component, namely $\sigma_q$, differs in its boundary conditions from the rest would result in a non-zero action for $T^q_\square$, otherwise if all components had the same boundary conditions (e.g. $-\pi/N$) at $z=0$ they would be linear functions interpolating between $-\pi/N$ at $z=0$ and zero at $z=J$ and when $J$ is taken to infinity would result in a zero action. This suggests that the main contribution to $T^q_\square$ would come from the components of $\sigma_a$ near\footnote{Note that due to the $Z_N$ symmetry of \eqref{eq:52} the $N$ components of $\sigma_a$, $1 \leq a \leq N$ can be considered similar to $N$ points on a circle corresponding to angles $\theta = {2 \pi a / N }$. The components near the $q$'th component are defined as the points (components) close to the $q$'th point on this circle.} the $q$'th (also, see Appendix \ref{sec:appxproduct}). Conversely, the components farther away from the $q$'th component would approach a linear configuration, similar to the case when all boundary conditions were the same, in order to minimize the action as much as possible and will have negligible effect on the value of the string tension action $T^q_\square$, with their contribution being suppressed by a power of $1/N$. 
 
 A similar picture is true for $T^{(ij)}_{\square \otimes \square}$ with $\mu^h_{\square \otimes \square} = \mu^{(ij)}_{{\square \otimes \square}}$.\footnote{In components, the weights of the product representation are $(\mu^{(ij)}_{\square \otimes \square})_p = \delta^i_p + \delta^j_p  -2/N$. The $\Z_N$ symmetry acts as $\mu^{(ij)}_{\square \otimes \square} \rightarrow \mu^{(i+1({\rm mod}  N),j+1({\rm mod} N))}_{\square \otimes \square}$, i.e. the $N^2$ weights of the product representation fall into $N$  $\Z_N$ orbits.} Due to the $Z_N$ symmetry of the action without loss of generality we can take $i = [(N - \Delta_{ij}+1) /2]$ and $j = [(N + \Delta_{ij}+1)/2]$ with $\Delta_{ij} = |i - j| \neq 0$; the square brackets refer to the integer part. For large $N$, all  components  $(\mu^{(ij)}_{\square \otimes \square})_p$  approach zero, except for the $i$-th and $j$-th components, which approach $1$. 
 
Consider now the calculation of the string tension $T^{(ij)}_{\square \otimes \square}$ for large $N$ with $|i - j| \gg 1$ and $|i - j| \ll N$. The components $\sigma_a$, $1 \leq a \leq N$ interpolate between $\pi (\mu^{(ij)}_{\square \otimes \square})_a$ at the origin and zero at infinity and therefore for large values of $N$ all the components would approach zero at $z=0$ except for $\sigma_i$ and $\sigma_j$ which approach $\pi$. Similar to the picture above for $T^q_\square$ it can be seen that the components of the dual photon fields $\sigma_a$ near the $\sigma_i$ and $\sigma_j$ components would make the main contribution to the string tension $T^{(ij)}_{\square \otimes \square}$. The components farther away from the $i$'th and $j$'th components would approach a linear configuration, similar to the case when all boundary conditions are the same, in order to minimize the action as much as possible and will have negligible effect on the values of the string tension action $T^{(ij)}_{\square \otimes \square}$, with their contribution being suppressed by a power of $1/N$. 
 Next, we divide the string tension action of $T^{(ij)}_{\square \otimes \square}$ into two parts, one part associated with the components $\sigma_a$ for $1 \leq a \leq N /2$ and one for $N/2 < a \leq N$; at large-$N$ and $|i - j| \gg 1$, $|i - j| \ll N$, these actions become independent of each other. In each part, the components closer to the $i$'th and $j$'th components of $\sigma_a$, which are relevant to the string tension value and their boundary conditions are similar to the components near the $q$'th component of $\sigma_a$ for  $T^q_\square$ for $SU([N/2])$.\footnote{In this regard, notice that the components of $\mu^{(ij)}_{\square \otimes \square}$ with $i \ne j$  can also be written as $-1/(N/2)$ or $1- 1/(N/2)$,  similar to the components of $\mu^i_\square$ for $SU([N/2])$ (when $N$ is odd the difference would be clearly negligible).}
For large $N$, these string tensions approach---as per our numerical results of Table~\ref{table:71} or from the analytic study, recall paragraph after \eqref{eq:4.39}---a nonzero value $T_1$  with their differences suppressed by a power of $1/N$. 
 From this observation, we  conclude  that for large $N$ and for $|i - j| \gg 1$ and $|i - j| \ll N$, $T^{(ij)}_{\square \otimes \square} \approx 2T_1$. Clearly, there are $O(N^2)$ such string tensions at large $N$. 
 
 On the other hand, the highest weight of the antisymmetric two index representation, $\mu_2$, see \eqref{eq:3.2}, which was argued and numerically found to give rise to the smallest $N$-ality two string tension, $T_2 < 2 T_1$, is obtained from $ \mu^{(ij)}_{{\square \otimes \square}}$, by taking $i = j+1$ (mod$N$). There are  $O(N)$ such string tensions, including the $\Z_N$-center orbit of the highest weight of the antisymmetric two-index representation.
 
We now combine the results of the previous two paragraphs to conclude, recalling  \eqref{eq:5.13'}, that at large $N$
\begin{equation}\label{eq:abd}
\langle W_{{\square \otimes \square}}(R,T) \rangle \sim  \sum^{d({\square \otimes \square})}_{h = 1} \text{exp}(- T^h_{{\square \otimes \square}}RT)~ \approx O(N^2) \; e^{- 2 T_1 RT} + O(N) \;e^{- T_2 RT}.
\end{equation}
 The first term in the last expression above represents the contribution of the $O(N^2)$ string tensions of weights $ \mu^{(ij)}_{{\square \otimes \square}}$ with $|i - j| \gg 1$ and $|i - j| \ll N$. The second  term is the contribution of the $O(N)$ $k=2$ strings in the $\Z_N$ orbit of the highest weight of the two-index antisymmetric representation. We now note that eqn.~\eqref{eq:abd}  exactly mirrors the situation described and discussed earlier in eqn.~\eqref{acd}, showing the subtlety of taking the large-$N$ vs. large-$RT$ limit.
  See also Appendix \ref{sec:appxproduct}, where \eqref{eq:abd} is recovered using the leading-order perturbative saddle point, evaluated analytically for large-$N$.
  
The discussion in this Subsection demonstrates that in the framework of a bounded large $N$ studied in this work (recall the preamble\footnote{At larger values of $N$, as mentioned in the preamble of  Section \ref{sec:largeN}, the virtual effects of the W-bosons become important which has not been taken into account in this work. We speculate, based on preliminary results, that with taking these effects into account the same picture, i.e. large $N$ factorization and interacting k-strings, persists at large $N$.} of  Section \ref{sec:largeN}) large $N$ factorization would not necessarily imply free k-strings and the leading exponential in the large area limit can come from the connected term of the correlator of two Wilson loops in the fundamental representation that is sub-leading in $N$ compared to the factorized term.
 We remind the reader that although the connected term is sub-leading in $N$ it would still contribute to the k-string tensions at large $N$---since, as already stressed at the beginning of this  Section, to find asymptotic k-string tensions, the large area limit must be taken first and the leading exponential in this limit should be considered. Thus, no matter how large $N$ is, the connected term, which contains the leading exponential in the large area limit, would contribute to the k-string tensions at large $N$.

 \subsubsection{A comment on ``holonomy-decorated''	 Wilson loops}

Here, we want to make a  point which gives additional  justification of our emphasis to study $k$-strings of minimal tensions, corresponding to  quark sources of a particular weight (e.g.  $ \mu^{(ij)}_{{\square \otimes \square}}$, with  $i = j+1$ (mod$N$) for $k=2$). 
So far,  we only  considered gauge invariant Wilson loop operators without insertions of the Higgs field (holonomy).
 In the small-$L$ abelianized regime of dYM theory,  one can isolate the contribution of individual components of the fundamental quarks by inserting powers of the holonomy inside the trace defining the Wilson loop. This gives rise to Wilson loops in $\R^3$ ``decorated'' by loops winding around the $\S^1$, similar to the construction of \cite{Cherman:2016jtu,Aitken:2017ayq}. The construction of these loops shows that the abelian strings of different tensions (due to quarks of a single weight) in product representations are physical, i.e.~they are created by gauge invariant operators.  
We now define a ``decorated'' Wilson loop as follows. The fundamental representation holonomy around the $\S_1$ is 
\begin{equation}
\label{eq:pol1}
\Omega(x)_F = {\cal{P}} e^{i \oint\limits_{\S_1} A^a_4(x,x_4)t^a_F dx_4}~, \ \ \ \ \ \  a =1, ..., N-1.
\end{equation}
The gauge invariant Wilson loop projecting on a single  component of a quark field can  then be written as
\begin{equation}
\label{eq:wilson1}
W^k_F ={\rm tr}_F {\cal P}   \left[{1 \over N} \sum\limits_{p=1}^N \omega_N^{-(N-k)p} (\Omega(x)_F)^p\right] \; e^{i \int\limits_{\text{x}}^{\text{x}} A_\mu d\text{x}^\mu} ~,
\end{equation}
 where $\omega_N = e^{i {2 \pi\over N}}$ and the integral $\int\limits_x^x$ is taken along a large $RT$ contour in $\R^3$, broken up at the point $x$ where the Higgs field is inserted. In the center symmetric vacuum  at weak coupling $\Omega$ can be replaced by its vacuum expectation value, $\langle \Omega \rangle$, given (for brevity, shown below only for odd $N$ and recalling (\ref{a4vev})), by
 \begin{equation}\label{evs12}
 \langle \Omega \rangle = {\rm diag}( \omega_N^{N-1}, \omega_N^{N-2},..., \omega_N, 1)~.
 \end{equation} Hence, the holonomy insertion and discrete Fourier transform in \eqref{eq:wilson1}project  (the term in square brackets inside the trace in \eqref{eq:wilson1})  the Wilson loop to an abelian component corresponding to a source given by the $k$-th component (weight) of the fundamental quark (in the ordering of eigenvalues as in (\ref{evs12})). Using \eqref{eq:wilson1}, one can construct sources of various weights in product representations.  
\appendix

\section{Derivation of $\mathbf{W}$-boson spectrum}
\label{app:wboson}

Consider two analogs of off-diagonal $SU(2)$ generators in $SU(N)$, namely $T_{(kl)}^1$ and $T_{(kl)}^2$ (analogs of $\tau^1/2$ and $\tau^2/2$ in $SU(2)$ respectively), $1 \leq k , l \leq N$, where  $k \neq l$, refer to the row and column of the non-zero components of these generators. We will work out the quadratic  {mass} terms associated with their corresponding gauge fields $A_i^1T_{(kl)}^1$ and $A_i^2T_{(kl)}^2$. The mass term comes from the $F^2_{i4}$ term in \eqref{eq:2.6} with $F_{i4} = \partial_i A_4 - \partial_4 A_i -i [A_i,A_4]$:
\begin{eqnarray}
\label{eq:2.9}
\{ {1 \over 2g^2} \text{tr}F^2_{i 4}(x) \}_{\text{quad} , A_i^{1,2}} &=& {1 \over 2g^2}\{ \text{tr}(\partial_4 A^{1,2}_i)^2 - \text{tr}([A^{1,2}_i,A^{vev}_4])^2  \\ \nonumber
&& + 2i \text{tr}(\partial_4 A^{1,2}_i [A^{1,2}_i,A^{vev}_4]) \} ~,
\end{eqnarray}
with $A_i^{1,2} = A_i^1T_{(kl)}^1 + A_i^2T_{(kl)}^2$. Noting that $[A^{1,2}_i,A^{vev}_4] = {2\pi |l-k| \over NL}i(A_i^2T_{kl}^1 - A_i^1T_{kl}^2)$, expanding each component in its Fourier modes using \eqref{eq:2.8}, and integrating over the compact $\text{x}_4$ direction we have:
\begin{flalign}
\label{eq:2.10}
\int^L_0 d\text{x}_4  \{ {1 \over 2g^2} \text{tr}F^2_{i 4}(x)  \}_{
quad, A_i^{1,2}} & =  {L \over 2g^2} \overset{+\infty}{\underset{m=0}{\sum}} \{({2 \pi m \over L})^2( A_{i,m}^1A_{i,m}^{1 \dagger} + A_{i,m}^2A_{i,m}^{2 \dagger} ) \nonumber
\\ & + ({2\pi |l-k| \over NL})^2 ( A_{i,m}^1A_{i,m}^{1 \dagger} +   A_{i,m}^2A_{i,m}^{2 \dagger} ) \\ \nonumber
& +2i {2\pi |l-k| \over NL}{2 \pi m \over L} (A^{1 \dagger}_{i,m}A^{2}_{i,m} - A^{2 \dagger}_{i,m}A^{1}_{i,m}) \} ~. &&
\end{flalign}
Expanding in real and imaginary parts of Fourier components we have:
\begin{flalign}
\label{eq:2.11}
\int^L_0 d\text{x}_4 \{ {1 \over 2g^2} \text{tr}F^2_{i 4}(x) \}_{
quad,A_i^{1,2}} & =  {{L \over 2g^2} \overset{+\infty}{\underset{m=0}{\sum}} \{ ({2 \pi m \over L})^2 [ (A_{i,m1}^1)^2 + (A_{i,m2}^1)^2 + (A_{i,m1}^2)^2 + (A_{i,m2}^2)^2 ] } \nonumber \\
 & {+4 {2\pi |l-k| \over NL}{2 \pi m \over L} ( A^{1}_{i,m2}A^{2}_{i,m1} - A^{2 }_{i,m2}A^{1}_{i,m1})}  \\ 
& \nonumber
{+ ({2\pi |l-k| \over NL})^2 [ (A_{i,m1}^1)^2 + (A_{i,m2}^1)^2 + (A_{i,m1}^2)^2 + (A_{i,m2}^2)^2 ] \}} ~, &&
\end{flalign}
with $A_{i,02}^1 = A_{i,02}^2 = 0$. The above mass terms can be diagonalized by defining the following fields:
\begin{equation}
\label{eq:2.12}
\begin{split}
\bar{A}^1_{i,m} \equiv (A^{1}_{i,m1} + A^{2}_{i,m2}) / \sqrt{2}, \bar{A}^2_{i,m} \equiv (A^{1}_{i,m2} + A^{2}_{i,m1}) / \sqrt{2} \\ \bar{A}^3_{i,m} \equiv (A^{1}_{i,m2} - A^{2}_{i,m1}) / \sqrt{2} , \bar{A}^4_{i,m} \equiv (A^{1}_{i,m1} - A^{2}_{i,m2}) / \sqrt{2}~,
\end{split}
\end{equation}
leading to the quadratic Lagrangian for the off-diagonal components:
\begin{equation}
\label{eq:2.13}
\begin{split}
\int^L_0 d\text{x}_4 \{ {1 \over 2g^2} \text{tr}F^2_{i 4}(x) \}_{
quad,A_i^{1,2}}  = & {L \over 2g^2} \overset{+\infty}{\underset{m=0}{\sum}} \{ ({2 \pi m \over L} - {2\pi |l-k| \over NL} )^2  {[(\bar{A}^1_{i,m})^2 + (\bar{A}^3_{i,m})^2]} \\ & + ({2 \pi m \over L} + {2\pi |l-k| \over NL} )^2 {[ (\bar{A}^2_{i,m})^2 + (\bar{A}^4_{i,m})^2 ]} \}~.
\end{split}
\end{equation}
Relation \eqref{eq:2.13} shows that there are W-bosons $ {W^{\pm}_1 = (\bar{A}^1_{i,m} \pm i \bar{A}^3_{i,m})/\sqrt{2}}$  and $ {W^{\pm}_2 = (\bar{A}^4_{i,m} \pm i \bar{A}^2_{i,m}) / \sqrt{2}}$ with masses $|{2 \pi m \over L} - {2\pi |l-k| \over NL}|$ and $|{2 \pi m \over L} + {2\pi |l-k| \over NL}|$ respectively for $m =0,1,2,...$ and $1\leq l < k \leq N$. 

\section{Error analysis}
\label{errorappendix}

\subsection{Truncation error}
\label{sec:A1}
In this  Section we will discuss relation \eqref{eq:3.13}. Consider \eqref{eq:3.10} at $m \rightarrow \infty$ and at its minimum solution:
\begin{equation}
\label{eq:A1}
\bar{T}^{m,J}_{k,\text{min}}=\bar{T}^{m,J}_{k1,\text{min}}+\bar{T}^{m,J}_{k2,\text{min}}~.
\end{equation}
The only explicit dependence on $J$ in \eqref{eq:A1} is through $\Delta z =  {J / m}$. Extracting this explicit dependence and suppressing the indices we have:
\begin{equation}
\label{eq:A2}
H^J={H_1\over J} + JH_2 , \ \ \ \bar{T}^{m,J}_{k,\text{min}} \equiv H^J, \ \ \ \bar{T}^{m,J}_{k1,\text{min}} \equiv {H_1\over J}, \ \ \ \bar{T}^{m,J}_{k2,\text{min}} \equiv {JH_2}~.
\end{equation}
Taking the derivative of $H^J$ with respect to $J$ gives:
\begin{equation}
\label{eq:A3}
{dH^J\over dJ}=-{H_1\over J^2} + H_2 + {1\over J}{\partial H_1\over \partial f_{jh}}{df_{jh}\over dJ} + J{\partial H_2\over \partial f_{jh}}{df_{jh}\over dJ}~.
\end{equation}
Since the partial derivatives at the minimum solution vanish we have:
\begin{equation}
\label{eq:A4}
{dH^J\over dJ}={1\over J}(-{H_1\over J} + JH_2)~.
\end{equation}
For $J_1 < J_2$ it can be shown that $H^{J_2} < H^{J_1}$. The minimum solution of $H^{J_i}$ is a path $P_i$ that connects the boundary point $\pi \mu_k$ at $\text{z} = 0$ to $0$ at $\text{z} = J_i$ for $i=1,2$ ( Section \ref{sec:3.3}). If we extend path $P_1$ on the z-axis from $\text{z} = J_1$ to $\text{z} = J_2$ we would obtain a path $\tilde{P_1}$ that connects the boundary point $\pi \mu_k$ at $\text{z} = 0$ to $0$ at $\text{z} = J_2$. But the value of the action of the paths $P_1$ and $\tilde{P_1}$ is the same since the portion of the path $\tilde{P_1}$ that is on the z-axis gives zero action. On the other hand the action of $\tilde{P_1}$ should be higher than the action of $P_2$ since $P_2$ is the minimizing path of $H^{J_2}$ hence $H^{J_2} < H^{J_1}$. Due to this $ {dH^J / dJ < 0}$. From \eqref{eq:A4} this gives ${H_1 / J} > JH_2$ and hence $\bar{T}^{m, J}_{k1,\text{min}} > \bar{T}^{m,J}_{k2,\text{min}}$ for all $0 < J < \infty$. Also we should have $\underset{{J \rightarrow \infty}}{\lim}  {dH^J / dJ} = 0$ since $H^J$ is a decreasing function of $J$ and it is bounded from below ($\underset{{J \rightarrow \infty}}{\lim} {H^J} = \bar{T_k}$). This shows that $\bar{T}^{m,\infty}_{k1,\text{min}} = \bar{T}^{m,\infty}_{k2,\text{min}}$. Also in the limit of $J \rightarrow 0$ the relations $\underset{{J \rightarrow 0}}{\lim} \bar{T}^{m,J}_{k1,\text{min}} = \infty$ and $\underset{{J \rightarrow 0}}{\lim} \bar{T}^{m,J}_{k2,\text{min}} = 0$ can be easily verified from \eqref{eq:3.4}. This can also be seen from relation \eqref{eq:A2}, $H_1$ and $H_2$ are finite quantities hence the previous limits follow. Summarizing the previous relations derived we have:
\begin{equation}
\label{eq:A5}
\begin{aligned}
& \bar{T}^{m, J}_{k1,\text{min}} > \bar{T}^{m,J}_{k2,\text{min}},\ \  \bar{T}^{m, J}_{k,\text{min}} > \bar{T}^{m, \infty}_{k,\text{min}}, \ \ \ \ \ \ \ \ \ \ \ 0 \leq J < \infty \\
& \bar{T}^{m,0}_{k1,\text{min}} = \infty ,\ \ \ \ \ \ \ \  \bar{T}^{m,0}_{k2,\text{min}} = 0, \ \ \ \ \ \ \ \ \bar{T}^{m, \infty}_{k1,\text{min}} = \bar{T}^{m, \infty}_{k2,\text{min}} = {1\over 2} \bar{T}^{m, \infty}_{k,\text{min}}~.
\end{aligned}
\end{equation}
From \eqref{eq:A5} it can be seen that $\bar{T}^{m,J}_{k2,min}$ starts from $0$ at $J=0$ and approaches ${1\over 2} \bar{T}^{m, \infty}_{k,\text{min}}$ at $J=\infty$. We conjecture that for $0<J<\infty$, $\bar{T}^{m,J}_{k2,\text{min}}<{1\over 2} \bar{T}^{m, \infty}_{k,\text{min}}$\footnote{ This behaviour has been verified in the numerical simulations up to $J=14$. It has also been verified for cases when an analytic solution is possible. For example expanding the cosine term and keeping only the quadratic term. For this case it would be possible to solve the saddle point analytically for a finite boundary condition at $\text{z} = J$ and see that $\bar{T}^{\infty,J}_{k2,min}$ starts from $0$ at $\text{z}=0$ and increases monotonically to ${1\over 2} \bar{T}^{\infty, \infty}_{k,\text{min}}$ at $\text{z} = \infty$.}. Also from \eqref{eq:A5} we have ${1\over 2} \bar{T}^{m, \infty}_{k,\text{min}} < {1\over 2} \bar{T}^{m, J}_{k,\text{min}}={1\over 2}(\bar{T}^{m, J}_{k1,\text{min}} + \bar{T}^{m,J}_{k2,\text{min}}) < \bar{T}^{m, J}_{k1,\text{min}}$ so we obtain the following inequalities:
\begin{equation}
\label{eq:A6}
\bar{T}^{m,J}_{k2,\text{min}} < {1\over 2} \bar{T}^{m, \infty}_{k,\text{min}} < {1\over 2} \bar{T}^{m, J}_{k,\text{min}} < \bar{T}^{m, J}_{k1,\text{min}}~.
\end{equation}
From \eqref{eq:A6} relation \eqref{eq:3.13} easily follows.
\subsection{Sample error calculation}
We have shown data for half $k$-string tensions of $SU(10)$, k = 5 in Tables \ref{table:5} and \ref{table:6} to perform a sample error calculation.
\begin{table}[ht]
\centering
\begin{tabular}{ |c|c|c|c|c|c|c|c|c|}
\hline
 m & n & $<T_{k1}>$ & $\sigma_{k1} \over \sqrt{R}$ &$<T_{k2}>$ & $\sigma_{k2} \over \sqrt{R}$ & $<T_{k}> $& $\sigma_{k} \over \sqrt{R}$\\
\hline
100 &18 & 5.82434259 & 6.9E-05  & 5.82600776 & 6.2E-07 & 11.6503503 & 6.2E-07\\
\hline
100 &19 & 5.82423417 & 1.4E-05 & 5.82610277 & 1.4E-05 & 11.6503369 & 1E-07\\
\hline
100 & 20 &5.82420397 & 3.9E-06 &5.82612949 & 3.9E-06 & 11.6503335 & 1.3E-08\\
\hline
200 & 20 & 5.82500582 & 1.20E-04 & 5.82415592 & 1.20E-04 & 11.6491617 & 6.90E-07\\
\hline
200 & 21 & 5.82486451 & 5.00E-05 & 5.82428283 & 5.00E-05 & 11.6491473 & 1.70E-07\\
\hline
200 & 22 & 5.82482467 & 2.30E-05 & 5.82431903 & 2.30E-05 & 11.6491437 & 4.00E-08\\
\hline
\end{tabular}
\caption{$SU(10)$, $k=5$ sample data with error in the mean ($\sigma \over \sqrt{R}$)}
\label{table:5}
\end{table}

\begin{table}[ht]
\centering
\begin{tabular}{ |c|c|c|c|c|c|c|c|c|}
\hline
 m & n & $<T_{k1}>$ & $\Delta_n$ &$<T_{k2}>$ & $\Delta_n$ & $<T_{k}> $& $\Delta_n$\\
\hline
100 &18 & 5.82434259 & -  & 5.82600776 & - & 11.6503503 & -\\
\hline
100 &19 & 5.82423417 & -(1E-04) & 5.82610277 & 1E-04 & 11.6503369 & -(1E-05)\\
\hline
100 & 20 &5.82420397 & -(3E-05) &5.82612949 & 3E-05 & 11.6503335 & -(3E-06)\\
\hline
200 & 20 & 5.82500582 & - & 5.82415592 & - & 11.6491617 & -\\
\hline
200 & 21 & 5.82486451 & -(1E-04) & 5.82428283 & 1E-04 & 11.6491473 & -1E-05\\
\hline
200 & 22 & 5.82482467 & -(4E-05) & 5.82431903 & 4E-05 & 11.6491437 & -(4E-06)\\
\hline
\end{tabular}
\caption{$SU(10)$, $k=5$ sample data with $\Delta_n=<X_n> - <X_{n-1}>$.}
\label{table:6}
\end{table}
\textbf{Sample minimization error calculation:}
\\
Sample calculation for $SU(10)$, $k = 5$ and $m = 100$:
\\
\ Min. \ Error \ for \ $<T_{k2}> = |\Delta_{20}| + {\sigma_k \over \sqrt{R}}$ = 3E-05 + 3.9E-06 $\sim$ 3E-05

From Tables \ref{table:5} and \ref{table:6} and the sample calculation above it is clear that the minimization errors are of order $10^{-5}$ and less so they can be safely neglected in comparison to the discretization and truncation errors.
\\
\\
\textbf{Sample discretization error calculation:}
\\
The difference between $<T_{k1}>$, $<T_{k2}>$ and $<T_{k}>$ for $m=100$ and $m=200$ is:
\begin{equation}
\label{eq:diserror}
\begin{split}
&<T_k>_{200} - <T_k>_{100} = 11.6491437 - 11.6503335 = -0.0011898
\\
&<T_{k1}>_{200} - <T_{k1}>_{100} = 5.82482467 - 5.82420397 = 0.0006207
\\
&<T_{k2}>_{200} - <T_{k2}>_{100} = 5.82431903 - 5.82612949 = -0.00181046~.
\end{split}
\end{equation}

Hence we predict that the continuum value of the string tensions of $SU(10)$, $k = 5$ for $J=14.0$ would be: 
\begin{equation}
\label{eq:A.8}
\begin{split}
&<T_k> = 11.6491_{-0.001}
\\
&<T_{k1}> = 5.8248^{+0.0006}
\\
&<T_{k2}> = 5.8243_{- 0.0018} ~.
\end{split}
\end{equation}
\textbf{Sample truncation error calculation:}
\\
\\
Relation \eqref{eq:3.13} gives an upper bound estimate for the truncation error. Based on \eqref{eq:A.8}:
\begin{equation}
\label{eq:truncerror1}
\begin{split}
|<T_{k1}> - <T_{k2}>| \lessapprox 5.8248 + 0.0006 - (5.8243 - 0.0018) = 0.0029
\\
 \Longrightarrow \text{Trunc.} \ E. \ \lessapprox 2\times 0.0029 = 0.0058~.
\end{split}
\end{equation}
Adding the truncation and discretization error and neglecting the minimization error we have:
\begin{equation}
\label{eq:tot}
\text{Total Error} = \text{Trunc. E.} + \text{Dis. E.} = 0.0058 + 0.001 = 0.0068 \approx 0.007~.
\end{equation}

Hence we predict the value of the half string tension for $SU(10)$ and $k = 5$ is: $11.6491_{- 0.007}$. The errors obtained by this method for different half $k$-string tensions varied from 0.005 to 0.007 therefore we have considered the average as an upper bound estimate for the value of the error for all half $k$-string tensions and included it in Table \ref{table:1}. It has to be noted that upper bound estimates for errors always overestimate the true value of the errors as can be seen from Table \ref{table:2}.

\section{Derivation of group theory results}

\subsection{Any representation of ${SU(N)}$ with ${N}$-ality ${1 \leq k \leq N-1}$ contains the fundamental weight ${\mu_k}$ as one of its weights}
\label{sec:B1}
The simple roots $\alpha_i$ and fundamental weights $\mu_k$ of $SU(N)$ are given by the following relations:
\begin{equation}
\label{eq:B1}
\begin{split}
& \alpha_i=(0,..,0,\overset {\text{i-th}}{\widehat{1}},-1,0,...,0), \ \ \ \ \ \ \ \ \ \ \ \ \ \ \ \ \ \ 1\leq i \leq N-1 \\
& \mu_k=({N - k \over N}, ..., \overset {\text{k-th}} {\widehat{{N-k \over N}}}, {-k \over N}, ..., {-k \over N}),\ \ \ \ \ \ 1\leq k \leq N-1
\end{split}
\end{equation}
An arbitrary representation of $SU(N)$ with $N$-ality $1 \leq k \leq N-1$ can be represented by its highest weight $w_k$:
\begin{flalign}
\label{eq:B2}
w_k = h_i \mu_i \ \text{with} \ h \equiv h_1 + 2h_2 + ... + (N-1)h_{N-1} = mN + k , \ m, h_i \in \mathbb{Z} , \ m, h_i \geq 0~,  && 
\end{flalign}
where $ {h_i \geq 0}$ are the $N-1$ Dynkin indices of the representation, which determine $k$,  its $N$-ality, by the mod($N$) relation given above.
The proof involves two steps. First we will prove the following lemma: \\
\textbf{Lemma} $w_k = \mu_k + a_i \alpha_i$ for $a_i \in \mathbb{Z} \ \text{and} \  a_i \geq 0$
\begin{proof}
It can be easily seen that:
\begin{equation}
\label{eq:B3}
\mu_k = k \mu_1 - \beta_k \ \  \text{with} \ \  \beta_k = (k-1)\alpha_1 + (k-2)\alpha_2 + ... + \alpha_{k-1}, \ \ \beta_1 = 0
\end{equation}
Hence $w_k$ can be written as:
\begin{equation}
\label{eq:B4}
 w_k = (h_1 + 2h_2 + ... + (N-1)h_{N-1}) \mu_1 - (h_2 \beta_2 + ... + h_{N-1} \beta_{N-1}) \\
\end{equation}
With knowing $N \mu_1 = (N-1) \alpha_1 + (N-2) \alpha_2 + ... + \alpha_{N-1}$ and \eqref{eq:B3} we have:
\begin{equation}
\label{eq:B5}
\begin{split}
 h \mu_1 = (mN + k) \mu_1 = & \ \mu_k + (m(N-1) + k-1) \alpha_1 + ... + (m(N-(k-1)) + 1) \alpha_{k-1} \\ & + m(N-k) \alpha_{k} + ... + m \alpha_{N-1}
\end{split}
\end{equation}
Therefore using \eqref{eq:B3} and \eqref{eq:B5}, $w_k$ in \eqref{eq:B4} can be written as:
\begin{equation}
\label{eq:B6}
w_k = \mu_k + b_i \alpha_i \ \ \ \ \text{with} \ \ \ \ b_i \in \mathbb{Z}
\end{equation}
We need to show that $b_i$ is greater than or equal to zero. Lets assume the contrary. First lets assume $b_k < 0$:
\begin{equation}
\label{eq:B7}
\begin{split}
&b_k < 0 \Longrightarrow b_k \leq -1 \\
&\eqref{eq:B6} \ \& \ \eqref{eq:B2} \Longrightarrow  \alpha_k \cdot w_k = 2b_k + 1 - b_{k+1} - b_{k-1} = h_k \geq 0,  \\
& b_k \leq -1 \ \& \ 2b_k + 1 - b_{k+1} - b_{k-1} \geq 0 \Longrightarrow b_{k+1} \leq b_k \ \text{or} \ b_{k-1} \leq b_k~.
\end{split}
\end{equation}
Lets assume $b_{k-1} \leq b_k$:
\begin{equation}
\label{eq:B8}
\begin{split}
& \eqref{eq:B6} \ \& \ \eqref{eq:B2} \Longrightarrow  \alpha_{k-1} \cdot w_k = 2b_{k-1} - b_{k} - b_{k-2} = h_{k-1} \geq 0, \\
& b_{k-1} \leq b_k \ \& \ 2b_{k-1} - b_{k} - b_{k-2} \geq 0 \Longrightarrow b_{k-2} \leq b_{k-1}~.
\end{split}
\end{equation}
Similarly, it can be concluded that $0 > b_k \geq b_{k-1} \geq b_{k-2} \geq b_{k-3} \geq ... \geq b_2 \geq b_1$. Here we will clearly have a contradiction since we have:
\begin{equation}
\label{eq:B9}
\begin{split}
& \eqref{eq:B6} \ \& \ \eqref{eq:B2} \Longrightarrow  \alpha_{1} \cdot w_k = 2b_{1} - b_{2} = h_{1} \geq 0 \\
& \text{But if} \ b_1 \leq b_2 < 0 \Longrightarrow 2b_1 - b_2 < 0~.
\end{split}
\end{equation}
Similarly, a contradiction occurs if it is assumed that $b_{k+1} \leq b_k$. Now, if any other $b_i < 0$ for $i \neq k$, similarly it can be argued that either $b_{i+1} \leq b_i$ or $b_{i-1} \leq b_{i}$ and concluded that $2b_1 - b_2 < 0$ or $2b_{N-1} - b_{N-2} < 0$ or $b_k \leq b_i < 0$, which would lead to contradictions similar to above. 
\end{proof}

The next step of the proof is to show that given a highest weight $w_k$ of a representation with $N$-ality $k$, it is always possible to lower with the simple roots to obtain $\mu_k$. Given a weight $\mu$ of a representation  of $SU(N)$, the master formula in \cite{09} can be applied:
\begin{equation}
\label{eq:B91}
{2\mu\cdot \alpha_i \over \alpha^2_i} = \mu\cdot \alpha_i = -(p_i - q_i)
\end{equation}
Where $p_i \in \mathbb{Z} \ \text{and} \ p_i \geq 0$ is the number of times which we can raise $\mu$ with $\alpha_i$ and $q_i \in \mathbb{Z} \ \text{and} \ q_i \geq 0$ is the number of times which we can lower $\mu$ with $\alpha_i$. Based on the above \textbf{Lemma}, we have $w_k = \mu_k + a_i \alpha_i$. If $a_i = 0$ for all $i$ then the representation contains $\mu_k$ as one of its weights but if at least one is greater than zero then we will show that for some $\alpha_i$ which $a_i > 0$, $w_k \cdot \alpha_i > 0$ which would imply that $q_i > 0$ and hence $w_k$ can be lowered with some $\alpha_i$ which $a_i > 0$. Let $a_j = \text{Max} \{ a_i | 1 \leq i \leq N-1 \} $ for some $1 \leq j \leq N-1$. Since we assumed at least one $a_i$ is greater than zero then $a_j > 0$. If $j = 1$, $j = N-1$ or $j = k$ then $w_k \cdot \alpha_j = 2a_1 - a_2 > 0$, $w_k \cdot \alpha_j = 2a_{N-1} - a_{N-2} > 0$ or $w_k \cdot \alpha_j = 2a_k +1  - a_{k-1} - a_{k+1} > 0$ respectively, since we assumed that $a_j > 0$ and is the maximum among others. Otherwise if $a_j > a_{j+1}$ or $a_j > a_{j-1}$ then $w_k \cdot \alpha_j = 2a_j - a_{j+1} - a_{j-1} > 0$. But if $a_j = a_{j+1} = a_{j-1}$ then $w_k \cdot \alpha_j = 0$. In this case $a_{j-1}$ and $a_{j+1}$ are greater than zero and both maximum among other $a_i$. Hence we can repeat what we did for $a_j$ for $a_{j+1}$ or $a_{j-1}$ for a number of steps until $j+r$ in $a_{j+r}$ for an $r \neq 0$ becomes $j+r = 1$, $j+r = N-1$ or $j+r = k$ or we would have $a_{j+r} > a_{j+r+1}$ or $a_{j+r} > a_{j+r-1}$ which in that case $w_k$ can be lowered with $\alpha_{j+r}$. So we proved that if at least one $a_i$ is greater than zero then $w_k$ can always be lowered with some $\alpha_h$ which $a_h > 0$. Hence we continue this process until all the $\alpha_i$ are removed from the highest weight $w_k = \mu_k + a_i \alpha_i$ and we reach $\mu_k$. This shows that any representation of $SU(N)$ with $N$-ality $k$ contains $\mu_k$ as one of its weights.

\subsection{Derivation of \eqref{eq:2.46}}
\label{sec:B2}

To derive \eqref{eq:2.46}, we will work out the steps for the contribution of one monopole with magnetic charge $q^1_m$. Consider the long distance behaviour of \eqref{eq:2.46} for generators in the fundamental representation of the gauge group, located at   $R \in \R^3$:
\begin{equation}
\label{eq:B11}
\begin{split}
\{ \text{tr} \ \text{exp}( i \oint_{R\times T} A^c_m t_F^c d\text{x}^m) \}_{\text{1 mon.}} & = \text{tr} \ \text{exp} (i \int_{S(R\times T)}\epsilon_{anm} \partial_n A^c_m t_F^c d\text{S}^a ) \\ & = \text{tr} \ \text{exp}( i \int_{S(R\times T)} d\text{x}_1d\text{x}_2\; {\text{R}_3 \over 2 |R - \text{x}|^{3}} \;Q^1)~.
\end{split}
\end{equation}
Here $c = 1, ..., N-1$ labels the abelian generators of $SU(N)$,  $Q^1 = \text{diag}(1,-1,0, ...,0)$, and we substituted the magnetic field of a monopole, \eqref{eq:2.21} converted to Cartesian coordinates. Next, we first write $Q^1$ as a linear combination of the abelian generators of the fundamental representation $Q^1 = \bar{V}_i t^i_F = \text{diag}(\bar{V}\cdot \bar{\mu}^1_F, ...,\bar{V}\cdot \bar{\mu}^N_F)$ for $i=1, ...,N-1$; here $\bar{\mu}^j_F$ for $j = 1, ...,N$ are the $(N-1)$-dimensional weight vectors of the fundamental representation of $SU(N)$ and $\bar{V}$ is an $(N-1)$-dimensional vector. In order to transform this to an arbitrary representation $r$ we replace $t^c_F$ by its corresponding generator in the representation $r$ and write $Q^1_r  {\equiv} \bar{V}_i t^i_r = \text{diag}(\bar{V}\cdot \bar{\mu}^1_r, ...,\bar{V}\cdot\bar{\mu}^{d(r)}_r)$. In order to write this in the $N$-dimensional form of weights used in this work \eqref{eq:B1}, we note that the weight vectors of the fundamental representation of $SU(N)$ in their $N$-dimensional form are:
\begin{equation}
\label{eq:B12}
\mu^j_{F} =(-{1 \over N}, ..., -{1 \over N}, \overset {\text{j-th}} {\widehat{1- {1 \over N}}}, - {1 \over N}, ..., - {1 \over N}),\ \ \ \ \ \ \ \ \ \ j = 1, ..., N~,
\end{equation}
which can be easily verified by lowering the fundamental weight $\mu_1$ in \eqref{eq:B1} with the simple roots. From \eqref{eq:B12}, the $N$-dimensional form of $\bar{V}$, named $V$, can be determined by requiring: $Q^1 = \text{diag}(\bar{V}\cdot\bar{\mu}^1_F, ...,\bar{V}\cdot\bar{\mu}^N_F) = \text{diag}(V\cdot \mu^1_F, ...,V\cdot\mu^N_F)$, which gives $V = q^1_m = (1,-1,0, ...,0)$. Hence $Q^1_r$, using the $N$-dimensional form of weights, becomes $Q^1_r = \text{diag}(V\cdot\mu^1_r, ...,V\cdot \mu^{d(r)}_r) = \text{diag}(q^1_m\cdot\mu^1_r, ...,q^1_m\cdot\mu^{d(r)}_r)$. Therefore \eqref{eq:B11}, for generators in an arbitrary representation $r$, becomes:
\begin{equation}
\label{eq:B13}
\begin{split}
\{ \text{tr} \ \text{exp}( i \oint_{R\times T} A^c_m t_r^c d\text{x}^m) \}_{\text{1 mon.}} & = \text{tr} \ \text{exp} (i \int_{S(R\times T)}\epsilon_{anm} \partial_n A^c_m t_r^c d\text{S}^a ) \\ & = \overset{d(r)}{\underset{j = 1}{\sum}} \ \text{exp}( i \int_{S(R\times T)} d\text{x}_1d\text{x}_2 \; \mu^j_r \cdot q^1_m\; {\text{R}_3 \over 2 |R - \text{x}|^{3}})~.
\end{split}
\end{equation}

\section{Perturbative saddle point $\mathbf{k}$-strings: leading order $\mathbf{+}$ leading correction}
\label{sec:C}
The following tables \ref{table:7} - \ref{table:15} compare the values of $k$-strings obtained from a perturbative saddle point calculation to their numerical values in Table \ref{table:1}. The ''Leading'' and ''Leading Corr.'' column give values for the coefficient of $- {RT \over \beta }$ in \eqref{eq:4.39} and \eqref{eq:4.43} respectively. If $T$ represents a half $k$-string value in Table \ref{table:1} and $T'$ its corresponding one in the ''Num. value'' column, they are related by:
\begin{equation}
T' = \text{ROUND}(\bar{T},3) \ \ \ \text{with} \ \ \  \bar{T} \equiv \text{ROUNDDOWN}(T,3)\times2/ \sqrt{2}
\end{equation}

We multiply the half $k$-strings by $2$ to obtain the full $k$-string then we divide it by $\sqrt{2}$ to normalize them similar to the perturbative saddle point method $k$-strings as in \eqref{eq:4.32} and \eqref{eq:4.33}. There is a high chance that the numerical half $k$-strings in Table \ref{table:1} will match the exact half $k$-strings rounded to the third decimal if they are rounded down to the 3rd decimal. This is due to the fact that the true value of half $k$-strings always lies below the values obtained in Table \ref{table:1} and the true value of the error is of order $0.001$ or less as can be seen from Table \ref{table:2}.
\\
\begin{table}[h]
\centering
\begin{tabular}{|c|c|c|c|c|c|}
\hline
$SU(N)$& Leading & Leading Corr. & Sum(Lead.+Lead. Corr.) & Num. value & Num. value - Sum\\
\hline 
2 & 9.870 & -2.029 & 7.841 & 8.000 & 0.159 \\
\hline 
3 & 11.396 & -2.343 & 9.053 & 9.238 & 0.185 \\
\hline 
4 & 11.913 & -2.396 & 9.517 & 9.699 & 0.182 \\
\hline
5 & 12.150 & -2.410 & 9.740 & 9.919 & 0.179 \\
\hline
6 & 12.277 & -2.415 & 9.862 & 10.041 & 0.179 \\
\hline
7 & 12.355 & -2.417 & 9.938 & 10.114 & 0.176 \\
\hline
8 & 12.405 & -2.418 & 9.987 & 10.163 & 0.176 \\
\hline
9 & 12.439 & -2.418 & 10.021 & 10.196 & 0.175 \\
\hline
10 & 12.463 & -2.418 & 10.045 & 10.221 & 0.176 \\
\hline
\end{tabular}
\caption{Comparison of $N$-ality $1$ $k$-strings for $SU(2 \leq N \leq 10)$}
\label{table:7}
\end{table}

\begin{table}[h]
\centering
\begin{tabular}{|c|c|c|c|c|c|}
\hline
$SU(N)$& Leading & Leading Corr. & Sum(Lead.+Lead. Corr.) & Num. value & Num. value - Sum\\
\hline 
3 & 11.396 & -2.343 & 9.053 & 9.238 & 0.185 \\
\hline 
4 & 13.958 & -2.870 & 11.088 & 11.314 & 0.226  \\
\hline
5 & 15.018 & -2.99 & 12.028 & 12.247 & 0.219 \\
\hline
6 & 15.568 & -3.029 & 12.539 & 12.751 & 0.212 \\
\hline
7 & 15.891 & -3.045 & 12.846 & 13.055 & 0.209 \\
\hline
8 & 16.097 & -3.052 & 13.045 & 13.253 & 0.208 \\
\hline
9 & 16.237 & -3.056 & 13.181 & 13.388 & 0.207 \\
\hline
10 & 16.337 & -3.058 & 13.279 & 13.485 & 0.206 \\
\hline
\end{tabular}
\caption{Comparison of $N$-ality $2$ $k$-strings for $SU(3 \leq N \leq 10)$ }
\label{table:8}
\end{table}

\begin{table}[h]
\centering
\begin{tabular}{|c|c|c|c|c|c|}
\hline
$SU(N)$& Leading & Leading Corr. & Sum(Lead.+Lead. Corr.) & Num. value & Num. value - Sum\\
\hline 
4 & 11.913 & -2.396 & 9.517 & 9.699 & 0.182 \\
\hline
5 & 15.018 & -2.990 & 12.028 & 12.247 & 0.219 \\
\hline
6 & 16.449 & -3.142 & 13.307 & 13.511 & 0.204 \\
\hline
7 & 17.248 & -3.197 & 14.051 & 14.247 & 0.196 \\
\hline
8 & 17.746 & -3.222 & 14.524 & 14.715 & 0.191 \\
\hline
9 & 18.078 & -3.234 & 14.844 & 15.032 & 0.188 \\
\hline
10 & 18.311 & -3.241 & 15.070 & 15.257 & 0.187 \\
\hline

\end{tabular}
\caption{Comparison of $N$-ality $3$ $k$-strings for $SU(4 \leq N \leq 10)$  }
\label{table:9}
\end{table}

\begin{table}[h]
\centering
\begin{tabular}{|c|c|c|c|c|c|}
\hline
$SU(N)$& Leading & Leading Corr. & Sum(Lead.+Lead. Corr.) & Num. value & Num. value - Sum\\
\hline
5 & 12.15 & -2.410   &   9.740 & 9.919 &0.179 \\
\hline
6 & 15.568 & -3.029 & 12.539 & 12.751 & 0.212 \\
\hline
7 & 17.248 & -3.197 & 14.051 & 14.247 & 0.196 \\
\hline
8 & 18.237 & -3.262 & 14.975 & 15.159 & 0.184 \\
\hline
9 & 18.876 & -3.292 & 15.584 & 15.761 & 0.177 \\
\hline
10 & 19.317 & -3.307 & 16.010 & 16.183 & 0.173 \\
\hline
\end{tabular}
\caption{Comparison of $N$-ality $4$ $k$-strings for $SU(5 \leq N \leq 10)$  }
\label{table:10}
\end{table}

\begin{table}[h]
\centering
\begin{tabular}{|c|c|c|c|c|c|}
\hline
$SU(N)$& Leading & Leading Corr. & Sum(Lead.+Lead. Corr.) & Num. value & Num. value - Sum\\
\hline
6 & 12.277 & -2.415 & 9.862 & 10.041 & 0.179 \\
\hline
7 & 15.891 & -3.045 &12.846 & 13.055 & 0.209 \\ 
\hline
8 & 17.746 & -3.222 & 14.524 & 14.715 & 0.191 \\
\hline
9 & 18.876 & -3.292 & 15.584 & 15.761 & 0.177 \\
\hline
10 & 19.629 & -3.326 &16.303 & 16.474 & 0.171 \\
\hline

\end{tabular}
\caption{Comparison of $N$-ality $5$ $k$-strings for $SU(6 \leq N \leq 10)$  }
\label{table:11}
\end{table}

\begin{table}[h]
\centering
\begin{tabular}{|c|c|c|c|c|c|}
\hline
$SU(N)$& Leading & Leading Corr. & Sum(Lead.+Lead. Corr.) & Num. value & Num. value - Sum\\
\hline
7 & 12.355 & -2.417 & 9.930  & 10.114 & 0.176 \\
\hline
8 & 16.097 & -3.052 & 13.045 & 13.253 & 0.208 \\
\hline
9 & 18.078 & -3.234 & 14.844 & 15.032 & 0.188 \\
\hline
10 & 19.317 & -3.307 & 16.010 & 16.183 & 0.173 \\
\hline

\end{tabular}
\caption{Comparison of $N$-ality $6$ $k$-strings for $SU(7 \leq N \leq 10)$ }
\label{table:12}
\end{table}

\begin{table}[h]
\centering
\begin{tabular}{|c|c|c|c|c|c|}
\hline
$SU(N)$& Leading & Leading Corr. & Sum(Lead.+Lead. Corr.) & Num. value & Num. value - Sum\\
\hline
8 & 12.405 & -2.418 & 9.987 & 10.163 & 0.176 \\
\hline
9 & 16.237 & -3.056 & 13.181 & 13.388 & 0.207 \\
\hline
10 & 18.311 & -3.241 & 15.070 & 15.257  & 0.187 \\
\hline

\end{tabular}
\caption{Comparison of $N$-ality $7$ $k$-strings for $SU(8 \leq N \leq 10)$ }
\label{table:13}
\end{table}

\begin{table}[h]
\centering
\begin{tabular}{|c|c|c|c|c|c|}
\hline
$SU(N)$& Leading & Leading Corr. & Sum(Lead.+Lead. Corr.) & Num. value & Num. value - Sum\\
\hline
9 & 12.439 & -2.418 & 10.021 & 10.196 & 0.175 \\
\hline
10 & 16.337 & -3.058 & 13.279 & 13.485 & 0.206 \\
\hline

\end{tabular}
\caption{Comparison of $N$-ality $8$ $k$-strings for $SU(9 \leq N \leq 10)$ }
\label{table:14}
\end{table}

\begin{table}[h]
\centering
\begin{tabular}{|c|c|c|c|c|c|}
\hline
$SU(N)$& Leading & Leading Corr. & Sum(Lead.+Lead. Corr.) & Num. value & Num. value - Sum\\
\hline
10 & 12.463 & -2.418 & 10.045 & 10.221 & 0.176 \\
\hline

\end{tabular}
\caption{Comparison of $N$-ality $9$ $k$-strings for $SU(10)$ }
\label{table:15}
\end{table}.
\\
\\
\\
\\
\\
\\
\\  
 \section{Large-$N$ limit of string tensions for product representations: a saddle point leading-order perturbative evaluation}
 \label{sec:appxproduct}
 
 Our starting point is \eqref{eq:4.39}. Recall that this equation gives the contribution to the expectation value of the Wilson loop of  quarks of charges (weight) $\mu$, evaluated to leading order using the perturbative saddle point method.  For convenience, we now reproduce the area-law part of that equation ($\hat{R}, \hat{T} \rightarrow \infty, \beta \rightarrow 0$):
 \begin{equation}
\label{eq:4.391}
{\{ Z^{\eta}_{g^4} \}_{\lambda =0} \over \{ Z^{\eta}_{g^4} \}_{b_q=\lambda =0} } = \text{exp}(- {1 \over \beta} \{ {1 \over 4} {\sum\limits_{q=1}^{N-1}\sqrt{\Lambda_q}b^2_q} \hat{R} \hat{T} \} )~.
\end{equation}
Here,   $b_q \equiv 2\pi (\mu)_j D_{jq}$ for a representation of weight $\mu$; $(\mu)_j$ denotes the $j$-th component, $j=1,...N$, of the weight vector. Recall also that $\Lambda_q = 4 \sin^2 {\pi q \over N}$ is the dimensionless mass of the $q$-th dual photon and that 
the components of the matrix $D_{jq}$, $1 \le j \le N$, are $D_{jq} = \sqrt{2 \over N} \sin {2 \pi q j \over N} $, for $1 \le q < {N\over 2}$ and  $D_{jq} = \sqrt{2 \over N} \cos {2 \pi q j \over N} $, for ${N\over 2} < q < {N}$; for brevity, we only give the values for odd $N$. The $q=N$ component of $D_{jq}$ does not contribute to \eqref{eq:4.391}.
 
 The main difference compared to the discussion in the main text is that we now consider also weights corresponding to product representations, for concreteness the ${\square \otimes \square}$ representation. 
 Recall, from \eqref{eq:abd}, that the expectation value of the Wilson loop in the product representation is given, in the abelianized regime of this paper, by a sum of exponentials, one for each weight of the product representation:
  \begin{equation}\label{eq:abd1}
\langle W_{{\square \otimes \square}}(R,T) \rangle \sim  \sum^{d({\square \otimes \square})}_{h = 1} \text{exp}(- T^h_{{\square \otimes \square}}RT) = \sum^{d({\square \otimes \square})}_{h = 1} \text{exp}(- {1 \over \beta} \hat{T}^h_{{\square \otimes \square}}\hat{R}\hat{T})   ~.\end{equation}
Where we have also written it in its dimensionless form (recall the relations $R = \hat{R} / m_{\gamma}$, $T = \hat{T} / m_{\gamma}$, $\beta = m_{\gamma}^3/ \tilde{\zeta}$ from the comment below \eqref{eq:4.32}). Comparing with \eqref{eq:4.391} $\hat{T}^h_{{\square \otimes \square}}$ to leading order (l.o.) is given by: 
\begin{equation}\label{eq:apxe1}
 \hat{T}^h_{{\square \otimes \square},\text{l.o.}}= {1 \over 4} {\sum\limits_{q=1}^{N-1}\sqrt{\Lambda_q}b^2_q}~, ~~ b_q \equiv 2\pi \sum\limits_{j=1}^N (\mu^h)_j D_{jq} ~,
\end{equation}
with $\mu^h$---the $h$-th weight of the ${\square \otimes \square}$ representation. 

The goal of this Appendix is to evaluate \eqref{eq:apxe1} for all weights of the ${\square \otimes \square}$ product representation, to leading order in the analytic perturbative saddle point method and in the large-$N$ limit. We shall see that the leading-order analytic considerations  support the findings discussed qualitatively after around Eqns.~(\ref{acd}) and \eqref{eq:abd} of the main text of the behaviour of the product-representation Wilson loop at large $N$. 

We begin by noting that the weights of the ${\square \otimes \square}$ representation are labeled by two integers $a, b=1,...N$ (there are $N^2$ weights) and are given by 
\begin{equation}\label{eq:weight}
(\mu^h)_j \rightarrow (\mu^{ab})_j = \delta^a_j + \delta^b_j -{2 \over N} \approx \delta^a_j + \delta^b_j . 
\end{equation}
The last equality is valid for sufficiently large $N$. From \eqref{eq:apxe1}, recalling that we consider odd-$N$, we find an explicit expression for the tension of strings sourced by quarks with weight $\mu^{ab}$,  \eqref{eq:weight}, of the product representation at leading order:\footnote{To obtain \eqref{eq:E.6}, we noted that for large and odd $N$,  for $1 \le q < {N\over 2}$:
\begin{equation}
\label{eq:E.6}
\begin{split}
{b_q \over 2\pi} = \sum\limits_{j=1}^N (\mu^h)_j D_{jq}   = \sqrt{{2 \over N}} \{ \sin{2 \pi a q \over N} + \sin{2 \pi b q \over N} - \sum\limits_{j=1}^N {2 \over N} \sin{2 \pi j q \over N} \}   \approx \sqrt{{2 \over N}} \{ \sin{2 \pi a q \over N} + \sin{2 \pi b q \over N} + 0 \},
\end{split}
\end{equation}
 and used the fact that the last term in the first line of \eqref{eq:E.6} for large $N$ can be approximated by $\approx -{1 \over \pi}\int_0^{2 \pi}d y  \sin q y = 0$. Also, a similar expression can be written for ${N\over 2} < q < N$.}
\begin{flalign}
\label{eq:appxe2}
\hat{T}^{ab}_{{\square \otimes \square},\text{l.o.}} = 4 \pi \left[ \sum\limits_{q=1}^{\left[{N\over 2}\right] } {\pi \over N} \sin {\pi q \over N} \left(\sin{2 \pi a q \over N} + \sin{2 \pi b q \over N}\right)^2 + \sum\limits_{q=\left[{N\over 2}\right]+1}^{N-1} {\pi \over N} \sin {\pi q \over N} \left(\cos{2 \pi a q \over N} + \cos{2 \pi b q \over N}\right)^2\right] &&
\end{flalign}
The sum in \eqref{eq:appxe2} can be evaluated exactly for arbitrary $N$, but to illustrate our point it suffices to consider {\it i.}) the results of a numerical evaluation and {\it ii.}) the evaluation of \eqref{eq:appxe2} at infinite $N$ by replacing the sum by an integral.

\begin{figure}[h]
\centering
\begin{minipage}{0.8\textwidth}
	\includegraphics[width=\textwidth]{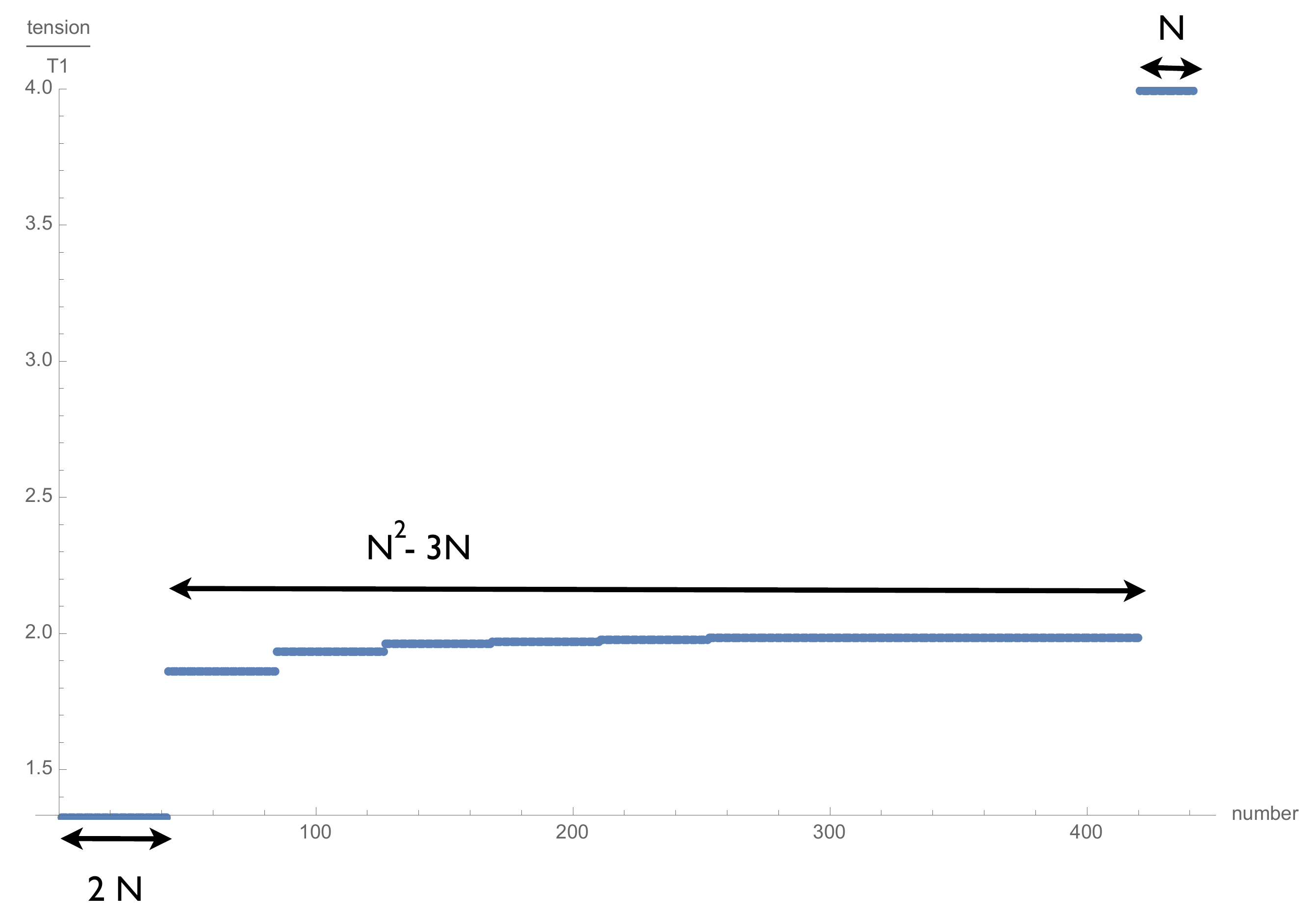}
	\caption{The  product  ${\square \otimes \square}$-representation string tensions $\hat{T}^{ab}_{{\square \otimes \square}}$, evaluated for $N=21$. They take values between $1.32$  and $3.99$ times the fundamental string tension. As the plot shows, $42 = O(N)$ string tensions take the minimal value (these correspond to $a = b \pm 1$(mod $N$)), $21 = O(N)$ tensions approximately equal  four times the fundamental string tension (corresponding to $a=b$) and the rest of the string tensions ($378 = O(N^2)$) are slightly less than twice the fundamental string tension }{\label{fig:productstrings}}
\end{minipage}
\end{figure}

We begin with a discussion of the numerical results for the $N^2$ product representation string tensions shown on Figure \ref{fig:productstrings}. The $N^2$ string tensions for $N=21$ are evaluated numerically using \eqref{eq:appxe2}. As the plot shows, most of the $N^2$ string tensions are of order $2 T_{1,\text{l.o.}}$, while 
 $2N$ of them are approximately equal to the minimal value $T_{2,\text{l.o.}}$, and $N$ are equal to approximately $4 T_{1,\text{l.o.}}$. 
Clearly, this is conforming to the discussion in the main text,  Section \ref{sec:5.2.2}..

 We can also evaluate \eqref{eq:appxe2} in the infinite-$N$ limit by replacing the sum by an integral, for $a,b$ fixed, i.e.,
\begin{flalign}
\label{eq:appxe3}
{\hat{T}^{ab}_{{\square \otimes \square},\text{l.o.}} \over 4 \pi} = \int\limits_{0}^{{\pi \over 2}} d x \sin x \left(\sin{2 x a } + \sin{2   b x}\right)^2 +  \int\limits_{{\pi \over 2}}^\pi d x \sin x   \left(\cos{2  a x} + \cos{2   b   x}\right)^2  ~.
\end{flalign}
Thus, the product representation string tensions, normalized to the fundamental string tension (equal to $4\pi$ in the leading saddle point approximation), becomes in the large-N limit
\begin{equation}
\label{eq:E.7}
{\hat{T}^{ab}_{{\square \otimes \square},\text{l.o.}}\over 4 \pi} = 2 - {2\over   4 (a - b)^2-1}~.
\end{equation}
 Due to the $Z_N$ symmetry of the string tension action we expect to obtain the same tensions for $|a-b| = n$ and $|a-b| =N - n$ for $1 \leq n \leq [N/2]$. This symmetry is lost in \eqref{eq:E.7} due to the infinite $N$ limit, therefore it is best to use this relation for $|a-b| \leq [N/2]$ only at large N and for $|a-b| > [N/2]$ make the replacement $|a-b| \rightarrow N - |a-b|$ in \eqref{eq:E.7}. In the limit $|a-b|\gg 1$, this relation approaches the value of $2$, while for $|a-b|=1$, we obtain the value ${4\over 3} \approx 1.33$; the value of $4$ for $a=b$ is also obtained. This distribution of the $N^2$ string tensions in the infinite-$N$ limit is consistent with the numerical result shown for $N=21$ and with the general discussion of  Section \ref{sec:5.2.2}.

At the end, we also acknowledge  an additional subtlety one might be worried about.  The calculation that led to eqn.~\eqref{eq:appxe2}---see \eqref{eq:4.39} as well as eqns.~(\ref{eq:4.7}--\ref{eq:4.14}) which directly lead to it---assumes that $RT$ is   larger than the inverse mass squared of all dual photons, including the lightest one. Thus, strictly speaking one expects \eqref{eq:appxe2} to pertain to the order of limits we advocated for here: infinite area at fixed N, followed by $N \rightarrow \infty$ which is the proper order of limits necessary for calculating k-strings at large $N$.

However for the discussion of large $N$ factorization in gauge theories the large $N$ limit is taken first. Now, if $RT$ is smaller than the mass of some dual photons, the area law due to these photons should be replaced with a perimeter law contribution. This remark is relevant because if the large-N limit is taken first, the masses of some dual photons vanish---recall that their masses are scale as $\sqrt{\Lambda_q} =2 \sin {\pi q \over N}$---and these dual photons do not lead to an area law. To take this into account, consider the integral \eqref{eq:appxe3} and omit contributions of dual photons of (dimensionless) mass $  2 \sin {\pi q\over N} = 2 \sin x < {1\over \sqrt{\hat R \hat T}}$, as they do not give rise to area law. Thus, the region of integration in \eqref{eq:appxe3}, instead of $(0, {\pi \over 2})$ and $({\pi\over 2}, \pi)$, should be replaced by, respectively,  $(\epsilon, {\pi \over 2})$ and $({\pi\over 2}, \pi-\epsilon)$, with $\epsilon \sim {1\over \sqrt{\hat{R} \hat{T}}}$. In the further large $\hat{R}\hat{T}$ limit, we have that $\epsilon \rightarrow 0$, showing that the contributions to the string tension of dual photons of mass vanishing at large-$N$ is negligible. Thus, we expect that if the order of limits is taken as described now ($N$ to infinity first, large area next), the factorization result analyzed above in terms of string tensions is recovered.

The discussion of large-$N$ factorization above and in the 2nd half of  Section \ref{sec:5.2.2} was carried out in terms of string tensions, since its more explicit and intuitive and allows for a qualitative analysis of large-$N$ factorization in terms of the full saddle point as was done in  Section \ref{sec:5.2.2}. For this analysis, the large area limit had to be taken to isolate the area law contribution and find the string tensions, as done in the previous paragraph. However one can show the large-$N$ factorization result in a more general and abstract setting without the need to refer to any large area limits or expressions for string tensions. Consider eq.~\eqref{eq:4.39'}, which is a general expression for the saddle point at leading order, without  reference to any large area limit:
\begin{equation}
\label{eq:E9}
{\{ Z^{\eta}_{g^4} \}_{\lambda =0} \over \{ Z^{\eta}_{g^4} \}_{b_q=\lambda =0} } = \text{exp}(- {1 \over \beta} \{ {\Lambda_q b^2_q \over 2} \underset{\text{A} \ \text{A}}{\iint} d^2 \text{x} d^2 \text{x}' P_q(\text{x} - \text{x}') + {b_q^2 \over 2} \underset{b(\text{A}) \  b(\text{A})}{\iint'} d\text{x}^l d\text{x}'^k \delta^{l k} ( P_q(\text{x}-\text{x}') - {1 \over 4 \pi |\text{x} -\text{x}'|}) \} )~,
\end{equation}
Using \eqref{eq:E.6} and noting that the integrals in \eqref{eq:E9} are finite quantities and a function of $\hat{R}, \hat{T}$ and $\sqrt{\Lambda_q} =2 \sin {\pi q \over N}$ with $x \equiv {\pi q \over N}$, the large $N$ limit of the leading saddle point (s.p.) result in \eqref{eq:E9} reduces to:
\begin{equation}
\label{eq:E10}
\hspace{-0.5cm} \text{s.p.}_{\square \otimes \square,\text{l.o.}} = \int^{\pi /2}_0 dx ( \sin{2 a x} + \sin{2 b x} )^2 F_{\hat{R}, \hat{T}}(\sin x) + \int^{\pi}_{\pi /2} dx ( \cos{2 a x} + \cos{2 b x} )^2 F_{\hat{R}, \hat{T}}(\sin x)~,
\end{equation}
where $F_{\hat{R},\hat{T}}(\sin {\pi q \over N})$ is given by:
\begin{equation}
\label{eq:E11}
\hspace{-1mm} F_{\hat{R}, \hat{T}}(\sin {\pi q \over N}) \equiv 4 \pi {\Lambda_q} \underset{\text{A} \ \text{A}}{\iint} d^2 \text{x} d^2 \text{x}' P_q(\text{x} - \text{x}') + 4 \pi \underset{b(\text{A}) \  b(\text{A})}{\iint'} d\text{x}^l d\text{x}'^k \delta^{l k} ( P_q(\text{x}-\text{x}') - {1 \over 4 \pi |\text{x} -\text{x}'|})
\end{equation}

The  expression corresponding to \eqref{eq:E10} in the fundamental representation of $SU(N)$ is:
\begin{equation}
\label{eq:E12}
\hspace{-0.5cm} \text{s.p.}_{\square,\text{l.o.}} = \int^{\pi /2}_0 dx ( \sin{2 a x})^2 F_{\hat{R}, \hat{T}}(\sin x) + \int^{\pi}_{\pi /2} dx ( \cos{2 a x})^2 F_{\hat{R}, \hat{T}}(\sin x)
\end{equation}
Making the change of variable $x \rightarrow \pi - x$ in the second integral of \eqref{eq:E10} and \eqref{eq:E12}, they can be simplified to:
\begin{flalign}
\label{eq:E13}
&\text{s.p.}_{\square \otimes \square,\text{l.o.}} = 2\int^{\pi /2}_0 dx F_{\hat{R}, \hat{T}}(\sin x) + 2\int^{\pi/2}_{0} dx \cos (2a x - 2b x) \; F_{\hat{R}, \hat{T}}(\sin x)~, \\ \label{eq:E14} & \text{s.p.}_{\square,\text{l.o.}} = \int^{\pi /2}_0 dx F_{\hat{R}, \hat{T}}(\sin x)~.
\end{flalign}
When $|a-b| \gg 1$ (and $|a-b| \ll N$ if the discrete form of \eqref{eq:E10} is considered for a finite but large $N$), the second integral in \eqref{eq:E13}, due to the rapid oscillations of $\cos (2a x - 2b x)$, is near zero therefore $O(N^2)$ weights of the product representation give approximately twice the value of the fundamental representation string tension in the leading saddle point approximation from \eqref{eq:E14}, which has the same value for all weights of the fundamental representation. Therefore relations \eqref{eq:E13} and \eqref{eq:E14} clearly show the large $N$ factorization result in dYM theory at leading order of the saddle point without any reference to a large area limit.

Although the calculations in this Appendix were done at the leading order saddle point level the same ideas and methods can be applied to show large $N$ factorization regarding the corrections (as in \eqref{eq:4.40}) to these leading order saddle point results.

\acknowledgments

We thank Mohamed Anber and Aleksey Cherman for comments on the manuscript and Adi Armoni for many enlightening discussions of k-strings. We acknowledge support by a NSERC Discovery Grant.
The computations for this paper were performed on the gpc supercomputer at the SciNet HPC Consortium. SciNet is funded by: the Canada Foundation for Innovation under the auspices of Compute Canada; the Government of Ontario; Ontario Research Fund - Research Excellence; and the University of Toronto.

\end{document}